\documentclass[useAMS,usenatbib]{mn2e}

\let\jnlstyle=\rm
\def\refjnl#1{{\jnlstyle#1}}

\def\apjl{\refjnl{ApJ}}                
\def\apjs{\refjnl{ApJS}}               
\def\ao{\refjnl{Appl.~Opt.}}           
\def\aap{\refjnl{A\&A}}                

\input{psfig.sty}

\usepackage{graphicx}
\usepackage{epsfig}
\usepackage{amssymb}
\usepackage{amsmath}
\usepackage{amsfonts}
\usepackage{txfonts}
\usepackage{multirow}
\usepackage{afterpage}
\usepackage{xcolor}

\definecolor{mypink1}{rgb}{0.858, 0.188, 0.478}
\definecolor{mygreen1}{rgb}{0.258, 0.788, 0.878}
\definecolor{myorange1}{rgb}{0.5, 0.2, 0.2}

\title[DES-Y1 Cosmology Beyond 2-Point Statistics]{Cosmic Shear Cosmology Beyond 2-Point Statistics:  A  Combined Peak Count and Correlation Function Analysis of DES-Y1}

\author[J. Harnois-D\'{e}raps, N. Martinet et al.]{Joachim Harnois-D\'{e}raps$^{1,2,3}$\thanks{E-mail: joachim.harnois-deraps@ncl.ac.uk}, Nicolas Martinet$^4$, Tiago Castro$^{5,6,7,8}$, Klaus Dolag$^{9,10}$, \newauthor 
Benjamin Giblin$^2$, Catherine Heymans $^{2,11}$, Hendrik Hildebrandt$^{11}$ \& Qianli Xia$^2$
\\
$^{1}$Astrophysics Research Institute, Liverpool John Moores University, 146 Brownlow Hill, Liverpool, L3 5RF, UK\\
$^{2}$Scottish Universities Physics Alliance, Institute for Astronomy, University of Edinburgh, Blackford Hill, Scotland, UK\\
$^{3}$School of Mathematics, Statistics and Physics, Newcastle University, Herschel Building, NE1 7RU, Newcastle-upon-Tyne, UK\\
$^{4}$Aix-Marseille Univ, CNRS, CNES, LAM, Marseille, France\\
$^{5}$Dipartimento di Fisica, Sezione di Astronomia, Universit\`a di Trieste, Via Tiepolo 11, I-34143 Trieste, Italy\\
$^{6}$INAF - Osservatorio Astronomico di Trieste, via Tiepolo 11, I-34131 Trieste, Italy\\
$^{7}$IFPU - Institute for Fundamental Physics of the Universe, via Beirut 2, 34151, Trieste, Italy\\
$^{8}$INFN - Sezione di Trieste, I-34100 Trieste, Italy\\
$^{9}$University Observatory Munich, Scheinerstr. 1, 81679 Munich, Germany\\
$^{10}$Max-Planck-Institut fur Astrophysik, Karl-Schwarzschild Strasse 1, 85748 Garching, Germany\\
$^{11}$Ruhr-University Bochum, Faculty of Physics and Astronomy, Astronomical Institute (AIRUB), German Centre for Cosmological Lensing, 44780 Bochum, Germany
}

\date{Accepted June 3$^{\rm rd}$, 2021. Received May 31$^{\rm st}$, 2021; in original form December 8$^{\rm th}$, 2020}

\pubyear{2020}

\begin{document}
\label{firstpage}
\pagerange{\pageref{firstpage}--\pageref{lastpage}}
\maketitle

\begin{abstract}
We constrain cosmological parameters from a joint cosmic shear analysis of peak-counts and the two-point shear correlation functions, as measured from the Dark Energy Survey (DES-Y1).    We find the structure growth parameter $S_8\equiv \sigma_8\sqrt{\Omega_{\rm m}/0.3} = 0.766^{+0.033}_{-0.038}$, which at 4.8\% precision, provides one of the tightest constraints on $S_8$ from the DES-Y1 weak lensing data.  In our simulation-based method we determine the expected DES-Y1 peak-count signal for a range of cosmologies sampled in four $w$CDM parameters ($\Omega_{\rm m}$, $\sigma_8$, $h$, $w_0$).  We also determine the joint covariance matrix with over 1000 realisations at our fiducial cosmology.  With mock DES-Y1 data we calibrate the impact of photometric redshift and shear calibration uncertainty on the peak-count, marginalising over these uncertainties in our cosmological analysis.  Using dedicated training samples we show that our measurements are unaffected by mass resolution limits in the simulation, and that our constraints are robust against uncertainty in the effect of baryon feedback.  Accurate modelling for the impact of intrinsic alignments on the tomographic peak-count remains a challenge, currently limiting our exploitation of cross-correlated peak counts between high and low redshift bins.  We demonstrate that once calibrated, a fully tomographic joint peak-count and correlation functions analysis has the potential to reach a 3\% precision on $S_8$  for DES-Y1.   Our methodology can be adopted to model any statistic that is sensitive to the non-Gaussian information encoded in the shear field.  In order to accelerate the development of these beyond-two-point cosmic shear studies, our simulations are made available to the community upon request.
\end{abstract}




\section{Introduction}
\label{sec:intro}

Over the last decade, weak gravitational lensing has emerged as one of the most promising techniques to investigate the  properties of our Universe on cosmic scales.  Based on the analysis of small distortions between the  shapes of millions of galaxies, weak lensing by large scale structures, or cosmic shear, can directly probe the total projected mass distribution between the observer and the source galaxies, as well as place tight constraints on a number of other cosmological parameters  \citep[for recent reviews of weak lensing as a cosmic probe, see][]{2015RPPh...78h6901K}. Following the success of the Canada-France-Hawaii Telescope Lensing Survey \citep{Heymans2012, 2013MNRAS.433.2545E}, a series of dedicated Stage-III weak lensing experiments, namely the Kilo Degree Survey\footnote{KiDS:kids.strw.leidenuniv.nl}, the Dark Energy Survey\footnote{DES:www.darkenergysurvey.org} and the Hyper Suprime Camera Survey\footnote{HSC:www.naoj.org/Projects/HSC/}, were launched and aimed at constraining properties of dark matter to within a few percent. These are now well advanced or have recently completed their data acquisition, and the community is preparing for the next generation of Stage IV experiments, notably the Rubin observatory\footnote{LSST: www.lsst.org}, and the {\it Euclid}\footnote{{\it Euclid}: sci.esa.int/web/euclid} and Nancy Grace Roman\footnote{WFIRST: roman.gsfc.nasa.gov} space telescopes. 

The central approach adopted by these surveys for constraining cosmology is based on two-point  statistics -- mostly either in the form of correlation functions  \citep[e.g.][]{Kilbinger2013, DESY1_Troxel,  HSCY1_2PCF, KiDS1000_Asgari} or its Fourier equivalent, the power spectrum, estimated using {\it pseudo}-$C_{\ell}$ \citep{HSCY1_Cell}, band powers \citep{2016PhRvD..94b2002B, 2017arXiv170605004V, KiDS1000_Joachimi}  and quadratic estimators \citep{Koehlinger2017}.  By definition, these two-point functions can potentially capture all possible cosmological information contained in a linear, Gaussian density field, and are thus highly efficient at analysing  large scale structure data. 
They have been  thoroughly studied in terms of signal modelling \citep{Kilbinger17}, measurement  \citep{TreeCorr, Schneider2002a, NaMaster} and  systematics  \citep{Mandelbaum18}.

With the improved accuracy and precision provided by current and upcoming surveys, it becomes increasingly appealing to probe small angular scales, where the signal is the strongest. In doing so, the measurements are intrinsically affected by the non-Gaussian nature of the matter density field, and it is natural to seek analysis techniques  that can extract the additional cosmological information that two-point functions fail to capture. A variety of alternative methods have been applied to lensing data with this in mind, including three-point functions \citep{Fu2014},  Minkowski functionals and lensing moments \citep{2015PhRvD..91j3511P}, peak count statistics \citep{2015MNRAS.450.2888L, 2015PhRvD..91f3507L, Kacprzak2016, Martinet18, Shan18},  density split statistics \citep{Gruen2017}, clipping of the shear field \citep{Giblin18}, convolutional  neural networks \citep{Fluri2019}  and neural data compression of lensing map summary statistics \citep{Jeffrey2021}. Other promising techniques are also being developed, notably  
the scattering transform \citep{Scattering}, persistent homology \citep{Homology}, lensing skew-spectrum \citep{Munshi20}, lensing minimas \citep{MinimaPeaks} and moments of the lensing mass maps \citep{VanWaerbeke2013, Gatti20}. 

While existing non-Gaussian data analyses revealed a constraining power comparable to that of the two-point functions, it is expected that the gain will drastically increase with the statistical precision of the data. For example, constraints on  the sum of neutrino mass ($\sum m_{\nu}$), on the matter density ($\Omega_{\rm m}$) and on the amplitude of the primordial power spectrum ($A_{\rm s}$),  in a tomographic peak count analysis of LSST, are forecasted to improve by 40\%, 39\%, and 36\% respectively, compared to a power-spectrum analysis of the same data \citep{MassiveNu1}. Upcoming measurements of the dark energy equation of state ($w_0$) will also benefit from these methods, with a forecasted factor of three improvement expected on the precision  when combining two-point functions with aperture mass map statistics \citep{Martinet20}. 
Similar results are found in the context of a final Stage-III lensing experiment such as the (upcoming) DES-Y6 data release, where the combination of non-Gaussian statistics with the power spectrum method reduces the error on the parameter combination $S_8 \equiv \sigma_8 \sqrt{\Omega_{\rm m}/0.3}$ by about 25\% compared to a two-point function \citep{Zuercher2020a}, where $\sigma_8$ is the normalisation amplitude of the linear matter power spectrum.

In the absence of accurate theoretical predictions for the signal, the covariance, and the impact of systematics, non-Gaussian statistics must be carefully calibrated on numerical simulations specifically tailored to the data being analysed, which are generally expensive to run. Faster approximate methods exist \citep[e.g.][]{ICE-COLA}, however they typically suffer from small scale inaccuracies exactly in the regime where the lensing signal is the strongest, introducing significant biases in the inferred cosmological parameters. Previous peak count analyses of the third KiDS data release (KiDS-450)  \citep[][M18 hereafter]{Martinet18} and of the DES Science Verification data \citep[][K16 hereafter]{Kacprzak2016} calibrated their signal on a suite of full $N$-body simulations spanning the $[\Omega_{\rm m} - \sigma_8]$ plane described in \citet{DH10}. The accuracy of this suite has however been later shown to be only  $\sim$10\% \citep{Giblin18}.  Significant improvements on the simulation side are therefore critical for the new generation of data analyses based on non-Gaussian statistics. 

This paper aims to address  this issue: we present a  cosmological re-analysis of the DES-Y1 cosmic shear data \citep{DESY1_data}, exploiting a novel simulation-based cosmology inference pipeline calibrated on state-of-the-art suites of $N$-body runs that are specifically designed to analyse current weak lensing data beyond two-point statistics. In this work, the incarnation of our pipeline is tailored for the peak count analysis of the DES-Y1 survey, however it is straightforward to extend it to alternative non-Gaussian probes.  Our pipeline first calibrates the cosmological dependence of arbitrary non-Gaussian measurements with the cosmo-SLICS \citep{cosmoSLICS}, a segment of the Scinet LIght-Cone Simulations suite that samples $\Omega_{\rm m}$, $\sigma_8$, $w_0$ and $h$ (the Hubble reduced parameter).  We next estimate the covariance from a suite of fully independent $N$-body runs extracted from the main SLICS sample\footnote{slics.roe.ac.uk} \citep{SLICS}. We further use the cosmo-SLICS to generate systematics-infused control samples that we use to model the impact of photometric redshift and shear calibration uncertainty. We study the impact of galaxy intrinsic alignment with dedicated mock data in which the ellipticities of central galaxies are aligned  (or not) with the shape of their host dark matter haloes, 
following the in-painting prescription of \citet{2006MNRAS.371..750H} \citep[see also][for a more recent application]{2013MNRAS.436..819J}. 
We finally use a suite of high-resolution simulations \citep[SLICS-HR, presented in][]{SLICS_1} to investigate the impact of mass-resolution on the non-Gaussian statistics, and full hydrodynamical simulation  light-cones from the {\it Magneticum Pathfinder}\footnote{www.magneticum.org}  to assess the effect of baryon feedback. All of the above are fully integrated with the {\sc cosmoSIS} cosmological inference pipeline \citep{cosmosis} and therefore interfaces naturally with the two-point statistics likelihood, enabling joint analyses with the fiducial DES-Y1 cosmic shear correlation function measurements presented in \citet[][T18 hereafter]{DESY1_Troxel}, with the $3\times2$pts analysis presented in \citet{2017arXiv170801530D}, or any other analysis implemented within {\sc cosmoSIS}.

 The current  document is structured as follow: In Sec. \ref{sec:data} we present the data and the simulation suites on which our pipeline is built; Sec. \ref{sec:methods} describes the theoretical background, the weak  lensing observables and the analysis methods. A detailed treatment of our systematic uncertainties is presented in Sec. \ref{sec:systematics}, the results of our DES-Y1 data analysis are discussed in Sec. \ref{sec:results},  and we conclude afterwards. The Appendices contain additional validation tests of our simulations and further details on our cosmological inference results.
  
\section{Data and Simulations}
\label{sec:data}

We present in this section the data and simulations included in our analysis.  We exploit multiple state-of-the-art simulation suites in order to conduct our cosmological analysis, including a {\it Cosmology training set} to model the response of our measurement to variations in cosmology, as well as a {\it Covariance   training set} and multiple  {\it Systematics  training sets}. These DES-Y1-specific simulation products are created from four suites of simulations, which we describe after introducing the data. 

The total computing cost of the SLICS, cosmo-SLICS and SLICS-HR are 12.3, 1.1 and 1.3 million {\sc cpu} hours, respectively. They were produced on a system of IBM iDataPlex DX360M2 machines equipped with one or two Intel Xeon E5540 quad cores, running at 2.53GHz with 2GB of RAM per core. Every simulation was split into 64 {\sc mpi} processes, each further parallelised with either four or eight {\sc openmp} threads. Modern compilers and {\sc cpu}s  would likely bring the total computing cost down if similar simulations had to be run again in the future.

\subsection{DES-Y1 data}
\label{subsec:data}

\begin{figure}
\begin{center}
\includegraphics[width=3.3in]{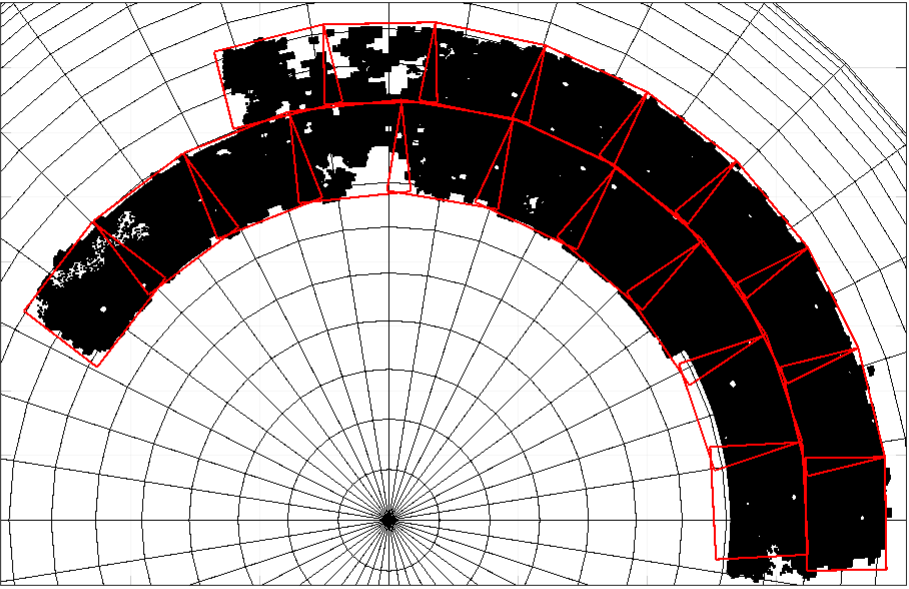}
\caption{Tiling strategy adopted to pave the full DES-Y1 data (black) with flat-sky $10\times10$ deg$^2$ simulations (red squares). 
The squares overlap owing to the sky curvature, hence we separate the data at the mean declination in the overlapping regions. 
In our pipeline, measurements are carried out in each tile separately, then combined at the level of summary statistics.}
\label{fig:DES_tiles}
\end{center}
\end{figure}

\begin{table}
   \centering
   \caption{Survey properties.  The effective number densities $n_{\rm eff}$ [in gal/arcmin$^{2}$] and shape noise $ \sigma_\epsilon$ listed here assume the definition of \citet{LSSTgal}.  The column `$Z_B$ range' refers to the photometric selection that defines the four DES-Y1 tomographic bins, while the mean redshift in each bin is listed under $\langle z_{\rm DIR} \rangle$. }
   \tabcolsep=0.11cm
      \begin{tabular}{@{} cccrrc @{}} 
      \hline
      \hline
      tomo & $Z_B$ range     & No. of objects &   $n_{\rm eff}$ & $ \sigma_\epsilon$  & $\langle z_{\rm DIR} \rangle$  \\
       \hline
       bin1    & $0.20 - 0.43$ &  6,993,471 &       1.45&0.26 & $0.403\pm0.008$ \\
       bin2    & $0.43 - 0.63$ &  7,141,911 &       1.43& 0.29 &$0.560\pm0.014$\\
       bin3    & $0.63 - 0.90$ &   7,514,933&       1.47& 0.26 &$0.773\pm0.011$\\
       bin4    & $0.90 - 1.30$ & 3,839,717  &       0.70& 0.27 & $0.984\pm0.009$\\
    \hline 
    \hline
    \end{tabular}
    \label{table:survey}
\end{table}

In this paper we present cosmological constraints obtained from a re-analysis of the public lensing catalogues of the Year-1 data release\footnote{des.ncsa.illinois.edu/releases/dr1} of the Dark Energy Survey \citep{DESY1_data}. These catalogues were obtained from the analysis of millions of galaxy images taken by the  570 megapixel DECam \citep{DES_CAM} on the Blanco telescope at the Cerro Tololo Inter-American Observatory,  observed in the  $grizY$ bands. The specific selection criteria of the DES-Y1 cosmic shear data used in this paper exactly match  those of the cosmic shear analysis presented in \citet{DESY1_Troxel}: they consist of 26 million galaxies that pass the {\sc flags\_select}, {\sc Metacal} and the {\sc redMaGiC} filters \citep{DESY1_shapes}, thereafter covering a total unmasked area of 1321 deg$^2$, for an object density of $5.07$ gal arcmin$^{-2}$. 
The footprint of the DES-Y1 data is presented in Fig. \ref{fig:DES_tiles}, which shows in black the galaxy positions from the selected sample.

The galaxy shears  in the DES-Y1 data are estimated by two independent methods, {\sc Metacalibration} \citep{METACAL} and {\sc IM3SHAPE} \citep{im3shape} that were both fully implemented \citep[see][for details]{DESY1_shapes}. While they provide consistent results, the former method has a larger acceptance rate of objects with good shape measurements, and thereby 
results in measurements with higher signal-to-noise. Following \citet{DESY1_Troxel}, we also adopt the {\sc Metacalibration} shear estimates  in our analysis. This method provides a {\it shear response} measurement per galaxy, ${\boldsymbol R}_{\gamma}$, a $2\times2$ matrix that must be included to calibrate any measured statistics \citep[we refer to][for more details on this calibration technique in the context of shear two-point correlation functions]{DESY1_shapes}.  Additionally, the galaxy selection itself can  introduce a selection bias, which can be captured by a second $2\times2$ matrix, labelled ${\boldsymbol R}_{S}$ in T18, which we choose not to include  due to the small relative contribution. We compute from these matrices the {\it shear response correction}, defined as $S = {\rm Tr}({\boldsymbol R}_{\gamma})/2$.  As explained in T18, the method imposes a prior on an overall multiplicative shear correction of $m\pm\sigma_m = 0.012 \pm 0.023$, which calibrates the galaxy ellipticities as ${\boldsymbol \epsilon} \rightarrow {\boldsymbol \epsilon} \left(1+m\right)$, with ${\boldsymbol \epsilon} \equiv \epsilon_1 + i \epsilon_2$.

The galaxy sample is further divided into four tomographic redshift bins based on the photometric redshift posterior estimated from $griz$ flux measurements \citep{DESY1_redshifts}. The redshift distribution in these bins, $n^{i}(z)$, must then be estimated, and a number of methods are proposed to achieve this.  The fiducial cosmic shear results presented in T18 are based on the Bayesian photometric redshift ({\sc bpz}) methodology described in \citet{BPZ}, which are consistent with a $n(z)$ estimated by resampling the COSMOS2015 field \citep{COSMOS15} with objects of matched flux and size \citep{DESY1_redshifts}. However, the accuracy of these two methods has been questioned in \citet[][J20 hereafter]{KiDS_DES_Joudaki}, where it is  argued that even though both the {\sc bpz }and COSMOS resampling estimates  account for statistical uncertainty, residual systematics effects could significantly  affect the inferred $n(z)$ distributions.In particular,  the COSMOS sample could be populated with outliers and/or an overall bias that would affect the calibration \citep[e.g. figure 11 of][]{Alarcon20}, and J20 proposes instead to calibrate with redshifts from matched  spectroscopic catalogues\footnote{Both the DIR  and the COSMOS resampling methods have been shown to be consistent with other $n(z)$  estimation techniques such as the cross-correlation between photometric and overlapping  spectroscopic surveys \citep{TheWiZZ, Johnson2017, KV450, Gatti20,DESY1_redshifts}.  J20 also show that the DIR method is robust 
against the specific choice of spectroscopic calibration sample, provided that the combination is sufficiently wide and deep.}. The direct reweighted estimation method \citep[][DIR hereafter]{DIR}  was  selected for the fiducial cosmic shear analysis of the third KiDS data release \citep{KiDS450, KV450}, and  for the DES-Y1 data re-analyses of J20 and \citet{Asgari_DES_KiDS_cosebi}, where it is found that this calibration brings both DES-Y1 and KV450 results in excellent agreement, affecting the constraints on $S_8$ by only $0.8\sigma$. It should be noted also that DIR has inherent systematic uncertainties that are hard to quantify. In particular, incomplete spectroscopy and colour pre-selection \citep{GruenBrimioulle} can potentially bias the DIR $n(z)$. Despite these issues that can in principle be addressed by a pre-selection of sources via the self organising map technique \citep{Wright_KV450_SOM}, we choose to adopt this DIR methodology for simplicity and to be able to easily relate our findings to previous work. We use the same  tomographic redshift distribution $n^i(z)$ and uncertainty about the mean redshift $\langle z_{\rm DIR}^i \rangle$ as in J20 here. In this method,  the uncertainties on the mean redshifts, $\sigma_z^i$,  are estimated from a bootstrap resampling of the spectroscopic samples. 
The density, the mean redshifts and the shape noise of the galaxies in individual tomographic bins are presented in Table \ref{table:survey}.

\subsection{Cosmology training set}
\label{subsubsec:cosmoSLICS}

\begin{table}
   \centering
   \caption{Summary of key properties from the four simulations suites used in our pipeline. $L_{\rm box}$ is the box side [in $h^{-1}$Mpc], $n_{\rm p}$ is the number of particles evolved, $N_{\rm sim}$ is the number of $N$-body runs, $N_{\rm LC}$ is the number of light-cones in the full training set and $N_{\rm cosmo}$ is the number of cosmology samples. The bottom section summarises the range in cosmological parameters that is covered by the cosmo-SLICS.}
   \tabcolsep=0.11cm
      \begin{tabular}{@{} rcrrrr @{}} 
      \hline
      \hline
       Sim. suite       &  $L_{\rm box}$ & $n_{\rm p}$ & $N_{\rm sims}$ & $N_{\rm LC}$  &$N_{\rm cosmo}$ \\
       \hline
       cosmo-SLICS & $505$ &   $1536^3$  &  52& 520 & 26\\
       SLICS             & $505$ &   $1536^3$  & 124& 124& 1\\
       SLICS-HR      & $505$ &   $1536^3$  & 5& 50 & 1\\
       {\it Magneticum}  2  & $352$ &   $2\times1583^3  $  & 1& 10& 1\\
       {\it Magneticum}  2b & $640$ &  $2\times2880^3 $   & 1& 10& 1\\    
       \hline
       parameter &  $\Omega_{\rm m}$& $S_8$\hspace{3mm} $ $& \multicolumn{3}{c}{$h$ \hspace{15mm}$w_0$}\\
       sampling &  [0.1, 0.55]& [0.6, 0.9] & \multicolumn{3}{c}{[0.6, 0.82]  [-2.0, -0.5]}\\
    \hline 
    \hline
    \end{tabular}
    \label{table:sims}
\end{table}

The training set is constructed from the cosmo-SLICS \citep[][HD19 hereafter]{cosmoSLICS}, a suite of $w$CDM $N$-body simulations specifically designed for weak lensing data analysis targeting dark matter and dark energy. These simulations cover a wide range of values in ($\Omega_{\rm m}, \sigma_8, h, w_0$). They sample the parameter volume at 25+1 coordinates organised in a latin hypercube (25 $w$CDM plus one $\Lambda$CDM point), and further include a  sample variance suppression technique, achieving a sub-percent to a few percent accuracy depending on the scales involved. This is comparable to the accuracy of many widely-used two-point statistics models based on non-linear power spectra from {\sc Halofit} \citep{Takahashi2012} or from {\sc HMcode} \citep{MeadFit,HMCode2020}. The full training range is detailed in  Table \ref{table:sims}, which also influences our choice of priors when sampling the likelihood (see Sec. \ref{subsubsec:likelihood}).

Each run evolved $1536^3$ particles inside a $505 h^{-1}{\rm Mpc}$ co-moving volume with the public {\sc cubep$^3$m}  $N$-body code \citep{CUBEP3M}, generating on-the-fly multiple two-dimensional projections of the density field. These flat-sky mass planes were subsequently arranged into past light-cones of 10 degrees on the side, from which lensing maps were extracted  at a number of redshift planes (see Sec. \ref{subsec:lightcone}). This process was repeated multiple times after the mass planes were randomly selected from a pool of six different projected sub-volumes, then their origins were randomly shifted. In total,  50 {\it pseudo}-independent light-cones  per cosmology are available for the generation of galaxy lensing catalogues (see HD19 for a complete description). In the end we include 10 light-cones per cosmology out of 50, after verifying that our results do not change when training on only five of them. Indeed, 1000 deg$^2$ is enough to reach convergence on our statistics, largely due to the sample suppression technique implemented in HD19. 

Two of these models (cosmology-fid and -00, see HD19) are used to infuse photometric redshift and shear calibration uncertainty, which we describe in Sec. \ref{subsec:syst_photoz} and \ref{subsec:syst_mcorr}, respectively.

\subsection{Covariance training set}
\label{subsubsec:SLICS}

Our covariance matrix is estimated from the SLICS \citep[][HD18 hereafter]{SLICS}, a public simulation suite in which the cosmology is fixed for every $N$-body run, but the random phases in the initial conditions are varied, offering a unique opportunity to estimate the uncertainty associated with sampling variance.  The volume and number of particles are the same as for the cosmo-SLICS, achieving a particle mass of $2.88 \, h^{-1} M_{\odot}$ (see the properties  summary in Table \ref{table:sims}).
 The light-cones are constructed in the same way as the cosmo-SLICS, except that in this case the mass sheets are sampled only once per $N$-body run, generating 124 truly independent realisations.  The accuracy of the SLICS has been  quantified in \citet{SLICS_1} by comparing their matter power spectrum  to that of the Cosmic Emulator \citep{Coyote3}, which match to within 2\% up to $k=2.0 \,h {\rm Mpc}^{-1}$; smaller scales progressively  depart from the emulator. The cosmo-SLICS have a similar  resolution.

\subsection{Systematics training set: mass resolution}
\label{subsubsec:SLICS_HR}

Numerical simulations are inevitably limited by their intrinsic mass and force resolution, and it is critical to understand how these affect any measurements carried out on the simulated data.
We employ for this purpose a series of `high-resolution' runs, first introduced in \citet{SLICS_1} and labelled `SLICS-HR' therein. These consist of 5 independent $N$-body simulations similar to the main SLICS suite, but in which the force accuracy of  {\sc cubep$^3$m} has been increased  significantly such as to resolve smaller structures, even though the particle number is fixed. These have been shown to reproduce the Cosmic Emulator  power spectrum to within 2\% up to  $k=10.0 \, h^{-1}{\rm Mpc}$, indicating that even those small scales are correctly captured by the simulations. The SLICS-HR are post-processed with a strategy similar to that adopted for the cosmo-SLICS, re-sampling the projected mass sheets in order to generate 10 {\it pseudo}-independent light-cones per run.

\subsection{Systematics training set: baryon feedback}
\label{subsubsec:magneticum}

Another important systematic we investigate in this analysis is the impact of strong baryonic physics that modifies the clustering property of matter.  As noted in multiple independent studies,  AGN feedback has a particularly important effect on the matter power spectrum but is challenging to calibrate. Simulations often struggle to reproduce
the correct baryon fraction in haloes of different masses, and these differences in turn cause major discrepancies in the clustering properties \citep[see][for example]{HorizonAGN}. 
In this paper, we examine one of these models and inspect which parts of our peak count measurements are affected by baryons.

We used for this exercise a series of light-cones ray-traced  from a subset of the {\it Magneticum Pathfinder} hydrodynamical simulations, which are designed to study the formation of cosmological structures in presence of baryonic physics and which were recently described in \citet{MagneticumBox2b}. These are based on the smoothed particle hydrodynamics (SPH) code {\sc p-gadget3} \citep{Gadget2}, in which a number of baryonic processes are implemented, including radiative cooling, star formation, supernovae, AGN and their associated feedback on the matter density field.  The  {\it Magneticum} reproduce a number of key observations such as statistical properties of the large-scale, inter-galactic and inter-cluster medium, but also  central dark matter fractions and stellar mass size relations \citep[see][for more details]{2014MNRAS.442.2304H,2015ApJ...812...29T,LensingPDF_baryons,MagneticumBox2b}. What is especially important in our case is that the total baryonic feedback on the matter field is comparable to that of the BAHAMAS cosmological hydrodynamical simulations \citep{BAHAMAS}, in particular in terms of the strength of the effect on the matter power spectrum. This derives from the similar baryon fractions produced by {\it Magneticum}  and BAHAMAS,  which are in reasonable agreement with observations. This validates the {\it Magneticum} as a good representation for the impact of baryon feedback, given the current uncertainty on the exact impact (see Sec. \ref{subsec:syst_baryons} for further discussion).

Among the various runs, we use a combination of the high-resolution {\it Run-2} \citep{2014MNRAS.442.2304H} and {\it Run-2b} \citep{2017A&C....20...52R}, which both co-evolve dark matter particles of mass $6.9\times10^8 \, h^{-1}M_{\odot}$ and gas particles with mass $1.4\times10^8 \, h^{-1}M_{\odot}$ in comoving volumes of side 352 and $640 \, h^{-1}$Mpc, respectively; the smaller (larger) box is used at lower (higher) redshift, and the transition occurs at $z=0.31$.  The input cosmology is consistent with the SLICS but slightly different, with  $\Omega_{\rm m} = 0.272$, $h = 0.704$,  $\Omega_{\rm b} = 0.0451$, $n_{\rm s} = 0.963$ and $\sigma_8=0.809$. Both {\it Run-2} and {\it Run-2b} also exist in pure gravity mode (i.e. dark matter-only) with otherwise identical initial conditions, allowing us to isolate the impact of the baryonic sector on our observables. 

\subsection{Simulation post-processing}
\label{subsec:post-process}

\subsubsection{Light-cones}
\label{subsec:lightcone}

The simulation suites used in this paper all work under the flat sky approximation, which assumes that the maps are far enough from the observer so that cartesian axes can be used instead of angles and radial distances. At preselected redshifts $z$, the  $N$-body/hydrodynamical codes assign the particles onto a three-dimensional grid, select a  sub-volume to be projected with pre-determined co-moving thickness, and collapse  the mass density along one of the axis. This procedure is repeated with different projection directions and sub-volumes, creating a collection of mass sheets at every redshift. These are next post-processed to generate a series of past  light-cone mass maps,   $\delta_{\rm 2D}({\boldsymbol \theta}, z)$, each of 100 deg$^2$, which are then used to generate convergence $\kappa({\boldsymbol \theta},z_{\rm s})$ and shear ${\boldsymbol \gamma}({\boldsymbol \theta},z_{\rm s})$ maps at multiple source redshift planes, $z_{\rm s}$, (see HD18 and HD19 for full details), where ${\boldsymbol \gamma} = \gamma_{1} + i\gamma_{2}$, the two components of the spin-2 shear field.
From these, mock lensing quantities ($\kappa,{\boldsymbol \gamma}$) can be computed for any galaxy position provided its (RA, Dec) coordinates and a redshift. 

\subsubsection{Assembling the simulated surveys}
\label{subsubsec:mosaic}

As for many non-Gaussian statistics, peak counts are highly sensitive to the noise properties of the data. As such the simulations need to reproduce exactly the position and shape noise of the real data, otherwise the calibration will be wrong. The solution, adopted in \citet{2015PhRvD..91f3507L}, K16 and M18 is to overlay data and simulated light-cones, and to construct mock surveys from the position and intrinsic shape of the former, and the convergence and shear of the latter.

Since the size of the full DES-Y1 footprint largely exceeds that of our individual light-cones, we connect the data and simulations with a `mosaic' approach, where the DES-Y1 galaxy catalogues are divided into smaller `tiles'\footnote{These tiles are sometimes called `patches' in the literature, e.g. in K16.} that all fit inside 100 deg$^2$ square areas. Each of these tiles  are then  overlaid with simulated light-cones from which lensing quantities are  extracted. In total, 19 tiles are required to assemble the full footprint with our mosaic, which is shown in Fig. \ref{fig:DES_tiles}. Every simulated light-cone from the {\it Cosmology}, {\it Covariance} or {\it Systematics training sets} is therefore replicated 19 times and associated with a full realisation of the survey. 

We emphasise that the simulated light-cones are discontinuous across tile boundaries, whereas data are not. To avoid significant calibration biases caused by this difference, no measurement whatsoever must extend over tile boundaries. Both data and mock data are separated in tiles at the catalogue level; these are then analysed individually, and the data vectors are combined at the end\footnote{Note that the tiles are identical for all simulations (SLICS, cosmo-SLICS, SLICS-HR and {\it Magneticum}) since their light-cones all have the same opening angle.}.

Another subtle difference that needs to be taken into account is that the position coordinates (RA, Dec) and the galaxy ellipticities ($\boldsymbol \epsilon$) from the data are provided on the (Southern) curved sky, whereas all of our simulations assume a $(X,Y)$ cartesian coordinate system. Since the physics are independent of our choice of coordinate system, and since we analyse every tile individually, we apply a coordinate transformation to centre every tile onto the equator, where both coordinate frames  converge\footnote{In this process, we rotate both the celestial coordinate and the ellipticities of every galaxy in the tile, to account for the modified distance to the South pole in the new coordinate frame. The exact transformation uses the method presented in the Appendix B of \citet{Qianli2020}, which rotates pairs of galaxies from any orientation on the sky onto the equator, placing one member at the origin. In our case we instead map to the equator the straight line that bisects every tile. Every tile has its unique rotation vector, which we also use to displace the galaxies and to recompute their ellipticities ($\epsilon_{1/2}$)  in this new coordinate frame.}. The weak lensing statistics of a given tile are unaffected by this rotation, a fundamental fact that we verify with two-point correlation functions in Sec. \ref{subsec:xipm}.

As easily noticed from looking at Fig. \ref{fig:DES_tiles}, some of the galaxies fall outside the tiles, which slightly affects the total number of galaxies in the sample. It is not ideal, but adding multiple simulated tiles for such a small fraction (1.9\%) of the data is arguably not worth the effort. The number of objects listed in Table \ref{table:survey} reflects this final selection and amounts to a total of 25.5 million of galaxies. 

\subsubsection{Mock galaxy shapes and redshifts}
\label{subsubsec:mockgal}

As mentioned above, the position and the intrinsic ellipticities of individual galaxies in the simulated catalogues are taken from the observations.  Redshifts are assigned to every object in a given tomographic bin `$i$' by sampling randomly the $n^i(z)$ described in Sec. \ref{sec:data}. Therefore, variations in survey depth  are not included in our training sets. This induces a systematic difference with the  data, but we expect that this has a minor effect on our cosmological measurement. Indeed, it was shown in \citet{Sven_SurveyDepth} that the impact of survey depth variability 
is subdominant for Stage-III surveys. At this stage, every galaxy has position and a redshift, which are used to extract the lensing quantities ($\kappa, {\boldsymbol \gamma}$) from the simulation light-cones.

We finally include the intrinsic galaxy shapes and {\sc Metacal} shear response correction in the simulations by randomly rotating the observed galaxy shapes  such as to undo the cosmological correlations from the data, and we then combine the new ellipticity ${\boldsymbol \epsilon}_{\rm int}$ with the simulated lensing signal as: 
\begin{eqnarray}
\boldsymbol \epsilon = \frac{{\boldsymbol \epsilon}_{\rm int} + {\boldsymbol g}}{1 + {\boldsymbol \epsilon}_{\rm int}^*{\boldsymbol g}} .
\label{eq:eps_obs}
\end{eqnarray}
Here $\boldsymbol g$ is the reduced shear, defined as ${\boldsymbol g} = {\boldsymbol \gamma}S/(1+ \kappa)$, and the bold-font symbols  $\boldsymbol g$, $\boldsymbol \gamma$, ${\boldsymbol \epsilon}$ and ${\boldsymbol \epsilon}_{\rm int}$ are again spin-2 complex quantities. 

We investigate in Sec. \ref{subsec:syst_IA} the impact of the intrinsic alignment of galaxies, where ${\boldsymbol \epsilon}_{\rm int}$ is no longer chosen at random and instead correlates with the shape of  dark matter haloes.

\section{Theory and  Methods}
\label{sec:methods}

Since we validate our simulation suites with cosmic shear correlation functions and lensing power spectra, we begin this section with a review of the theoretical modelling and the measurement strategies related to these quantities. We next move to the primary focus of this paper and describe our peak count statistics pipeline,  detailing our treatment of the data, our approach to modelling the signal and estimating the covariance matrix, and we finally describe our cosmological inference methods.

\subsection{$\xi_{\pm}$ statistics}
\label{subsec:xipm}

\begin{figure*}
\begin{center}
\includegraphics[width=6.6in]{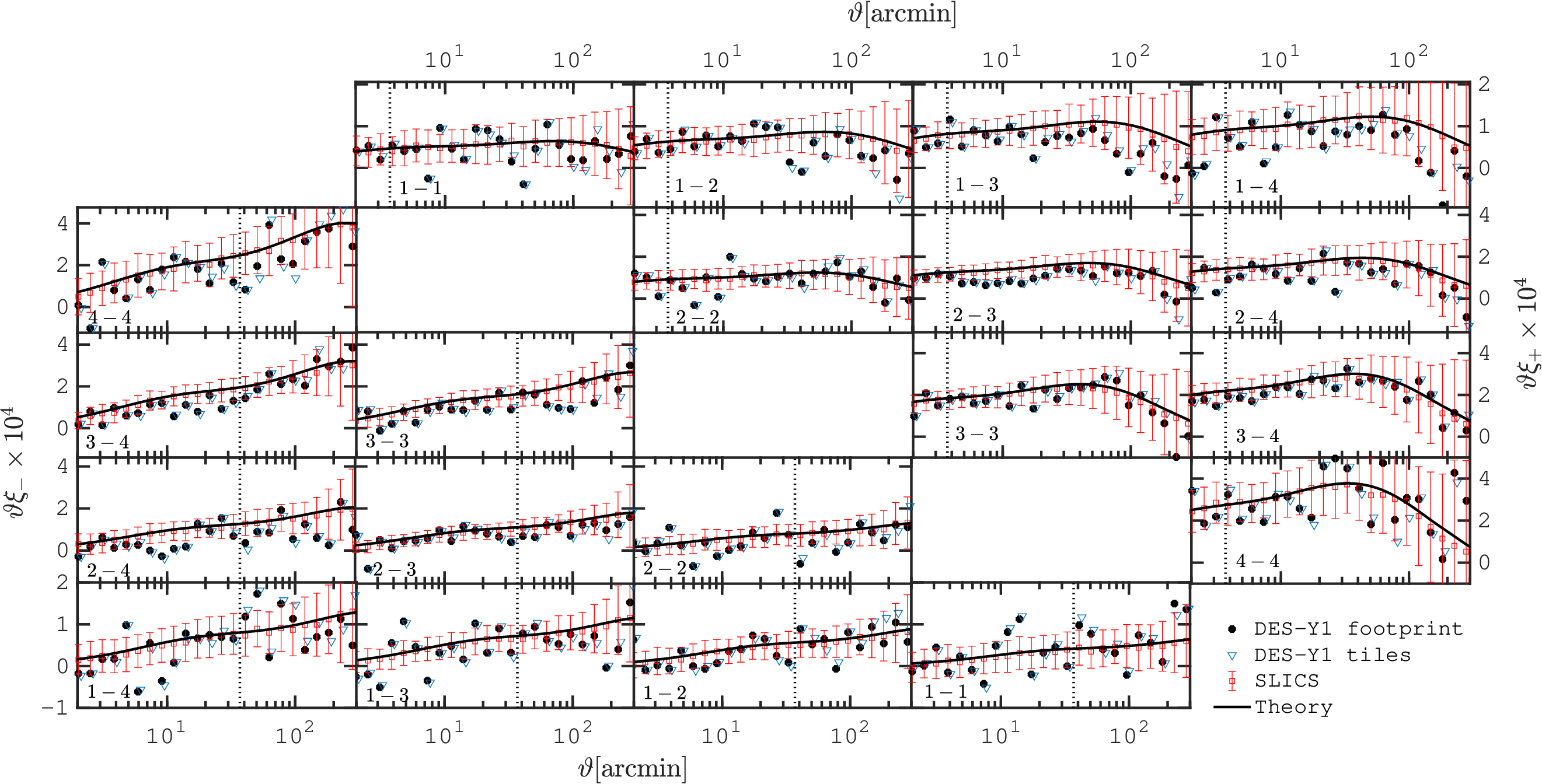}
\caption{Two-point correlation functions measured in the DES-Y1 data (filled circles  and opened triangles present measurements on the full survey footprint and from a weighted mean over the tiles presented in Fig. \ref{fig:DES_tiles}, respectively) and in the SLICS simulations (red squares, with error bars showing the statistical error on the DES-Y1 data), compared to the analytical model computed at the input SLICS cosmology (solid lines). The left- and right-hand side ladder plots present the $\xi_-$ and $\xi_+$ statistics respectively, and the sub-panels in each correspond to  different combinations of tomographic bins. The vertical dotted lines indicate the angular scales excluded in the cosmological analysis, which match those of T18.}
\label{fig:xi}
\end{center}
\end{figure*}

Two-point correlation functions (2PCFs) are well studied and have a key advantage over other measurement techniques:  as for all lensing two-point statistics, their modelling can be accurately related to the matter power spectrum, $P(k,z)$, whose accuracy is reaching the percent level far in the non-linear regime when calibrated with $N$-body simulations at small scales, and in absence of baryonic physics \citep{Coyote3, EuclidEmulator}.  From this $P(k,z)$, the lensing power spectrum between  tomographic bins `$i$' and `$j$'  is computed in the Limber approximation as:
\begin{eqnarray}
C_{\ell}^{ij} = \int_0^{\chi_{\rm H}}  \frac{q^i(\chi) \,q^j(\chi)}{\chi^2} \, P\, \bigg(\frac{\ell+1/2}{\chi},z(\chi)\bigg) \ {\rm d}\chi,
\label{eq:C_ell}
\end{eqnarray}
where $\chi_{\rm H}$ is the co-moving distance to the horizon, and the lensing kernels $q^{i}$ are computed from the redshift distributions $n(z)$ as:
\begin{eqnarray}
q^i(\chi) = \frac{3}{2}\Omega_{\rm m} \, \bigg(\frac{H_0}{c} \bigg)^2 \frac{\chi}{a(\chi)} \int_{\chi}^{\chi_{\rm H}} n^i(\chi')\frac{{\rm d}z}{{\rm d} \chi'}\frac{\chi' - \chi}{\chi'}{\rm d}\chi',
\label{eq:q_lensing}
\end{eqnarray}
where $c$ and $H_0$ are the speed of light and the Hubble parameter, respectively. 
The cosmic shear correlation functions $\xi_{\pm}^{ij}$ are computed from the $C_{\ell}^{ij}$ as:
 \begin{eqnarray}
\xi_{\pm}^{ij}(\vartheta) = \frac{1}{2\pi}\int_0^{\infty} C_{\ell}^{ij} \, J_{0/4}(\ell \vartheta) \, \ell \, {\rm d}\ell,
\label{eq:xipm_th}
\end{eqnarray}
with $J_{0/4}(x)$ being Bessel functions of the first kind.
Following T18, Eq. \ref{eq:xipm_th} is solved with the cosmological parameter estimation code {\sc cosmoSIS}\footnote{bitbucket.org/joezuntz/cosmosis/wiki/Home} \citep{cosmosis}, in which the matter power spectrum is calculated by the {\sc Halofit} model of \citet{Takahashi2012}. The $\xi_{\pm}^{ij}$ predictions for the DES-Y1 measurements are shown by the black lines in Fig. \ref{fig:xi} for all pairs of tomographic bins, at the SLICS input cosmology.

The measurements of $\widehat{\xi_{\pm}^{ij}}$ from simulations and data are carried out with {\sc Treecorr} \citep{TreeCorr},  a fast parallel tree-code that computes shape correlations between pairs of  galaxies `$a,b$' separated by an angle $\vartheta$ as:
 \begin{eqnarray}
\widehat{\xi_{\pm}^{ij}}(\vartheta) = \frac{\sum_{ab} W_a W_b \, \Bigg[ \epsilon_{a,{\rm t}}^{i}({\boldsymbol \theta}_a) \, \epsilon_{b,{\rm t}}^{j}({\boldsymbol  \theta}_b)  \pm  \epsilon_{a,\times}^{i}({\boldsymbol \theta}_a) \, \epsilon_{b,\times}^{j}({\boldsymbol \theta}_b)  \Bigg]\Delta \vartheta_{ab}}{\sum_{ab} W_a W_b \, S_a S_b} . 
\label{eq:xipm_data}
\end{eqnarray}
In the above expression, the sums are over all galaxies `$a$' in tomographic bin $i$ and  galaxies `$b$' in tomographic bin $j$; $\epsilon_{a,{\rm t}}^{i}$ and $\epsilon_{a,\times}^{i}$ are the tangential and cross components of the  ellipticity of galaxy $a$ in the direction of galaxy $b$; $W_{a/b}$ are weights attributed to individual galaxies, which are set to unity in the {\sc Metacalibration} shear inference method; $S_{a/b}$ are the `shear response correction' per object mentioned in Sec. \ref{subsec:data} and provided in the DES-Y1 catalogue; $\Delta \vartheta_{ab}$ is the binning operator, which is equal to unity if the angular separation between the two galaxies falls within the $\vartheta$-bin, and zero otherwise. Our raw measurements are organised in 32 logarithmically-spaced $\vartheta$-bins, in the range [$0.5 - 475.5$] arcmin, but not all angular scales are used in this work\footnote{Note that T18 used 20 logarithmic bins in the range 2.5-250 arcmin.}.

We present in Fig. \ref{fig:xi} our measurements of $\widehat{\xi_{\pm}^{ij}}$ on the DES-Y1 data,  showing with the black solid points the measurement on the full footprint, and with the open  blue triangles the measurements on the 19 data tiles  described in Sec. \ref{subsubsec:mosaic}, which are combined with a weighted mean using the {\sc Treecorr} $N_{\rm pairs}(\vartheta)$ per tile as our weights. We see that the two results are similar, with differences that are everywhere at least twice as small as the statistical error measured from the covariance mocks (see below) and evenly scattered about the black points, validating our tiling method. We further verified that the difference on the inferred cosmology is negligible (see Sec. \ref{sec:results}).

We also show, with the red squares in Fig. \ref{fig:xi}, the mean and expected $1\sigma$ error  on the DES-Y1 data as estimated from the {\it Covariance training set}. The agreement between theory and simulations is excellent at all scales, and the slight differences are well under the statistical precision of the data. We can observe a slight loss of power at large angular scales in the $\xi_+$ statistics, a finite box effect that we forward model (see Appendix \ref{subsec:2pcf_validation}). For every simulated light-cone we generate a total of 10 realisations of the shape noise by rotating, as many times, every galaxy in the catalogue, and recomputing new observed ellipticities (with Eq. \ref{eq:eps_obs}) and the correlation functions (Eq. \ref{eq:xipm_data}). The red squares in Fig. \ref{fig:xi}, as well as their associated error bars, correspond to one of these realisations; we observe no significant change in the  other nine realisations, and recover the error bars reported by T18 to within 5-15\% over most angular scales, further demonstrating the robustness of our training set. We do not expect a perfect match due to the slightly different binning scheme. 

Our cosmological analyses exclude the same angular scales as in T18, removing the elements of the data vector where T18 conclude that the uncertainty on the baryonic feedback and in the non-linear matter power spectrum is non-negligible. These scales are indicated by the vertical lines in Fig. \ref{fig:xi}. 

The variation of  $\xi_{\pm}$ with cosmology are well captured by  {\sc cosmoSIS}, and so are the responses to photometric redshift and shear calibration uncertainties (see Sec. \ref{sec:systematics}). We therefore do not measure this statistic in the {\it Cosmology} nor the {\it Systematics training sets}, and use instead the public modules provided in the latest {\sc cosmoSIS} release to calculate these.  

\subsection{Shear peak count statistics}
\label{subsec:peaks}

As mentioned in the introduction, the peak count statistic is a powerful alternative method to extract cosmological information from weak lensing data. It consists of measuring the `peak function', i.e. the number of lensing peaks as a function of their signal-to-noise, which is very sensitive to cosmology and robust to systematics \citep[see][for recent comparisons with other lensing probes]{Zuercher2020a, Martinet20}. 

Our measurement technique closely follows that described in K16 and M18, which we review here.   Peaks are identified from local maxima in the signal-to-noise maps of the mass within apertures \citep{Schneider1996}, $\mathcal{M}_{\rm ap}(\boldsymbol \theta)$, searching for pixels with values higher than their 8 neighbours. This is one of many ways to estimate the projected mass map from galaxy lensing catalogues, and was chosen primarily for its local response to data masking. This is to be contrasted with e.g. the Fourier methods of \citet{KaiserSquires} in which masking introduces a complicated mode-mixing matrix that can affect all scales. Other techniques such as Bayesian mass reconstruction \citep{Price2020} or wavelets transforms \citep{Leistedt2017} are also promising and merit to be explored in the future \citep[see as well][and references therein]{Gatti20}.

From a lensing  catalogue containing the position, ellipticity ${\boldsymbol \epsilon}_a$ and shear response correction $S_a$ per galaxy, we construct an  aperture mass map on a grid by summing\footnote{In practice, we use a link-list to loop only over nearby galaxies.} over the tangential component of the ellipticities from galaxies surrounding every pixel at coordinate ${\boldsymbol \theta}$, weighted by an aperture filter $Q$. More precisely, we compute:
 \begin{eqnarray}
M_{\rm ap}(\boldsymbol \theta) = \frac{1}{n_{\rm gal}(\boldsymbol \theta) \sum_a S_a}\sum_a \epsilon_{a,{\rm t}}({\boldsymbol \theta}, {\boldsymbol \theta}_a) Q(|{\boldsymbol \theta} - {\boldsymbol \theta}_a|, \theta_{\rm ap}, x_c),
\label{eq:Map}
\end{eqnarray}
where $n_{\rm gal}(\boldsymbol \theta) $ is the  galaxies density  in the filter centred at $\boldsymbol \theta$, and $\boldsymbol \theta_a$ is the position of galaxy $a$. 
The tangential ellipticity with respect to the aperture centre  is computed as $\epsilon_{a,{\rm t}}({\boldsymbol \theta}, {\boldsymbol \theta}_a)= -[\epsilon_1({\boldsymbol \theta}_a)\ {\rm cos}(2\phi({\boldsymbol \theta}, {\boldsymbol \theta}_a))+\epsilon_2({\boldsymbol \theta}_a)\ {\rm sin}(2\phi({\boldsymbol \theta}, {\boldsymbol \theta}_a))]$,  where $\phi({\boldsymbol \theta}, {\boldsymbol \theta}_a)$ is the angle between both coordinates. Our filter $Q(\theta, \theta_{\rm ap}, x_c)$, abridged to $Q(\theta)$ to shorten the notation, is identical to that in \citet{Schirmer2007}, which is optimal for detecting haloes following an NFW profile (but faster than solving the actual numerical NFW equation):
 \begin{eqnarray}
Q(x) = \frac{{\rm tanh}(x/x_c)}{x/x_c} \big[1 + {\rm exp}(6 - 150x) + {\rm exp}(- 47 + 50x)\big]^{-1}.
\label{eq:Q}
\end{eqnarray}
In the above expression,  $x=\theta/\theta_{\rm ap}$, where $\theta$ is the distance to the filter centre, and we adopt $x_c = 0.15$ as in previous works, a choice that maximises the sensitivity of the signal to the massive haloes, which carry the majority of the cosmological information.
 The filter size of $\theta_{\rm ap}=12.5$ arcmin is adopted as in M18, however we also consider 9.0 and 15.0 arcmin, and report results for these where appropriate.
 Hereafter, Eq. (\ref{eq:Map}) defines the signal of our aperture mass map, which we compute at every pixel location.

The variance about this map is calculated at every pixel location from:
 \begin{eqnarray}
\sigma^2_{\rm ap}(\boldsymbol \theta) = \frac{1}{2 n_{\rm gal}^2(\boldsymbol \theta)  \bigg[ \sum_a S_a \bigg]^2} \sum_a |{\boldsymbol \epsilon}_{a}|^2 Q^2(|{\boldsymbol \theta} - {\boldsymbol \theta}_a|) ,
\label{eq:MapNoise}
\end{eqnarray}
where again the sum runs over all galaxies in the filter.
Note that the magnitude of the measured galaxy ellipticities that enters this equation must also be calibrated by the shear response correction \citep[see the Appendix A of][]{Asgari_DES_KiDS_cosebi}, hence the term $[\sum_a S_a ]^2$ in the denominator. The signal-to-noise map from which peaks are identified, $\mathcal{M}(\boldsymbol \theta)\equiv\mathcal{S}/\mathcal{N}$, is computed by taking the ratio between Eq. (\ref{eq:Map}) and the square root of Eq. (\ref{eq:MapNoise}) at every pixel location, e.g.
\begin{eqnarray}
\mathcal{M}(\boldsymbol \theta) \equiv M_{\rm ap}(\boldsymbol \theta)  / \sigma_{\rm ap}(\boldsymbol \theta).
\end{eqnarray}

Peaks catalogues are first constructed from the galaxy catalogues separated in tomographic bins (which we label 1, 2, 3 \& 4), and then from every combination of pairs of tomographic catalogues  (which we label 1$\cup$2, 1$\cup$3, 1$\cup$4 ... 3$\cup$4).  As detailed in  \citet{Martinet20}, analysing these `cross-tomographic' catalogues provides additional information that is not contained within the `auto-tomographic' case. They went further and also included triplets (1$\cup$2$\cup$3, 1$\cup$2$\cup$4, 1$\cup$3$\cup$4...) and quadruplets (1$\cup$2$\cup$3$\cup$4), showing that these also contained additional information, but this gain is not as significant in our case, where the noise levels are much higher. 

\subsubsection{Masking}

\begin{figure}
\begin{center}
\includegraphics[width=3.3in]{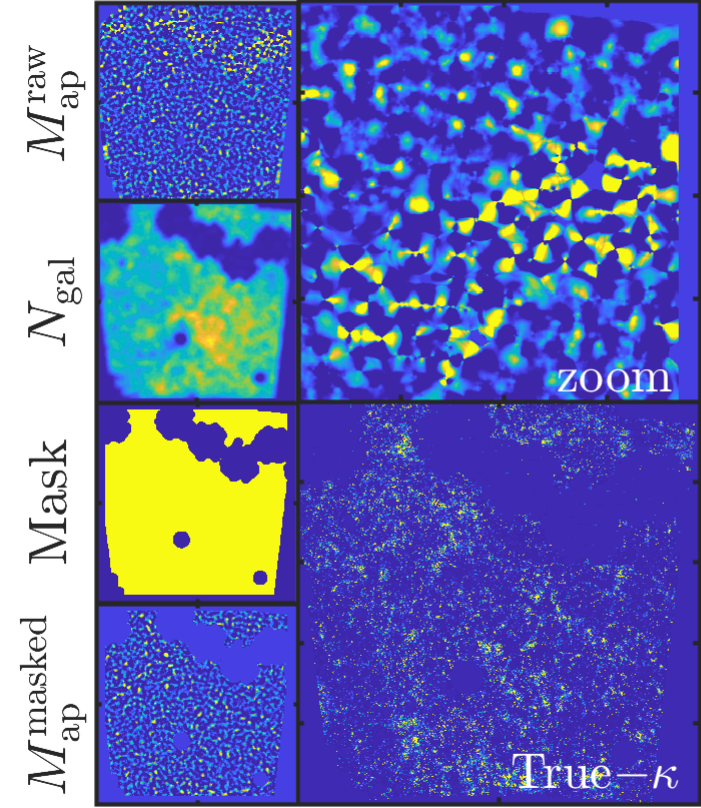}
\caption{Example of the $M_{\rm ap}$ mass-reconstruction pipeline over one of our $10\times 10$ deg$^2$ tiles. The larger panel on the bottom right presents the true $\kappa$ values at the position of the galaxies in this field, extracted from the cosmo-SLICS model-00. The raw $M_{\rm ap}$ map is shown in the top left panel in the noise-free case. The number of galaxies in the filter (second panel) are then used to construct a mask (third panel), which we apply on the raw $M_{\rm ap}$ maps (bottom panel). The top right panel shows a zoom-in of the top left panel, highlighting the effect of masking on the raw reconstructed $M_{\rm ap}$ map.}
\label{fig:M_ap}
\end{center}
\end{figure}

\begin{figure*}
\begin{center}
\includegraphics[width=6.8in]{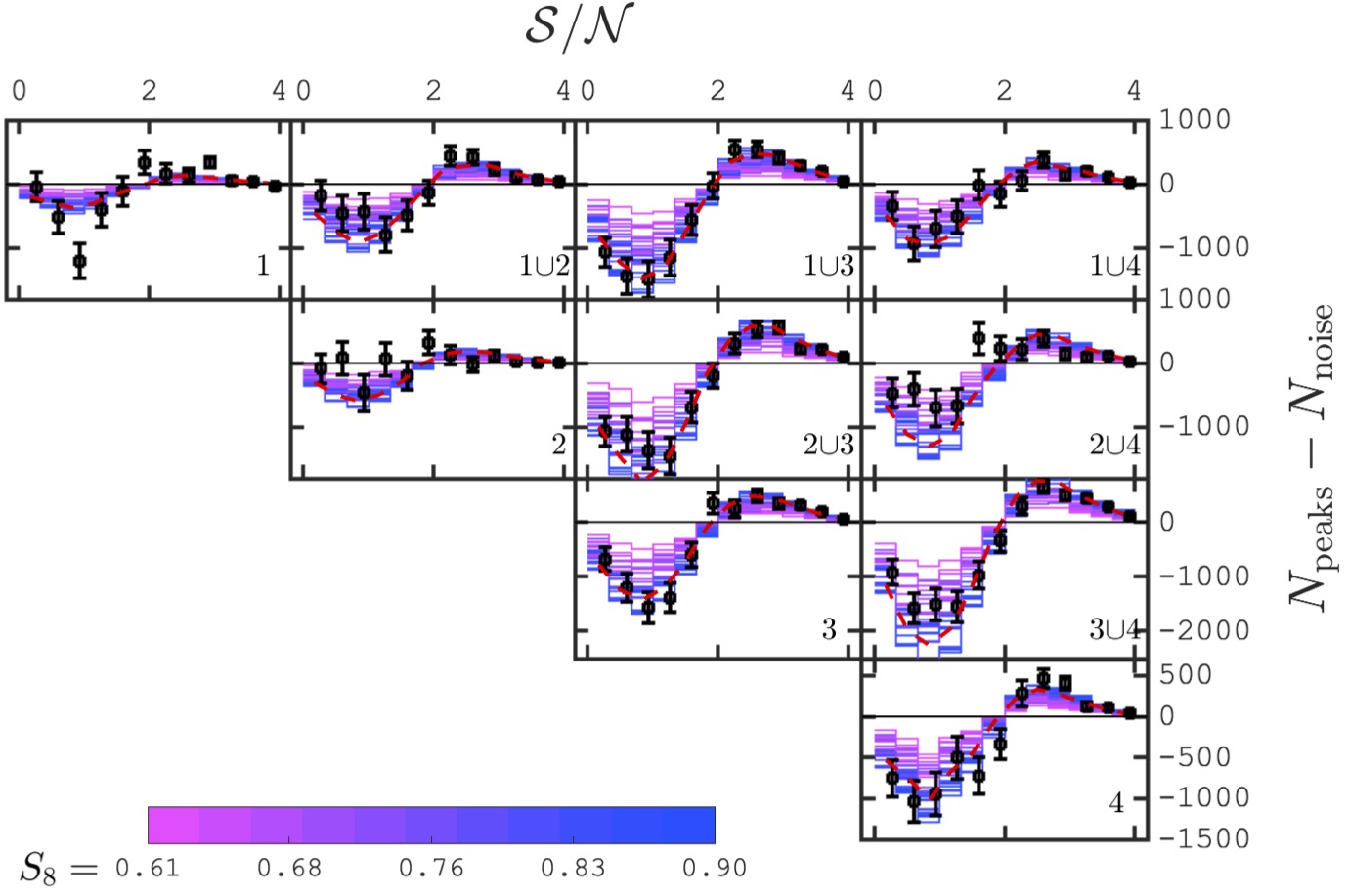}
\caption{Peak function $N_{\rm peaks}(\mathcal{S}/\mathcal{N})$ in the DES-Y1 data (black squares) and simulations (coloured histograms), from which the expectation from pure shape noise $N_{\rm noise}(\mathcal{S}/\mathcal{N})$ has been subtracted. The panels show different tomographic bin combinations, as labelled in their lower-right corners. The predictions are colour-coded by their  $S_8$ value,  with the red dashed line showing the best-fit value. The DES-Y1 error bars are estimated from the {\it Covariance training set}.}
\label{fig:N_peaks}
\end{center}
\end{figure*}

Weak lensing data are taken inside a survey footprint, and parts of the images are removed in order to mask out satellite tracks, bright stars, saturated foreground galaxies, etc. 
The effect of data masking on the aperture mass map can be significant: the signal and the noise are coherently diluted  in apertures that strongly overlap with masked regions, generating regions where $\mathcal M$ is overly smooth. Therefore the survey mask must be included in the simulations and in the estimator such as to avoid biasing the statistics. 

If the masked pixels are known, this can be taken into account by avoiding pixels for which e.g. more than half of the filter overlaps with masked areas. Alternatively, one can examine the object density in the aperture filter, $n_{\rm gal}(\boldsymbol \theta)$, and require that it exceeds a fixed threshold in order to down-weight or reject heavily masked apertures. In this method, pixels with little or no galaxies are treated as masked. We opted for the second method, setting the threshold to $1/\pi^2$ gal/arcmin$^2$  after a few different trials, which directly identifies regions with very low galaxy counts. We further augment the masking selection with an apodisation step that flags as `also masked' any pixel within a distance $\theta_{\rm ap}$ of a masked region found in the first step.  

Fig. \ref{fig:M_ap} illustrates this procedure for one of the tiled catalogues, for an idealised noise-free case. For our fiducial  choice of filter $\theta_{\rm ap}$ = 12.5 arcmin, we show in the upper two left panel the `raw'  $M_{\rm ap}(\boldsymbol \theta)$ map (e.g. before masking, computed directly from Eq. \ref{eq:Map}) as well as $N_{\rm gal}(\boldsymbol \theta)$. The masked regions are clearly visible in the latter but not so much in the former. A close inspection (top right panel) however reveals overly smooth features  in $M_{\rm ap}(\boldsymbol \theta)$, in regions where there are no galaxies (i.e. in the blue regions of the $N_{\rm gal}(\boldsymbol \theta)$ map).  The third left panel shows the masked regions constructed from our pipeline, which is finally applied on the raw aperture map, resulting in the masked map shown on the bottom panel. All choices of $\theta_{\rm ap}$ result in aperture maps that closely recover the true convergence (shown in the bottom right panel). 

It is clear from Fig. \ref{fig:M_ap} that the unmasked area of our final maps is affected by the aperture filter size. Indeed, larger filters  can be blind to small features in the mask,  while the survey edges are more severly excluded. This does not bias our cosmological inference since we apply the same filter to the data and the simulations, but it does slightly affect the signal-to-noise of our measurement, which increases with the area of the survey. The net unmasked area in our  final maps are (1426, 1408, 1366, 1327, 1284) deg $^2$ for  $\theta_{\rm ap}$ = (6.0, 9.0, 12.5, 15.0,  18.0)', respectively.

\subsubsection{Peak function}

Peaks found in the (masked) $\mathcal{M}$ maps are counted and binned as a function of their pixel value, thereby measuring   the peak function $N_{\rm peaks}(\mathcal{S}/\mathcal{N})$. We use 12 bins covering the range $0<\mathcal{S}/\mathcal{N}\leq4$ in our main data vector, which was found in K16 and M18 to avoid scales where multiple systematics uncertainties such as the effects of baryon feedback and intrinsic alignments  of galaxies become large (we extend this range to higher $\mathcal{S}/\mathcal{N}$ values in some of our systematics investigations). 12 bins  is also a good trade-off between our need to capture most cosmological information from  $N_{\rm peaks}(\mathcal{S}/\mathcal{N})$, while keeping a small data vector for which the covariance matrix will be  less noisy.  A number of recent studies \citep[M18,][]{LensingVoids, MinimaPeaks, Zuercher2020a, Martinet20, Davies20} have shown that cosmological information is contained in peaks of negative $\mathcal{S}/\mathcal{N}$ or in lensing voids, however, as noted in Appendix B of M18, the peaks with negative $\mathcal{S}/\mathcal{N}$ strongly correlate with those of positive $\mathcal{S}/\mathcal{N}$ value and only marginally improve the constraints from peak statistics in the case of Stage III surveys. We therefore focus only on the positive peaks in this DES-Y1 analysis.

We show in  Fig. \ref{fig:N_peaks} the peak function measured from the {\it Cosmology training set} with $\theta_{\rm ap} = 12.5$ arcmin, for all pair combinations of the four redshift bins and colour-coded as a function of the input $S_8$. A pure noise case ($N_{\rm noise}$), obtained from the average peak function after setting $\boldsymbol \gamma=0$ on 10 full survey realisations, has been subtracted to highlight the cosmological variations. Off-diagonal panels present the cross-tomographic measurements. The colour gradient is clearly visible in all tomographic bins; more precisely, all cosmologies present an excess of large $\mathcal{S}/\mathcal{N}$ peaks and a depletion of low $\mathcal{S}/\mathcal{N}$ peaks compared to pure noise. This is caused by the gravitational lensing signal, which create peaks and troughs in the $M_{\rm ap}$ map and smooths out the smallest peaks. Importantly, these differences are accentuated for high-$S_8$ cosmologies. Also shown with black squares are the measurements from the DES-Y1 data, with error bars estimated from the {\it Covariance training set}. These demonstrate that most of the constraining power comes from the  auto-tomographic bins 3 and 4 and from the cross-tomographic bins. Some additional information is contained in the highest $\mathcal{S}/\mathcal{N}$ peaks of the redshift bins 1 and 2, whereas the low $\mathcal{S}/\mathcal{N}$ peaks of bin 2 mostly contribute noise.

\subsection{Analysis pipeline}
\label{subsec:pipeline_overview}

In this analysis we extend multiple aspects of the K16 and M18 methodologies. Here is a summary of these improvements:
\begin{enumerate} 
\item{We include a tomographic decomposition of the data, including the cross-redshift pairs inspired by
the method presented in  \citet{Martinet20};}
\item{Our {\it Cosmology training set}  (see Sec. \ref{subsubsec:cosmoSLICS}) now includes four parameters ($\Omega_{\rm m}, \sigma_8, h$ and $w_0$), and it would be straightforward to increase that parameter list with additional training samples. Additionally, the cosmo-SLICS simulations are more accurate than those of \citet{DH10}, which were used in both K16 and M18: they resolve smaller scales, and suffer less from finite box effects, having a volume almost 8 times larger;}
\item{We deploy a fast emulator  (see Sec. \ref{subsec:GPR}) that can model the signal at arbitrary cosmologies within the parameter volume included in the training. In contrast with a likelihood interpolator, emulating the data vector directly allows us to combine the summary statistics with other measurement methods such as the two-point correlation functions, to better include systematic uncertainties, and to easily interface with most  likelihood samplers;}
\item{We generate a {\it Covariance  training set} from a larger ensemble of independent survey realisations (see Sec. \ref{subsubsec:SLICS}), and feed it into a novel hybrid internal resampling technique that improves the accuracy and precision of  lensing covariance matrices estimated from the suite (see Sec. \ref{subsec:covmat}). Moreover, the covariance training set is shown  to closely reproduce the published DES-Y1 cosmological constraints of T18 when analysed with two-point correlation functions (see Table \ref{table:cosmo_pipeline_test}), thereby validating both the simulations themselves and the covariance estimation pipeline. Our method is also compatible with joint-probe measurements; }
\item{We construct a series  of dedicated {\it Systematics  training sets} specifically tailored to our data, in which the most important cosmic shear-related systematics are infused. Specifically, we investigate the impact of photometric redshift uncertainty (Sec. \ref{subsec:syst_photoz}), of multiplicative shear calibration uncertainty (Sec. \ref{subsec:syst_mcorr}), of baryonic feedback (Sec. \ref{subsec:syst_baryons}), of possible intrinsic alignment of galaxies (Sec. \ref{subsec:syst_IA}), and of limits in the  accuracy of the non-linear physics (Sec. \ref{subsec:highres}). These tests   allow us to flag the elements of our data vector that are affected, and in some case to model the impact. Following K16, we construct a linear response model to the photometric redshift and multiplicative shear calibration uncertainties, but calibrate our models on a sample of 10  deviations from the mean of the distribution (respectively $\Delta z^i$ and $\Delta m^i$, with $i=1..10$) as opposed to one;}
\item{We implement the emulator, the covariance matrix and the linear systematic models  within the cosmology inference code {\sc cosmoSIS}  \citep{cosmosis}, allowing us to  carry out a joint likelihood analysis based on peak statistics and shear two-point correlation functions, while coherently marginalising over the nuisance parameters that affect both of these measurement methods. 
}
\end{enumerate}

With this new pipeline, we are fully equipped to investigate the impact of different  measurement and modelling methods, of different systematics mitigation strategies, but also  of analyses choices related to the likelihood sampling, such as the prior ranges, the specific combination of parameters to be sampled, or the manner in which maximum likelihoods and confidence intervals are reported. Indeed, it has been shown that these have a non-negligible impact on the final cosmological constraints, specifically in the context of weak lensing cosmic shear analyses \citep{2017MNRAS.471.1259J, 2019MNRAS.482.3696C, KiDS_DES_Joudaki, KiDS1000_Asgari, KiDS1000_Joachimi}\footnote{In particular, \citet{KiDS1000_Joachimi} demonstrates that reporting the projected maximum likelihood value and the associated confidence interval can introduce biases when collapsing a high-dimensional hyper-volume into a one-dimensional space. Instead, it is argued therein that a more accurate 1D inference is obtained by reporting  the  multivariate maximum a posteriori (MAP) distribution, along with a credible interval calculated using the projected joint highest posterior density (PJ-HPD) of the full likelihood.}.  It turns out that these approaches do not make too much of an impact for current cosmic shear data.  Moreover, it is not our primary goal here to optimise these choices, as we are rather interested in establishing our simulation-based inference method as being robust, accurate and flexible. We therefore opted for an overall analysis pipeline that maximally resembles that of the fiducial DES-Y1 cosmic shear data, and leave some additional tuning for future work.

Aside from a different choice of $n(z)$ calibration (see Sec. \ref{subsec:data}),  the key differences between our current  pipeline and that of T18 are the impossibility of ours to vary and marginalise over the other cosmological  parameters --  the power spectrum tilt parameter $n_{\rm s}$, the baryon density $\Omega_{\rm b}$ and the sum of neutrino mass $\sum m_{\nu}$. These would require more light-cone simulations  such as the {\it Mira-Titan} \citep{MiraTitan} or the {\it MassiveNuS} \citep{MassiveNuS}, which are not folded in our training set  at the moment, but form a natural extension to this work. Also missing is a cosmology-dependent model for the effect of intrinsic alignment, which  could be necessary in future analyses.

\subsection{Covariance matrix}
\label{subsec:covmat}

\begin{figure*}
\begin{center}
\includegraphics[width=5.3in]{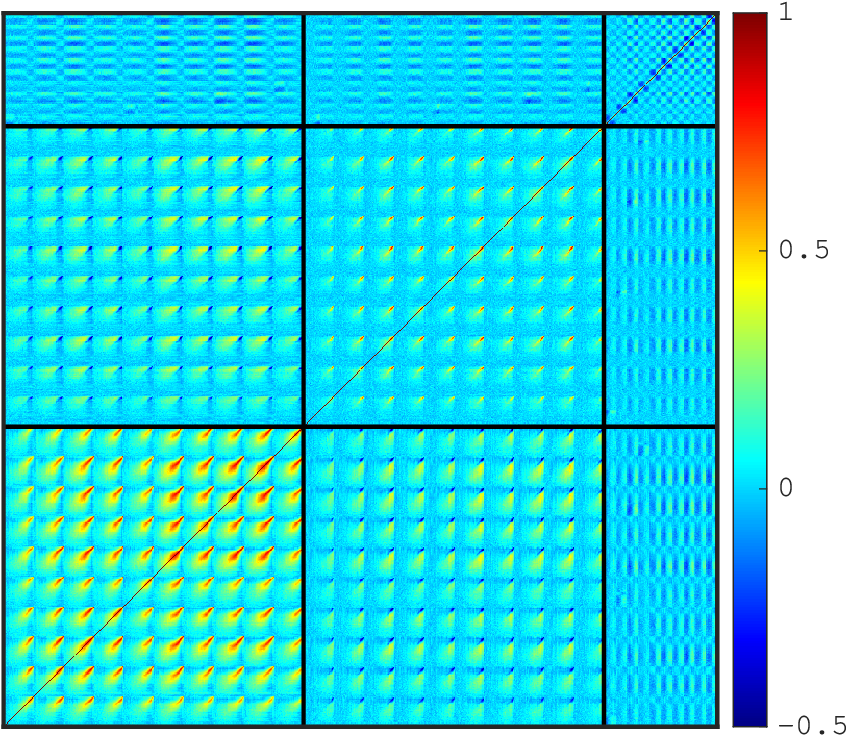}
\caption{This cross-correlation matrix highlights the correlations between the different elements of the data vector. From left to right, the first 10 blocks show the $\xi_+$ tomographic measurements, followed by the 10  $\xi_-$ blocks, while the last 10 blocks show the tomographic peak count. Not all elements are used in the analysis, see the main text for more details.}
\label{fig:CovMatrix}
\end{center}
\end{figure*}

The covariance matrix is a central ingredient to our cosmological inference as it describes the level of correlation between different elements of our data vector, and its inverse directly enters in the evaluation of the likelihood. In our analysis, it is estimated from the {\it Covariance training set}, which is based on 124 independent light-cones, each replicated onto the 19 survey tiles such as to fully cover the DES-Y1 footprint. For each of these survey realisations, we further generate 10 shape noise realisations  by randomly  rotating the ellipticity measurements from the data, which increases the number of {\it pseudo}-independent realisations to $N_{\rm sim}=1240$ and is largely enough for the current analysis. 

The next step consists in combining the measurements obtained in the 19 different tiles into a final measurement  of the [$\xi_\pm^{ij}(\vartheta); N_{\rm peaks}^{ij}(\mathcal{S}/\mathcal{N})$] covariance. To achieve this, we mix the light-cones at the survey construction stage, such that for each of the 124 full survey assembly, the 19 tiles are extracted from 19 different light-cones selected at random. This mixing suppresses an unphysical large-scale mode-coupling  caused by the replication, which otherwise results in an overall variance on  $\xi_{\pm}$ that is an order of magnitude too large. (Note that the $\xi_{\pm}$ covariance block is identical to an alternative estimation based on computing the matrix for individual tiles, which are subsequently  combined  with an area-weighted average, but the $N_{\rm peaks}^{ij}(\mathcal{S}/\mathcal{N})$ block in that latter case becomes inaccurate in this case so we reject this approach.) We repeat this for each of the 10 noise realisations and use the average matrix as our final estimate.

 The net effect of averaging over multiple shape noise realisations is to significantly lower the noise in the matrix, especially over the terms where shape noise dominates. A similar technique is applied in M18, who also find a negligible impact on the cosmological results from KiDS-450 data whether they average over  5 or 20 noise realisations. Fig. \ref{fig:CovMatrix} shows the resulting matrix, normalised to unity of the diagonal, e.g. $r(x,y) = {\rm Cov}(x,y)/\sqrt{{\rm Cov}(x,x)\ {\rm Cov}(y,y)}$. Whereas the $\xi_+$ block shows the highest level of correlation, the off-diagonal blocks are mostly uncorrelated, which is promising for the prospect of learning additional information from the joint analysis.

\subsection{Peak function emulator}
\label{subsec:GPR}

\begin{figure}
\begin{center}
\includegraphics[width=3.3in]{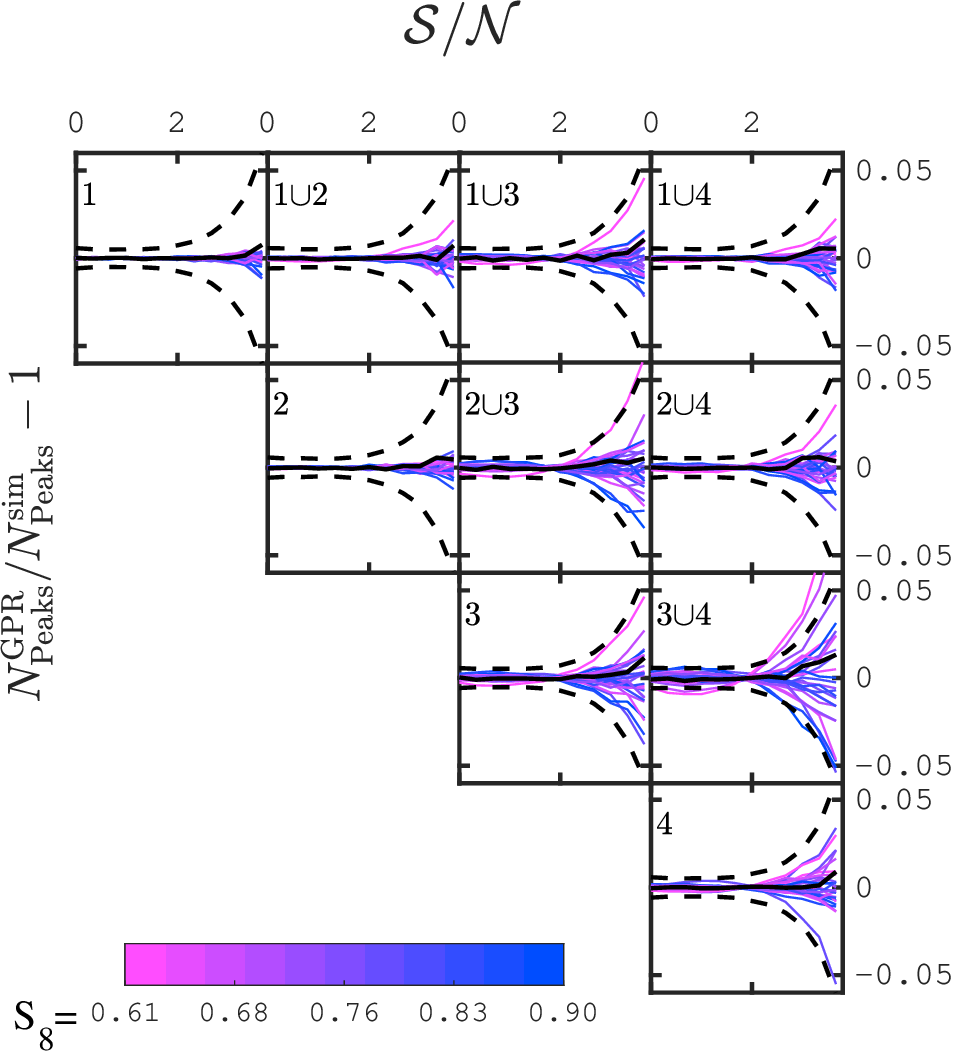}
\caption{Accuracy of the GPR emulator, computed with a leave-one-out cross-validation test. The results are colour-coded with the input $S_8$ value of the removed training point, and compared with the statistical uncertainty on the measurement (shown with the black dashed lines). The black solid line shows the accuracy at the $\Lambda$CDM node, and the different panels  show the ten combinations of tomographic bins.}
\label{fig:GPR_CV}
\end{center}
\end{figure}

The peak count statistics measured from the {\it Cosmology training set} (shown in Fig. \ref{fig:N_peaks} with the coloured histograms) is computed at 26 points in a wide 4-dimensional volume. 
From these we train a Gaussian Process Regression (GPR) emulator that can model the peak function  given an input set of cosmological parameters $[\Omega_{\rm m}, S_8, h, w_0]$ at any point within the training volume. Directly adapted from the public cosmo-SLICS emulator\footnote{github.com/benjamingiblin/GPR\_Emulator/}  described in Appendix A of HD19, we train our GPR emulator on the individual elements of the $N_{\rm peaks}$ data vector, first optimising the hyper-parameters from an MCMC analysis that includes 200 training restarts, then `freezing' the emulator once the best-fit solution has been found. As described in  HD19, the training can also involve a PCA decomposition and a measurement error; we include the former but find that the modelling is more accurate without the latter.  

We evaluate the accuracy of the emulator from a leave-one-out cross-validation test: the emulator is trained on all but one of the training nodes, then generates a prediction of the peak function at the removed cosmology, which is finally compared with the actual measurement. This test is performed for all nodes and provides an upper bound on the interpolation error, since in this case the distance between the evaluation point and all other training nodes is significantly larger than if all points had been present. Moreover, many of these points lie at the edge of the training volume, hence removing them for this test requires the emulator to extrapolate from the other points, which is significantly less accurate than the interpolation that is normally performed. As discussed in HD19, the node at the fiducial cosmology was added by hand close to the centre of the $w$CDM Latin hypercube, hence the cross-validation test performed at that single $\Lambda$CDM point is more representative of the actual emulator's accuracy. 

The results from this accuracy test are  presented in Fig. \ref{fig:GPR_CV}, again colour-coded with $S_8$. We achieve sub-percent interpolation accuracy over data points with $\mathcal{S}/\mathcal{N}<3$, and for all points when testing the $\Lambda$CDM model (shown in thick black). We observe  in some other models a scatter of up to a few percent  for peaks with $\mathcal{S}/\mathcal{N}>3$, but this scatter over-estimates the true interpolation error for reasons explained above.  In term of accuracy target for the model and other systematics effects, we generally aim for  an impact on the cosmological inference that is less than $0.5 \sigma_{\rm stat}$; all dominant effects are documented in Secs. \ref{sec:systematics} and \ref{subsec:results_GPR}. Generally speaking, most systematic effects that have a  $<5\%$ impact on a small number of elements in the data vector will satisfy this criteria.
When compared to the statistical error on the DES-Y1 measurement, the GPR emulator's  error is always subdominant (see the  thick dashed lines in Fig. \ref{fig:GPR_CV}). We conclude from this that the accuracy of our model  is high enough, given our current data, and that it should introduce no noticeable bias in the cosmological inference.

\subsection{Likelihood}
\label{subsubsec:likelihood}

The GPR emulator is embedded within the {\sc cosmoSIS} cosmological inference package, which allows us to evaluate the likelihood at  any cosmology within the cosmo-SLICS training range, given measurements of $\xi_{\pm}^{ij}(\vartheta)$ and $N_{\rm peaks}^{ij}({\mathcal{S}/\mathcal{N}})$ from the data, plus our joint covariance matrix.
At the moment we can only provide predictions of the peak function for the $\theta_{\rm ap}$ values on which the GPR was trained, but in the future this could also be treated as a free parameter to be emulated, providing even more flexibility to the prediction code and optimisation avenues.

The predictions $\boldsymbol{x}$ at cosmology $\boldsymbol{\pi}$ are then compared with the data $\boldsymbol{d}$ using a multivariate $t$-distribution likelihood following \citet{SellentinHeavens}:
\begin{eqnarray}
\mathcal{L}(\boldsymbol{\pi}|\boldsymbol{d}) \propto   \frac{N_{\rm sim} }{2}  {\rm ln}\bigg[1 + \chi^2 / (N_{\rm sim} - 1)\bigg],\mbox{\hspace{5mm} with} 
\label{eq:like}
\end{eqnarray}
\begin{eqnarray}
\chi^2 = \sum [\boldsymbol{x}(\boldsymbol{\pi}) - \boldsymbol{d}]^{\rm T}{\rm Cov}^{-1} [\boldsymbol{x}(\boldsymbol{\pi}) - \boldsymbol{d}] .
\end{eqnarray}
This likelihood correctly takes into account the residual noise in the covariance matrix that stems from its sampling with a finite number of simulations, and reduces to the standard multivariate Gaussian likelihood when $N_{\rm sim}\rightarrow \infty$. Since there are, at the very least, hundreds of peaks in each of our bins, adopting this  likelihood is justified.

For our first pipeline validation exercise, our choice of priors matches that of  T18 in our two-point statistics-only analysis (see Table  \ref{table:priors}), allowing us to investigate  the effect of replacing the fiducial (analytic) covariance matrix with our simulation-based matrix on the parameter inference.  At the same time this serves to  validate our  simulations. 

In our three fiducial analyses (2PCFs, peaks and joint), the priors reflect the parameter range probed by the cosmo-SLICS, and hence we assign a flat prior  on $\Omega_{\rm m}$, $\sigma_8$, $h$ and $w_0$  (summarised in Table \ref{table:priors}). All other parameters (i.e. the baryon density, the tilt in the primordial power spectrum  and the sum of neutrino masses) are kept fixed to $\Omega_{\rm b}=0.0473$, $n_{\rm s}=0.969$ and $\sum m_{\nu}=0.0$eV, respectively. The lensing constraints on these are very weak at the moment, hence we do not expect that holding them fixed should significantly affect our results.
 
We finally include the same 10 nuisance parameters as in  T18: a shear calibration $\Delta m^i$ and a photometric redshift calibration $\Delta z^i$ per tomographic bin, plus two parameters associated with the modelling of the intrinsic alignments (IA) in the non-linear alignment model \citep[NLA]{2007NJPh....9..444B}. The latter two are not included in the peak count case for which we conduct instead a simulation-based assessment of the impact of IA.  
 We sample the likelihood with the {\sc MultiNest} sampler \citep{multinest}, set with a tolerance parameter of 0.1 and an efficiency of 0.3. We refer the interested reader to T18 and \citet{DESY1_Methods} for more details about the DES-Y1 cosmology inference pipeline.

We validate our implementation with a series of  likelihood sampling analyses where the `data' is taken from the mean measurement extracted from the {\it Covariance training set}, for which the cosmology and the systematic biases are known. We detail  these results in Appendix \ref{sec:syst_pipeline}, and compare the 2PCFs and the peaks $w$CDM performance on these mocks as well. We further validate our $\Lambda$CDM 2PCFs segment both against the T18 and J20 results in Sec. \ref{subsec:results_2PCF}.  It is worth mentioning here that \citet{Jeffrey2019} have proposed a correction to the likelihood calculation when the model is inferred from noisy estimates, which we could have used the residual noise in our training sample had been judged too large, however this is not the case, with 1000 deg$^2$ of light-cone data per cosmology, times 10 noise realisation.

\section{Systematics}
\label{sec:systematics}

 \begin{figure*}
\begin{center}
\includegraphics[width=7.0in]{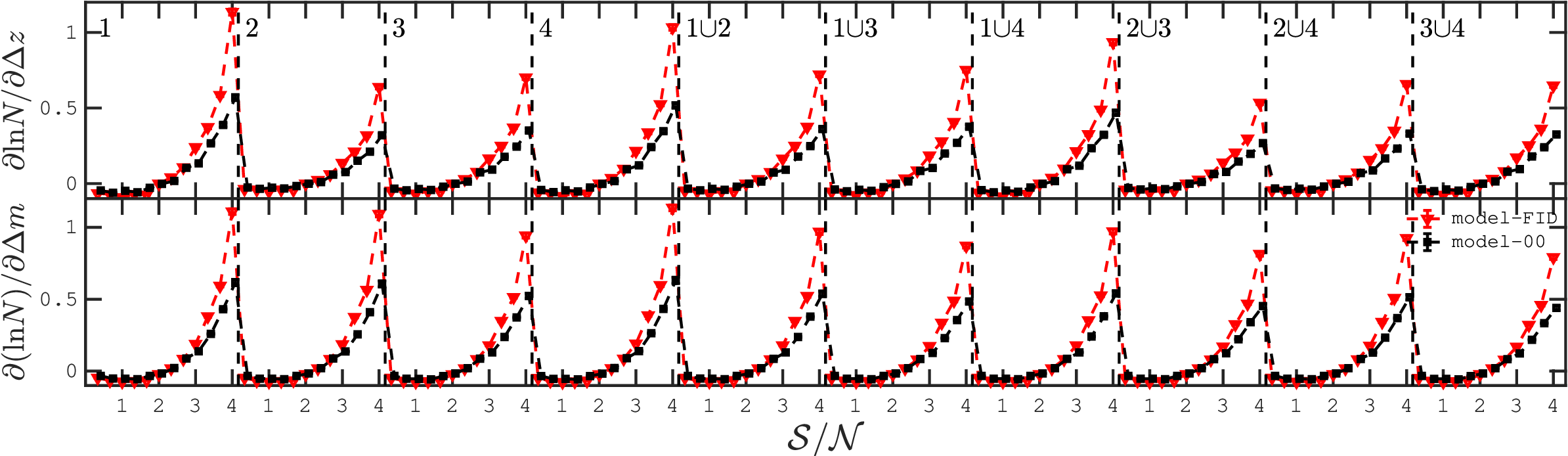}
\caption{Derivative of the  (log of the) peak function $N_{\rm peaks}$ with respect to shifts in the mean of the galaxy redshift distribution, $\Delta z$ (upper) and in the mean shear calibration $\Delta m$ (lower). These derivatives are shown as a function of $\mathcal{S}/\mathcal{N}$ bin, for every tomographic  case and are computed from 10 deviation points (see main text for details). The cosmo-SLICS model-FID is in red, model-00 in black, and the error bars are obtained from the scatter over 10 realisations of the full survey. Other filter sizes yield slightly different derivatives, but exhibit  a similar level of agreement between the two cosmological models. The FID derivatives are used in the cosmology pipeline to marginalise over these two nuisance parameters.}
\label{fig:dNpeaks_dz}
\end{center}
\end{figure*}

The likelihood modules within {\sc cosmoSIS} are equipped with an infrastructure that allows us to define nuisance parameters and to marginalise over them. In particular, for the two-point correlation function sector, the photometric redshift errors $\Delta z^i$ are included by shifting the tomographic redshift distributions, e.g. $n^i(z) \rightarrow n^i(z+\Delta z^i)$, after which new predictions for $\xi_{\pm}^{ij}$ are computed from Eqs. (\ref{eq:C_ell} -- \ref{eq:xipm_th}). Errors on the shear calibration, $\Delta m^i$, are included directly on the statistics as $\xi_{\pm}^{ij}\rightarrow \xi_{\pm}^{ij}(1+\Delta m^i)(1+\Delta m^j)$. Finally, we include  the two-parameters model  of intrinsic alignments of galaxies that was used in T18. We keep the T18  priors on the IA and shear calibration nuisance parameters, but use the J20 priors for the redshift uncertainty in our fiducial analyses. These are all summarised in Table \ref{table:priors}. Inaccuracies at small scales due to uncertainty in the non-linear physics and in the baryonic feedback are controlled with the angular scale cuts applied on $\xi_{\pm}$, rejecting elements of the data vector that vary by more than 2\% in presence of these systematics. As pointed out in T18 (see their table 3), even with these stringent cuts, strong feedback mechanisms could shift the inferred $S_8$ value by more  than 0.5$\sigma$. 

We expand the existing  {\sc cosmoSIS}  infrastructure to include systematics models  of the peak statistics based on our simulation training sets. Specifically, we added modules to include the effect of photometric and shear calibration errors, which we parameterised by the same $\Delta z^i$ and $\Delta m^i$ nuisance parameters as for the 2PCFs, allowing us to marginalise coherently over these in a joint $[\xi_{\pm};N_{\rm peaks}]$ analysis. These two models are detailed in Secs. \ref{subsec:syst_photoz} and \ref{subsec:syst_mcorr}. All other sources of uncertainty are identified and controlled by removing $\mathcal{S}/\mathcal{N}>4$ bins, which are identified from our {\it Systematics training set} as being contaminated beyond an acceptance threshold, or by verifying that they do not impact the best-fit cosmological parameters.  The following sub-sections detail our treatment of these sources of  systematic uncertainty for the peak count measurements;  the reader hungry for results can skip ahead to Section \ref{sec:results}.

\subsection{Photometric redshifts}
\label{subsec:syst_photoz}

As there is no analytic prescription to model the effect of photometric redshift uncertainty  on the peak function, we investigate its impact directly from the simulations: we infuse different shifts $\Delta z^i$ in the galaxy distribution of the four tomographic bins (similar to the treatment in the 2PCFs),  generate new mock light-cones and galaxy catalogues, and compute the peak statistics from these. Our approach is similar to the linear model adopted by K16, which computes a two-point numerical derivative from simulations produced with a (single) shift in the mean $n(z)$ by $\Delta z$. Our model is slightly more sophisticated: we sample 10 $\Delta z^i$ values  in every $\mathcal{S}/\mathcal{N}$ bin, drawing numbers randomly from Gaussian distributions with means of zero and standard deviations $\sigma_z$ given by J20 priors on the nuisance parameters (see Table \ref{table:priors}). In the case of cross-tomographic bins, we use the mean between the two $\sigma_z$ values.

For every shift,  we generate five light-cones and use these to cover the full survey (with the tiling strategy described in Sec. \ref{subsubsec:mosaic}); we next measure  $N_{\rm peaks}(\mathcal{S}/\mathcal{N})$ as a function of $\Delta z^i$, and  fit a straight line through these 10 points in order to extract the numerical derivative $\left(\partial N_{\rm peaks}/\partial \Delta z\right)$ for every $\mathcal{S}/\mathcal{N}$ and tomographic bin.  The linear fit captures well the response to changes in $\Delta z^i$, reaching signal-to-noise ratios between 5 and 40 depending on the bin,  except for the $6^{\rm th}$ bin where the peak function and the derivatives are consistent with pure noise. We carry out this calculation for the different aperture filters investigated in this paper, but also at two different cosmological models\footnote{The cosmo-SLICS model-FID has $\Omega_{\rm m} = 0.2905$, $S_8=0.8231$, $h=0.6898$ and $w_0=-1.0$, while  model-00 has $\Omega_{\rm m} = 0.3282$, $S_8=0.6984$, $h=0.6766$ and $w_0=-1.2376$.} (model-FID and -00) in order to assess the stability of the derivative with respect to cosmology. The results are shown in the upper panels of Fig. \ref{fig:dNpeaks_dz}. 

We note that the results for the two cosmological models are in qualitative agreement, where the response of high (low) $\mathcal{S}/\mathcal{N}$ peaks to an increased survey depth is positive (negative). This is caused by the fact that a greater depth  increases the shear signal, which shifts the peak function towards higher $\mathcal{S}/\mathcal{N}$ values. Some differences are observed  towards the large $\mathcal{S}/\mathcal{N}$ values. With model-00 being quite distant from model-FID -- notably a 12\% lower  value of $S_8$ -- we do not expect the derivatives to be identical, 
 but the impact of this difference is highly suppressed by the tight priors on $\Delta z^{ij}$. 
Given the current statistical uncertainty of the measurements however, we therefore ignore this cosmology dependence, but this will need to be revisited in the future. 
Similarly to K16, we choose to model the redshift uncertainty by scaling the measurement in each of the $\mathcal{S}/\mathcal{N}$ bins with this linear model prior to marginalising over
$\Delta z$ in the likelihood inference. Namely, we compute $N_{\rm peaks}^{ij}(\Delta z^{ij}) = N_{\rm peaks}^{ij}(\Delta z=0) + \left(\partial N_{\rm peaks}^{ij}/\partial \Delta z\right)\Delta z^{ij}$, using the derivative extracted from the model-FID cosmology, where $\Delta z^{ij}= \left( \Delta z^{i} + \Delta z^{j} \right) /2$.

\subsection{Shear calibration}
\label{subsec:syst_mcorr}

The uncertainty in the shear calibration is forward-modelled with a similar method, except  that no additional ray-tracing is required.
Instead we include a uniform $(1+\Delta m^i)$ correction factor at the catalogue level, which multiplies every observed ellipticities in Eqs. (\ref{eq:Map}) and (\ref{eq:MapNoise}). We next de-bias the peaks measurement with the original shear response  $S_a$ but deliberately ignore these additional  $\Delta m^i$ factors, resulting  in a net residual bias caused by an incorrect shear calibration, which is exactly what we wish to model. We repeat this process for 10 values of  $\Delta m^i$ sampled from the priors on the shear calibration errors (a Gaussian with width $\sigma_m= 0.023$, see Table \ref{table:priors}), we measure the peak function for each of these samples on the full survey, average over 5 light-cones, and fit a straight line to these points in every $\mathcal{S}/\mathcal{N}$ bin and tomographic bin, allowing us to compute derivatives and model $N_{\rm peaks}^{ij}(\Delta m^{ij})=N_{\rm peaks}^{ij}(\Delta m=0) + \left(\partial N_{\rm peaks}^{ij}/\partial \Delta m\right) \Delta m^{ij}$.  

The partial derivatives are calibrated this way  for cosmological models-FID and -00 and reported in the lower panels of Fig. \ref{fig:dNpeaks_dz}. We observe a global agreement between the two cosmologies, similar in shape to the effect of increasing the survey depth,  with the amplitudes of the derivative being slightly larger  towards the high $\mathcal{S}/\mathcal{N}$ bins for model-FID, which is  linked to its higher $S_8$ value. We ignore these differences in this work due to the small size of $\sigma_m$ relative to the statistical precision of the DES-Y1 data, and use solely the model-FID derivatives in the  likelihood sampling. However, this could be easily addressed with our approach: the derivatives $\left(\partial N_{\rm peaks}/\partial \Delta z\right)$ and $ \left(\partial N_{\rm peaks}/\partial \Delta m\right)$ could be estimated at our 26 cosmological nodes, from which we could train a Gaussian process emulator the same way we model our signal. This improved accuracy treatment of the derivatives will  likely need to be included in future analyses.

We finally note that in contrast with the 2PCFs, where shifts in  $\Delta m$ affect all scales equally, the  derivative presented above exhibit a more complicated structure, caused by the fact that the $m$-calibration affects both the shear estimate and the noise in a non-trivial manner (see Eqs. \ref{eq:Map} and \ref{eq:MapNoise}).

We present in Appendix \ref{subsec:GPR_validation} a comparison between a cosmological inference in which  $\Delta z^i$ and $\Delta m^i$ are allowed to vary, and one where these two are held fixed at zero, and notice that while no biases on the preferred parameters are observed, the uncertainty about $S_8$ almost doubles in the former case.

\subsection{Baryon feedback}
\label{subsec:syst_baryons}

The uncertain impact of baryonic feedback on the peak count statistics has received an increasing degree of attention over the last few years \citep{Osato2015,  Weiss_Peaks_Baryons, MinimaPeaks}. The current interpretation  can be summarised as follows: radiative pressure from sustained stellar winds, combined with supernovae explosions and AGN activity combine to expel gas towards the outer regions of the haloes. These mechanisms are maximally efficient on medium-size (e.g. $10^{14}M_{\odot}$) clusters \citep[e.g.][]{BAHAMAS}, since light haloes generally do not host AGNs, while heavier haloes manage to keep the material inside due to their deeper gravitational  wells. This redistribution of matter tends to reduce the number of high $\mathcal{S}/\mathcal{N}$ peaks, and possibly augment that of smaller $\mathcal{S}/\mathcal{N}$ values, however the exact size of this effect is highly uncertain and depends on the feedback model adopted. Just as for cosmic shear two-point correlation functions, its significance depends on the noise level of the data. We note that for less massive haloes, stellar feedback is also important, however this occurs at significantly smaller scales not probed by our filter. Also, radiative cooling at high redshift should produce more concentrated haloes and could enhance the lensing signal, acting in the opposite direction of the AGN feedback.

\begin{figure}
\begin{center}
 \includegraphics[width=3.3in]{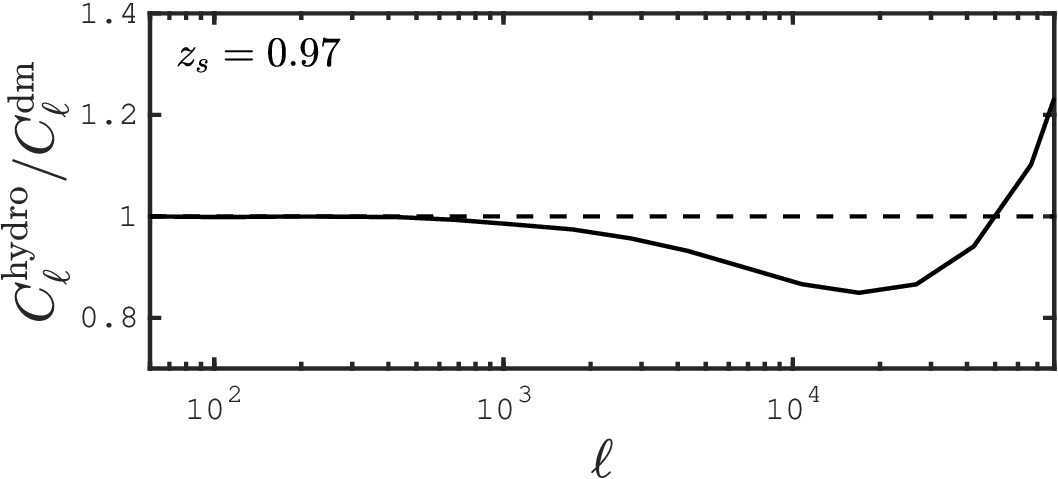}
\caption{Effect of baryon feedback on the convergence power spectrum measured from the {\it Magneticum} simulations, assuming a fixed redshift for the sources.  }
\label{fig:BaryonFeedback}
\end{center}
\end{figure}

The impact on the lensing power-spectrum computed on a single redshift slice (taken to be $z_{\rm s} = 0.97$ here) of the {\it Magneticum} reaches 15\% at high-$\ell$, as seen in Fig. \ref{fig:BaryonFeedback}, an amplitude that is similar to those of the BAHAMAS simulations mentioned in Sec. \ref{subsubsec:magneticum}, and which are consistent with the PCA constraints.

To investigate the relative impact of baryons on the peak function specifically for our analysis, we tile the full DES-Y1 survey with the {\it Magneticum} light-cones introduced in Sec. \ref{subsubsec:magneticum} either with the full hydrodynamical simulations or with the gravity-only solution, then evaluate and compare the peak functions.  We repeat this process and average the results over the 10 {\it pseudo}-independent light-cones, each further sampled with 10 shape noise realisations. We also extend the data vector to higher $\mathcal{S}/\mathcal{N}$ values in order to verify where the baryons start to play an important role. We additionally repeated the process in a non-tomographic set-up, where the catalogues from the four redshift bins are merged before producing the aperture mass map and counting the peaks; this reduces the impact of the pure noise peaks to highlight subtle effects that occur in the underlying matter  distribution.

\begin{figure*}
\begin{center}
\includegraphics[width=7.0in]{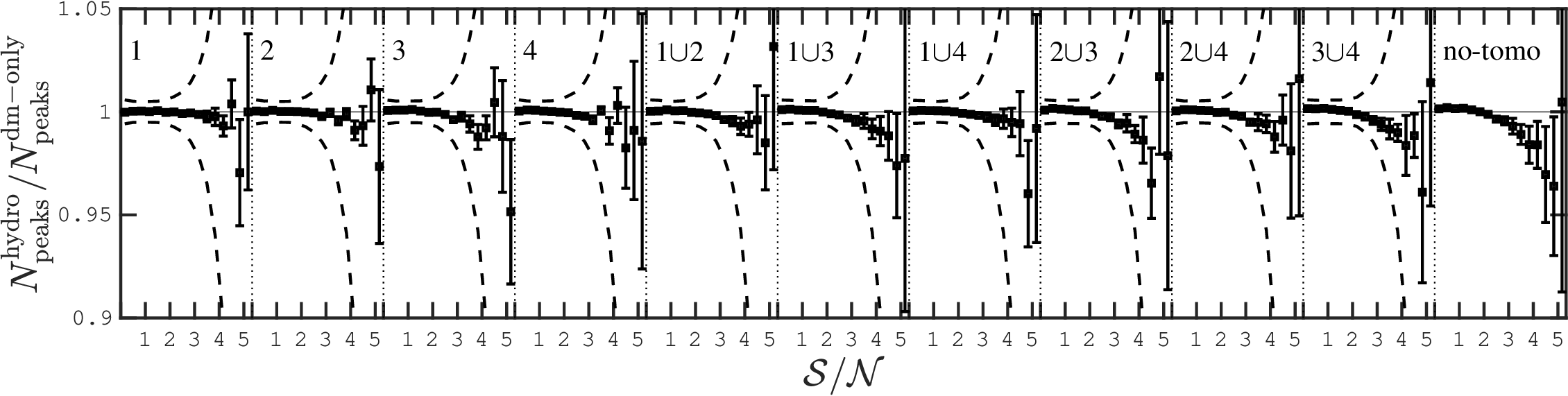}
\caption{Ratio between the number of peaks measured in the {\it Magneticum} light-cones with and without including the baryonic physics. Results are shown for different tomographic bins, and for an aperture filter size of $\theta_{\rm ap}=12.5$ arcmin; other filters show a similar relative effect. The dashed lines represent the statistical precision, also plotted in Fig. \ref{fig:GPR_CV}. }
\label{fig:BaryonFeedback2}
\end{center}
\end{figure*}

The results are shown in Fig. \ref{fig:BaryonFeedback2} for all tomographic bins, as well as for the case where no tomography is applied. We see that the effect is generally under a percent for $\mathcal{S}/\mathcal{N}<4$, and in the no-tomographic case reaches five percent for $4<\mathcal{S}/\mathcal{N}<5$. The statistical precision is also reported on these plots, which shows that the impact of baryons is everywhere sub-dominant compared to the uncertainty on the GPR emulator.  This reinforces our choice of selecting $\mathcal{S}/\mathcal{N}\le4$ to ensure a measurement mostly free of uncertainty related to baryons, although this suggests that we could push the upper limit to higher $\mathcal{S}/\mathcal{N}$ in the tomographic case without much contamination, and that modelling the effect could be relatively straight-forward. This follows a logic similar to  \citet{2020arXiv200715026H}, where the impact of baryonic feedback on the DES-Y1 2PCFs is modelled and captured with a PCA decomposition, allowing them to include smaller angular scales in the data vector and increase the constraints.

We use the ratios presented in Fig. \ref{fig:BaryonFeedback2} to construct a multiplicative correction factor that is optionally applied to the data vector during the cosmology inference pipeline, from which we can estimate the impact of baryon feedback on the recovered best-fit parameters, as modelled with the {\it Magneticum} baryon model. 
We note that  a similar approach is adopted in the context of a Stage-IV survey in   \citet{Weiss_Peaks_Baryons} and in \citet{Martinet21}, where the increased galaxy density and overall statistical precision accentuates the bias caused by the baryons.

\subsection{Galaxy intrinsic alignment}
\label{subsec:syst_IA}

\begin{figure}
\begin{center}
 \includegraphics[width=3.3in]{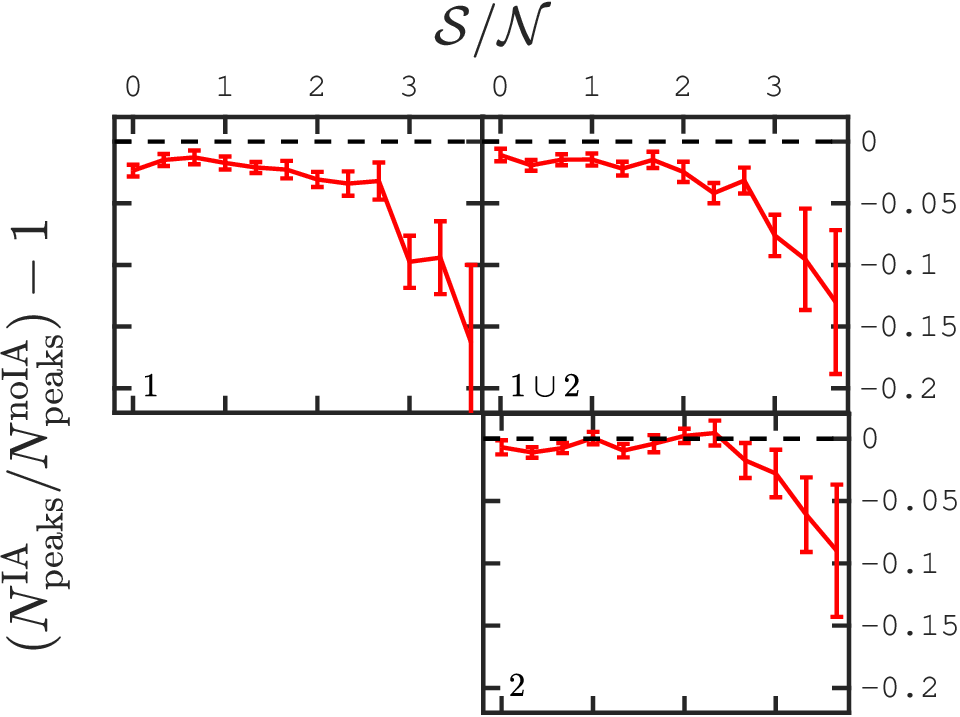}
\caption{Effect of intrinsic alignment on the peak count statistic, measured in the lowest two redshift bins of our dedicated IA training set. The error bars show the error on the mean. Full details of this model are provided in Appendix \ref{sec:syst_IA}.}
\label{fig:peaks_IA}
\end{center}
\end{figure}

The intrinsic alignment of galaxies is an astrophysical systematic signal that mimics weak lensing measurements, and arises from the fact that the intrinsic  orientation of galaxies is not exactly random \citep[see][for a review]{Kirk2015,2015SSRv..193...67K}. Indeed, it has been shown in multiple hydrodynamical simulations that the formation of galaxies, and thus their final shape and alignment, is affected by their environment, notably by the  neighbouring large scale structures \citep{Chisari_IA2, 2015MNRAS.448.3391C} and tidal fields \citep{Catelan_IA_Tidal}, and by a complex relation with their host haloes \citep{Chisari_IA}. The observed galaxy shape is therefore a combination of the intrinsic ($I$) and the shear ($G$) term, which both contribute to the measured weak lensing signal. 

Intrinsic alignments have been directly measured and constrained, notably  in the COSMOS galaxies by \citet{2013MNRAS.431..477J}, who detect the signal for early-type (e.g. red galaxies) but hardly any signal  for late-type galaxies. The WiggleZ blue galaxies were also found to be consistent with no IA in \citet{BlueIA}, while a significant IA signal was found in the BOSS LOWZ  sample \citep{Singh_IA_LOWZ}. \citet{Johnston_IA} found similar results from the KiDS, SDSS and GAMA surveys, with a null detection from the blue galaxies and a $9\sigma$ detection for an IA signal for red galaxies, consistent with a value of $A_{\rm IA} = 3.18_{-0.46}^{+0.47}$ when interpreted in the NLA model. 

The IA signal has also been indirectly inferred  from cosmic shear measurements, although with some dispersion in the results. For example, T18 finds a $2.5\sigma$ detection of the signal from the DES-Y1 data, with $A_{\rm IA}=1.3^{+0.5}_{-0.6}$, the KV450 analysis by \citet{KV450} found an $A_{\rm IA}$ value consistent both with unity and with zero, while \citet{HSCY1_Cell}   and \citet{HSCY1_2PCF} prefer a values of  $A_{\rm IA}=0.38\pm0.70$ and $0.91^{+0.27}_{-0.32}$  from the power spectrum and 2PCFs analyses of the HSC-Y1 data, respectively. Similar variations of the best fit IA amplitudes  are observed in the KiDS-1000 cosmic shear analysis by  \citet{KiDS1000_Asgari},  who found $A_{\rm IA}$ values between $0.26_{-0.33}^{+0.42}$ and $0.97_{-0.38}^{+0.29}$ depending on the estimator.  These $A_{\rm IA}$ measurements are not expected to agree perfectly given the differences in the modelling of the IA signal, but also in the redshift, in the physical scales that are probed  and in the selection of galaxies in these surveys. It is also worth pointing out  that the IA nuisance parameter marginalisation is degenerate with the photo-$z$ errors as well \citep[see, for example, the discussion in Appendix C of][]{KiDS1000_Heymans,Wright_KV450_SOM}. For our 2PCFs analyses, we use the same two-parameters NLA model as in T18, with priors on  $A_{\rm IA}$ and $\eta$ listed in Table \ref{table:priors}.

The impact of IA on peaks statistics has not been well studied in the literature so far, and a percent level calibration will require a significant level of development beyond what has been done so far.
Nevertheless we present here a first step in this direction, with a measurement of the effect based on in-painting galaxies with intrinsic shapes determined by properties of the dark matter haloes contained inside the lensing light-cones. Again, the  amplitude of the IA signal measured from peaks is not expected to be the same as for the two-point shear correlation function, largely because the physical scales and the number of galaxies involved in each estimator calculations are different.

Within the NLA model  of \citet{2007NJPh....9..444B} with  $A_{\rm IA}=1.0$ for example, intrinsic alignments can modify by up to 40\% the $\xi_{\pm}$ signal in the DES-Y1 bin 1, 20\% in bin 2, but only about 5\% in bins 3 and 4. Even if the effective $A_{\rm IA}$ increases with redshift \citep[see the discussion in Appendix A of][]{ACTxKiDS}, the lensing kernels of the $GI$ and $II$ terms are suppressed compared with the $GG$ signal. Considering  the IA model of \citet{Fortuna2020}, we note that the IA effect is more significant at small physical scales, which are only well resolved at low redshift. For these reasons, we choose to only model the peaks IA signal in the low redshift bins, more precisely on bins  1, 1$\cup$2 and 2. This likely leaves residual, unmodelled $GI$ and $II$ terms present in the  bins 1$\cup$3,  1$\cup$4, 2$\cup$3 and 2$\cup$4. Within the NLA model the $II$ term is completely sub-dominant in the cross-tomographic bins and can be neglected. We therefore estimate the impact of unmodelled  $GI$ contributions in an analysis variation in which we remove all cross-tomographic bins. As it will become clear in Sec. \ref{subsec:results_GPR}, this is currently a limiting factor in our data analysis, which we will seek to improve with a better IA model in the future.

Our IA model is inspired from the methods of \citet{2006MNRAS.371..750H} and \citet{2013MNRAS.436..819J}, which assign an intrinsic ellipticity ${\boldsymbol \epsilon}_{\rm int}$ to the galaxies based on the shape
 of their host dark matter halo (we summarise this method and detail our implementation in Appendix \ref{sec:syst_IA}). The model requires both light-cone haloes and in-pasted galaxies, two intensive post-processing steps that have not yet been completed on the cosmo-SLICS. It is therefore not possible at the moment to explore this in a cosmology-dependent manner.  Instead, we use 26 light-cones from the KiDS-HOD galaxy sample described in HD18, which have these properties and have been downsampled to closely match the $N(z)$ in the four DES-Y1 tomographic bins. These also cover 100 deg$^2$ each, and since we are only interested in the relative effect, we do not tile the full footprint and work instead directly on the light-cone galaxy samples. The effect of masking is hence not included, but it should equally impact the measurements with and without IA in these mocks.

For every galaxy, the model outputs ${\boldsymbol \epsilon}_{\rm int}$; this quantity is then inserted in Eq. (\ref{eq:eps_obs}) alongside a randomly rotated version, from which we compute  observed ellipticities with or without IA. We finally run our peak finder on these catalogues, count the peaks as for the other training sets, and compare the measurements in Fig. \ref{fig:peaks_IA}.
We observe an important (10-15\%) suppression of the number of peaks with $\mathcal{S}/\mathcal{N}>3$, which clearly exceeds the statistical uncertainty in our measurement and therefore  needs to be accounted for. Moreover, in the top two panels, peaks at all $\mathcal{S}/\mathcal{N}$ values are suppressed by a few percent. We note that the results from Fig. \ref{fig:peaks_IA} align well with those found in K16 (see their figure C3), even though a direct comparison is impossible due to differences in the source distribution of the samples. When examined with 2PCFs measurements, we find that our IA prescription is bracketed by the NLA model with $A_{\rm IA} \in [1.0 - 2.0]$, providing a consistent but slightly larger IA signal than that preferred by the data (see Appendix \ref{sec:syst_IA}).

We recognise that our simple IA model is unlikely to represent accurately the real physical effect, and at the moment has no free parameter, which means that we cannot yet marginalise over different IA strengths like we do for 2PCFs. Further development on the IA model will be required to achieve this in the future. However, we can get a sense of the impact of IA by infusing the relative effect observed in Fig. \ref{fig:peaks_IA} in the model returned by the emulator, and record the deviation from the no-IA case on the best-fit parameters. It turns out that this results in a $\sim$0.1$\sigma$ shift on the cosmological parameters, hence we do not include it in the fiducial peak count pipeline.

\subsection{$N$-body force resolution}
\label{subsec:highres}

\begin{figure}
\begin{center}
\includegraphics[width=3.3in]{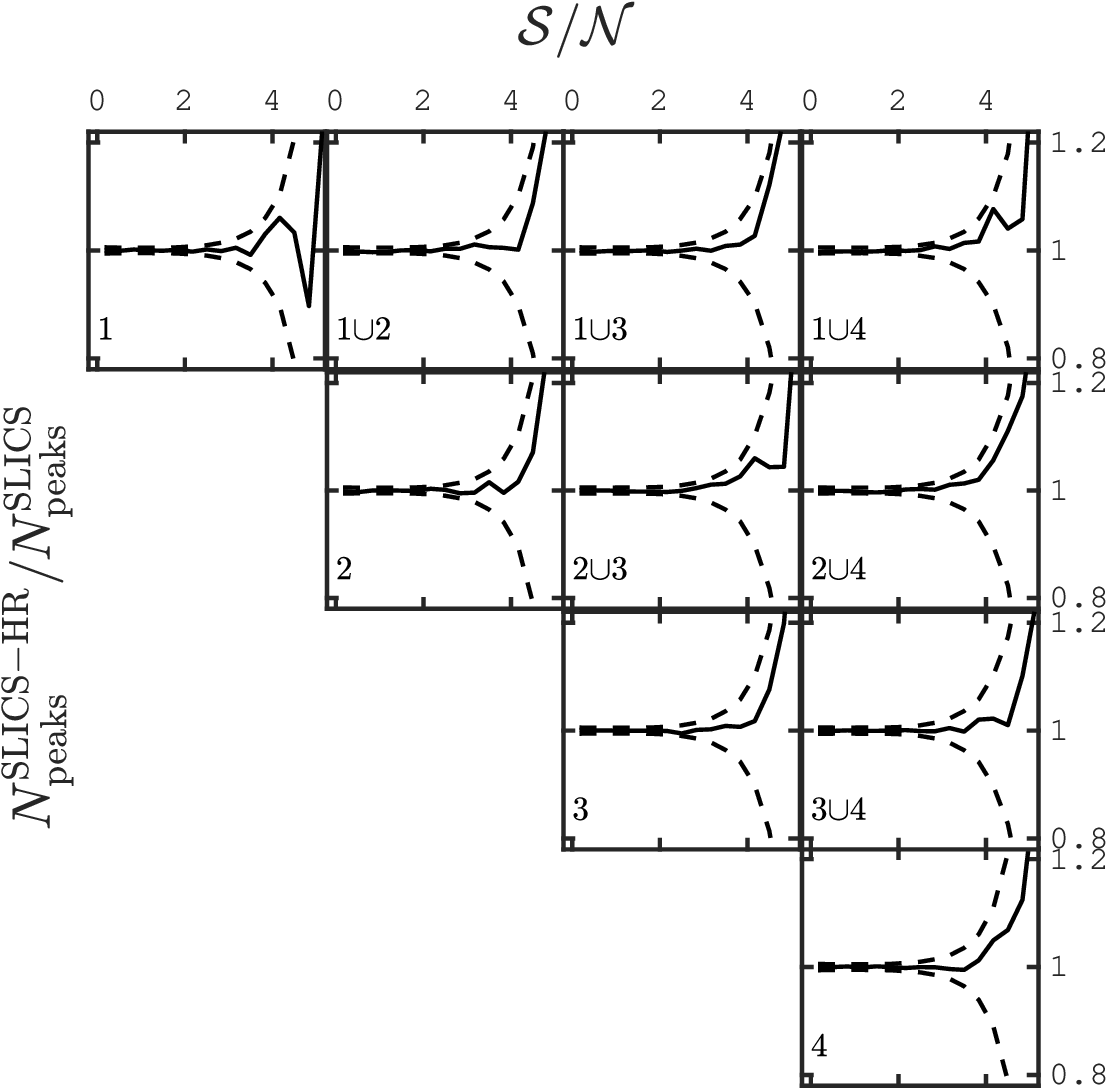}
\caption{Impact of the $N$-body force resolution on the peak statistics, measured from the ratio in $\mathcal{S}/\mathcal{N}$ between the SLICS-HR and the covariance training set.
The dashed-lines show the statistical error of the measurement. Deviations for $\mathcal{S}/\mathcal{N}\le4$ peaks are generally under $0.2\sigma$.}
\label{fig:HR_vs_LE}
\end{center}
\end{figure}

A large fraction of the weak lensing signal receives a contribution from small scales, which are difficult to model accurately. Even in simplified gravity-only scenarios, different methods and codes to estimate the amount of small-scale structures vary significantly for wave vectors larger than $k=1.0 \,h^{-1} {\rm Mpc}$. Fitting functions such as  {\sc Halofit} \citep{Smith03, Takahashi2012}, simulation-based emulators \citep{Coyote3, MiraTitan, DarkEmulator} and halo model calculations \citep{MeadFit, HMCode2020} provide fast predictions for two-point function statistics, but these disagree sometimes at the 5-10\% level, depending on the scales, redshifts and cosmological parameters. Recent efforts approach the 1\% accuracy   on the matter power spectrum \citep{EuclidEmulator, HMCode2020}, at least for a subset of the cosmological parameter space. 

It is possible to protect the shear 2PCFs measurement against residual non-linear  uncertainty by rejecting angular scales in $\xi_{\pm}(\vartheta)$ where differences between these models affect the cosmological inference beyond some threshold\footnote{There is a caveat to this argument, see the discussion in \citet{Asgari_DES_KiDS_cosebi} about the small $k$-scale power leaking into $\xi_\pm(\vartheta)$ to some level at all $\vartheta$.}. This is one of the main drivers, along with baryon feedback, for the choice of angular scales in the T18 2PCFs analysis (and hence ours). 

For weak lensing probes beyond two-point statistics that are calibrated directly on  simulations however, one must additionally understand the impact of  small-scale smoothing caused by limits in the mass/force resolution of the $N$-body code.  Specifically, higher-resolution simulations better resolve the highly non-linear, small-scale structures that describe the concentrated inner regions of massive clusters, which are responsible for the high SNR peaks. Therefore, degrading the mass resolution directly affects both the high-$k$ limit of the matter $P(k)$ and the number of large lensing  peaks,  leading to a potential mis-calibration.

To assess this, we rely on the SLICS-HR simulations introduced in Sec. \ref{subsubsec:SLICS_HR}, in which the force resolution has been significantly augmented, thereby resolving scales almost an order of magnitude smaller. A comparison between the peak function of the SLICS-HR measured on 10 realisations of the full DES-Y1 footprint and that of our main  {\it Covariance training set} is presented in Fig. \ref{fig:HR_vs_LE}; it reveals slight differences for the $\mathcal{S}/\mathcal{N}$ peaks we are probing, but strong deviations are observed for peaks with $\mathcal{S}/\mathcal{N} > 4$. These objects are rare, which explains the increased noise in the ratio towards $\mathcal{S}/\mathcal{N}=5$, but the trend is clear: there is an overall shortcoming of  large $\mathcal{S}/\mathcal{N}$ objects in the training set  compared to the SLICS-HR, which justifies our choice of restricting the data vector to $\mathcal{S}/\mathcal{N}\le4$. Upon closer examination, the average size of the small deviations seen in that range correspond to  no more than 20\% of   the statistical error,  and are never higher than 0.5$\sigma_{\rm stat}$. If we wanted to include these peaks in a future analysis, we would possibly need a new generation of training sets (for cosmology, and possibly covariance) with an increased mass resolution.

Just as for IA and baryons, we investigate the impact of mass resolution by extracting a correction factor from the black curves shown in Fig. \ref{fig:HR_vs_LE}, which we consequently  apply to the signal during the cosmology inference.

 \subsection{Source-lens clustering}
 
\begin{figure}
\begin{center}
\includegraphics[width=3.3in]{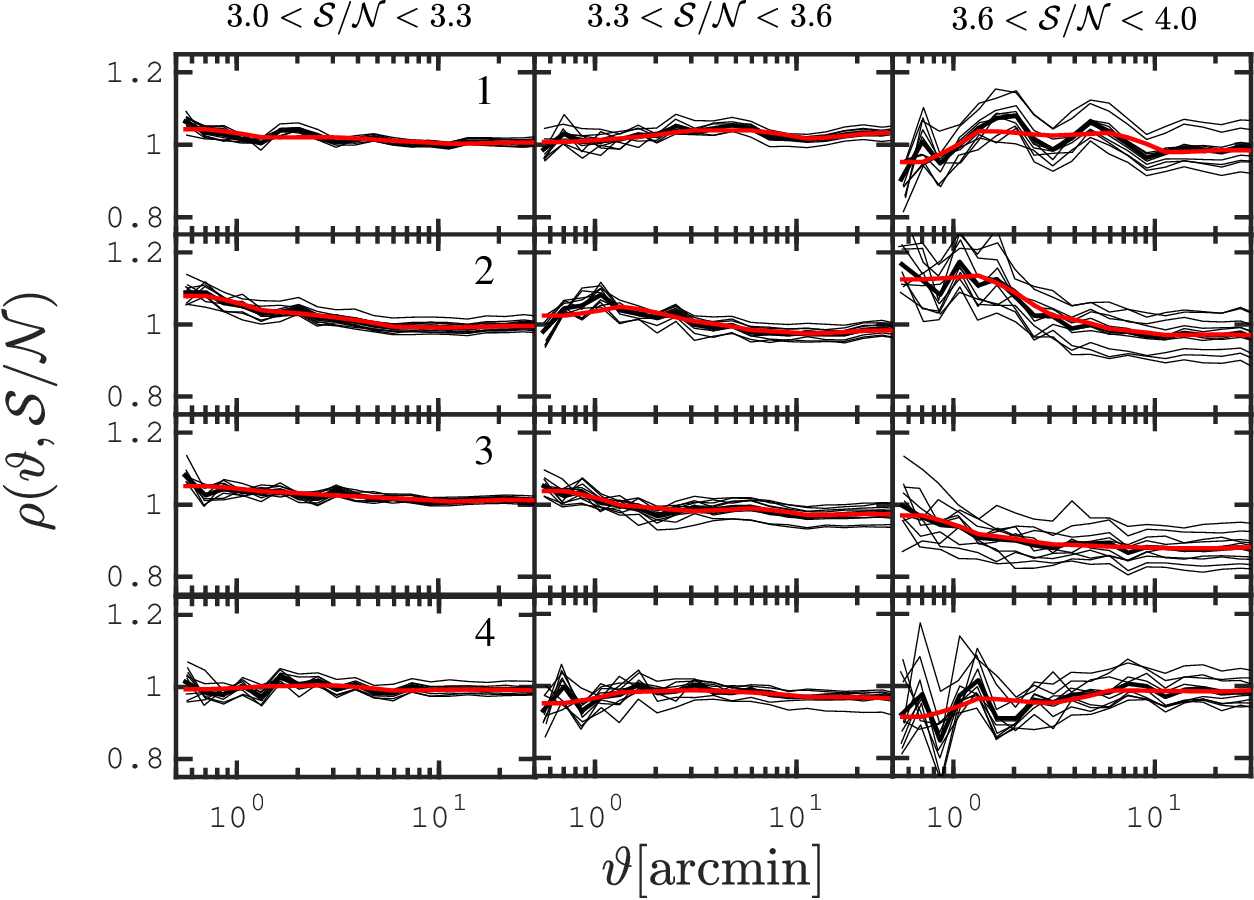}
\caption{Profile of the excess galaxy clustering, as a function of angular separation. The different columns show the profiles for  three different $\mathcal{S}/\mathcal{N}$ bins,  while the rows present the results for the four tomographic bins. These profiles (red lines) are averaged over 10 independent survey realisations and enter in the boost factor correction  (Eq. \ref{eq:filter_boost_profile}, see main text). }
\label{fig:profiles}
\end{center}
\end{figure}

 \begin{figure}
\begin{center}
\includegraphics[width=3.3in]{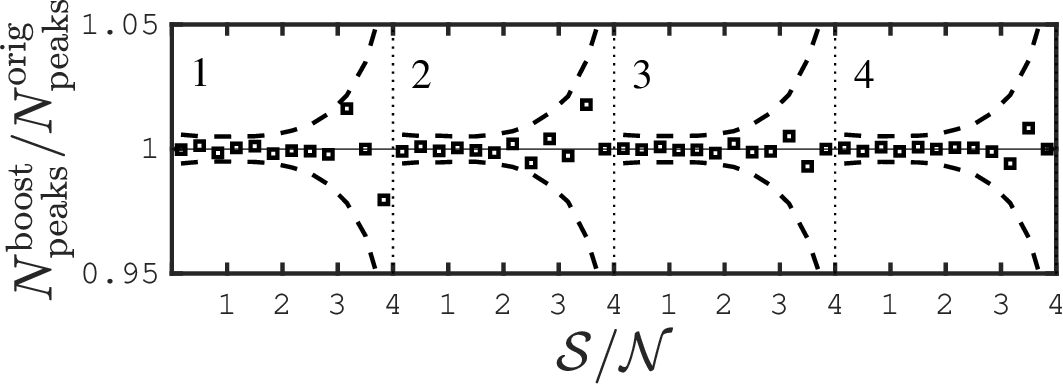}
\caption{Impact of the boost factor on the peak statistics, measured from the ratio in $N_{\rm peaks}$ between the boost-corrected  and the original peak count. The dashed-lines show the statistical error of the measurement. Similar results are obtained for the cross-tomographic bins.}
\label{fig:N_peaks_boost}
\end{center}
\end{figure}

\begin{figure*}
\begin{center}
\includegraphics[width=7.0in]{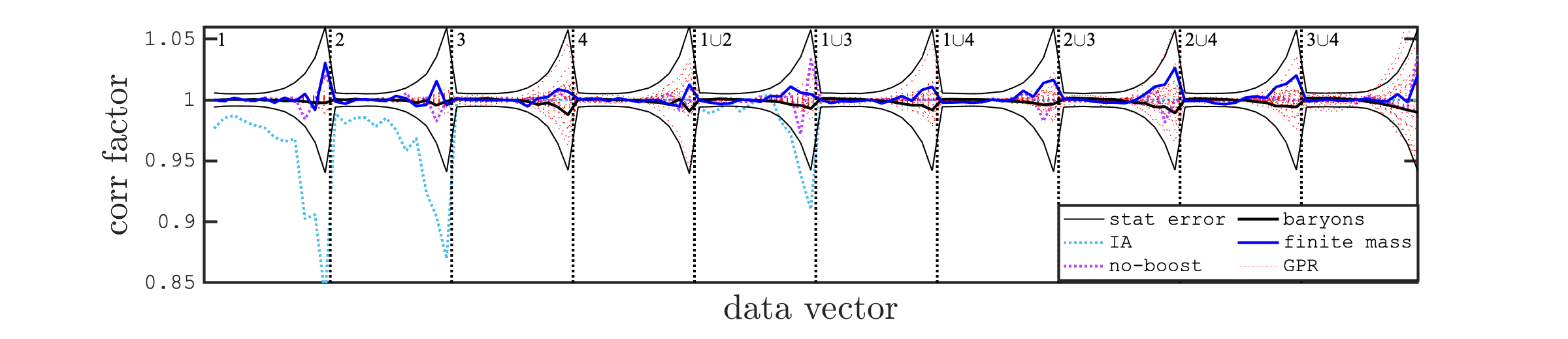}
\caption{Impact on the peak statistics of various sources of systematic uncertainty (IA, mass resolution, baryons and boost), presented as a ratio between the measurement on mocks with and without the effect. These are used as optional correction factor applied to the model in the cosmology inference, as described in Sec. \ref{subsec:results_GPR}. Also shown is the scatter in the GPR cross-validation test, as well as the statistical precision on the measurements.}
\label{fig:corr_factors}
\end{center}
\end{figure*}

 One of the key differences between the mock DES-Y1 simulations from our {\it Cosmology training set} and the DES-Y1 data is the presence of source-lens coupling. In real data, the source density is not homogeneous and in fact increases around foreground clusters. As explained in Appendix C of K16, this introduces a coupling between the peak positions and the amplitude of the measured shear relative to the expected shear -- the sources that are associated with the cluster dilute the observed signal. Furthermore, these regions of high galaxy densities will  have a larger blending rate, meaning that source galaxies behind clusters are more likely to be missed,  while residual errors in the photometric redshift can wrongly assign cluster members to background sources. When combined, these effects can result in a mis-calibration between the data and the simulations, which can be corrected with a `boost factor' \citep{2005MNRAS.361.1287M}. 

Boost factors for peak statistics can be evaluated in different ways. K16  estimate the  fractional over-crowding and over-blending rates in peaks of different $\mathcal{S}/\mathcal{N}$ from the {\sc Balrog} catalogue \citep{Balrog}, a separate image simulation that matches the DES-SV $n(z)$  and blending properties. These rates are computed as a function of distance to peaks centres, and a correction factor is used to correct the peak function found in their cosmological training set as a function of $\mathcal{S}/\mathcal{N}$. They found that by restricting their measurement to $\mathcal{S}/\mathcal{N}<4$, the impact is minimal (a shift in $S_8$ of about 0.01) and could be neglected. 

\citet{Shan18} instead use the \citet{Radovich17} cluster catalogue that overlaps with the KiDS-450 survey, and evaluated the boost factor from the excess source density around these massive objects. They  found  that the contamination to the peak function  reaches 27\% for peaks with  $\mathcal{S}/\mathcal{N}=5$, however it caps at   less than 6\%  for $\mathcal{S}/\mathcal{N}<4$.
 In their analysis, this effect of source-coupling is about twice the size of that of their baryonic feedback model, and acting in the same direction, e.g. suppressing the number of high $\mathcal{S}/\mathcal{N}$ peaks. If overlooked, this   mis-calibration could lead to a best-fit inference with a $S_8$ that is too low.

We account for source-lens coupling by estimating the effect in the DES-Y1 data and recalibrating our measurements, leaving the simulations unchanged.
We first extract $w_{\rm data}^{ij}(\vartheta, \mathcal{S}/\mathcal{N})$, the clustering of source galaxies along the line-of-sight of peaks identified in the data, as a function of their peak height, 
for each combination of auto- and cross-tomographic bins $`ij'$. 
These are next compared with a similar measurement carried on ten survey realisations sampled from the {\it Cosmology training set} at the fiducial cosmology, and the ratio of the two reveal clustering profiles  in excess of random:
\begin{eqnarray}
\rho^{ij}(\vartheta, \mathcal{S}/\mathcal{N}) \equiv \frac{w_{\rm data}^{ij}(\vartheta, \mathcal{S}/\mathcal{N})}{\langle w_{\rm sim}^{ij}(\vartheta, \mathcal{S}/\mathcal{N}) \rangle} \, \times \,\left(\frac{\langle N_{\rm peaks}^{\rm sim}\rangle}{N_{\rm peaks}^{\rm data}}\right),
\end{eqnarray} 
which are shown in Fig. \ref{fig:profiles} for a sample of tomographic bins and $\mathcal{S}/\mathcal{N}$ bins. The brackets refer to the mean measurement in the above expression, and the right term involving $N_{\rm peaks}^{\rm data/sim}$ normalises the profiles.  It is clear from this figure that the largest peaks are generally more severely affected by this boost factor correction, however the  size of effect varies across redshift in a non-trivial manner.  For example, the fourth tomographic bin is less affected than the second or the third.
In absence of source-lens coupling these profiles would be flat. The excess of galaxies in these profiles are for the most part cluster members; their shapes are therefore not sheared by the foreground matter over-density and only dilute the lensing signal. We compensate for this by up-weighting the shear signal following the profile, which is most efficiently done by modifying the filter $Q(r)$ in identified peaks as:
\begin{eqnarray}
Q(r) = Q_{\rm orig}(r) \, \times \,\rho^{ij}(r, \mathcal{S}/\mathcal{N}),
\label{eq:filter_boost_profile}
\end{eqnarray} 
and re-evaluate the peak height with Eq. (\ref{eq:Q}). Fig. \ref{fig:N_peaks_boost} shows the ratio between the corrected and original peak function in the four auto-tomographic bins (similar results are obtained with the cross-bins). The effect if generally small, however it approaches the $1\sigma$ level in some isolated data elements.  The boost factor is included in our fiducial analysis, and has been applied to the data points shown in Fig. \ref{fig:N_peaks},   bypassing the need to forward-model the source-lens coupling at all cosmological points. Following the method used in the previous sections, we isolate its impact  on our cosmological inference by optionally removing this correction factor.

Fig. \ref{fig:corr_factors} summarises all the correction factors we can include in our cosmology inference, from IA, mass resolution, baryons and no-boost. Our fiducial analysis includes none of them, since their overall impact is relatively minor  and many of these effects act in opposite directions.

\subsection{Cosmology inference}
\label{subsec:syst_cosmo_inference}

\begin{table}
   \centering
   \caption{Priors used in the likelihood sampling. The ranges for the four cosmological parameters  are determined by the cosmo-SLICS simulations, the photometric redshift ranges and priors are taken from the DIR errors found in J20, while those of the shear calibration and intrinsic alignments are taken from T18. Gaussian priors are characterised by a mean and a standard deviation $(\mu,\sigma)$.}
   \tabcolsep=0.11cm
      \begin{tabular}{@{} cccccccc @{}} 
      \hline
      \hline
       Parameter       &  range & prior \\
       \hline
       Cosmology\\ 
       $\Omega_{\rm m}$ &[0.1, 0.55] & Flat\\
       $\sigma_8$ & [0.53, 1.3] & Flat\\
       $h$ & [0.6,  0.82] & Flat\\
       $w_0$ & [-2.0, -0.5]& Flat\\
       \hline
       Nuisance\\
       $\Delta z^1\times10^2$ & [-10, 10] & $\mathcal{G}(0,0.8)$\\                    
       $\Delta z^2\times10^2$ & [-10, 10]&  $\mathcal{G}(0,1.4)$\\                    
       $\Delta z^3\times10^2$ & [-10, 10]& $\mathcal{G}(0,1.1)$\\                    
       $\Delta z^4\times10^2$ & [-10, 10]&  $\mathcal{G}(0,0.9)$\\                    
       $\Delta m^i\times10^2$ & [-10, 10]& $\mathcal{G}(1.2,  2.3)$ \\  
        \hline
       IA\\
       $A_{\rm IA}$ & [-5, 5]&  Flat\\                    
       $\eta$ & [-5, 5] &  Flat\\                    
     \hline
    \hline 
    \end{tabular}
    \label{table:priors}
\end{table}

We test our cosmology inference pipelines on mock data vectors taken to be the mean value from the {\it Covariance training set}, providing a measurement that is almost  noise-free. We  present our results in Appendix \ref{sec:syst_pipeline}, notably in Fig. \ref{fig:pipeline_validation_wCDM_2pcf_GPR}. This exercise reveals a high degree of similarity in the constraints between the two-point functions, the peak statistics and the fiducial T18 analyses, despite major differences in our covariance matrix estimation techniques and in the observation data vector (DES-Y1 data vs SLICS, 2PCFs vs peaks). The best-fit cosmological parameters are also consistent with the input at the $1\sigma$ level, with no noticeable shift between the probes, and the sizes of all contours closely match that of the DES-Y1 analysis, two properties that  respectively validate our cosmology calibration and our covariance matrix.

 \subsection{Others sources of uncertainty}

Our method relies on a certain number of well-justified approximations, which could potentially contribute to the error budget in addition to the systematic effects described above. 
In this section we introduce these effects, and  justify our choice to neglect them in this study.

Our simulated light-cones are constructed with mass planes of constant thickness set to $256.5 \,h^{-1}$Mpc (see Sec. \ref{subsec:lightcone}), a choice which has an effect on the reconstructed lensing planes compared to a construction made of hundreds of finer shells. This has been recently quantified  in \citet{LensingHyperparameters}, where it is shown that the impact on peaks with $\mathcal{S}/\mathcal{N}\le4$ is below the one percent level, regardless of the thickness and of the source redshift. 

Correlations between the mass planes in our light-cones are explicitly suppressed by randomly shifting and rotating the mass sheets, which breaks the long line-of-sight correlations that exist in the data. It was shown in \citet[][see their Appendix B]{HSCmocks} that this affects the projected power-spectrum, reducing the power at intermediate scales by a few percent on the sheets, however, the lensing kernels project an even larger volume and mixes these scales, which makes our measurements relatively immune to this.

The lensing plane construction has been carried out under the Born approximation (see HD18), whereby light bundles record the convergence and shear along straight lines and ignore the deflection angles in this calculation. It has been shown  \citep{Hilbert_LensSimAccuracy} that  the difference between these methods induces  variations smaller than 0.2\% on the lensing power spectrum up to scales of $\ell=2\times 10^4$. It is therefore reasonable to expect that Born approximation plays a minor role  in the peak statistics as well, however \citet{LensingPDF_baryons} find that the impact  on the PDF of the lensing maps is of a few percent. We ignore this effect at the moment, but it will need to be revisited in the future.

Finite box effects are also known to plague the estimation of 2PCFs and of their covariance matrix \citep{SLICS_1}, being sensitive to Fourier modes larger than the simulation box.
This has an impact on the $\xi_{\pm}$ covariance matrix estimated from our {\it Covariance training set}, however it has a minor impact on the cosmological contour, as shown by the good match between our analysis on the mocks and that of T18. Furthermore, since the peak statistics measure quantities in local apertures, it is not sensitive to these large scales, and hence are protected against this. Similar conclusions can be drawn regarding the incomplete account of the super-sample covariance \citep[][SSC hereafter]{SSC} captured by our simulations: HD19 found that the SLICS light-cones capture more than 75\% of this SSC term, yielding two-dimensional constraints on $(\Omega_{\rm m}, \sigma_8, h, w_0)$ that match  to better than 10\% those of an analytical covariance matrix. Given the suppressed sensitivity of the peak statistics to these large-scale modes, we expect the residual missing SSC contribution to play a minor role on the full uncertainty, although this may need to be better quantified in the future.

It has been recently shown that the depth variability across a lensing survey can impact the cosmic signal and variance \citep{Sven_SurveyDepth, KiDS1000_Joachimi}. This will need to  be the subject of future investigations. Given the findings of \citet{KiDS1000_Joachimi}, we expect the impact of unmodelled variable depth to   be negligible, given the statistical power of DES-Y1.

\section{Results}
\label{sec:results}

\begin{table}
   \centering
   \caption{Properties of the different pipelines discussed in this paper.}
   \tabcolsep=0.11cm
      \begin{tabular}{@{} ccccc @{}} 
      \hline
      \hline
                    & prior on   &  Cov. &$n(z)$ \\
      pipeline &   $\sigma_8/A_{\rm s}$  &   Matrix & method  \\

       \hline
        This work                                  & $\sigma_8 \in [0.53 - 1.3]$  &SLICS &     DIR \\
        DIR-$w{\rm CDM}$                & $A_{\rm s} \in [0.5 - 5.0]\times10^{-9}$             &SLICS &     DIR \\
       T18                                      & $A_{\rm s} \in [0.5 - 5.0]\times10^{-9}$  &Analytic &     Stacked PDF \\
       J20                                    & ${\rm ln} \left(10^{10} A_{\rm s}\right) \in [1.5 - 5.0]$  &Analytic &     DIR \\
    \hline 
    \hline
    \end{tabular}
    \label{table:pipelines}
\end{table}

\begin{table}
   \centering
   \caption{Cosmological pipeline comparison. The values used in the T18 comparison are taken from their Table 3, using a fixed neutrino mass density. Details on the different pipelines are summarised in Table \ref{table:pipelines}. Most posteriors on $h$ and $w_0$ are prior-limited, so no constraints are reported here. The same applies to $\Omega_{\rm m}$ in many cases. Tests on the mock data are presented in Appendix \ref{sec:syst_pipeline}.}
   \tabcolsep=0.11cm
      \begin{tabular}{@{} llcc @{}} 
      \hline
      & pipeline&            $S_8$    &      $\Omega_{\rm m}$\\
      \hline
   					\hline		
   \multirow{3}{*}{Fiducial}	& Peaks & $0.780^{+0.019}_{-0.056} $ & - \\
   					& 2PCFs  & $0.753^{+0.043}_{-0.043}   $ & $0.254^{+0.033}_{-0.056}$\\
    					& Joint   & $ 0.766^{+0.033}_{-0.038}$ & - \\
					\hline
 \multirow{10}{*}{Variations}& 2PCFs (T18, $w$CDM) & $0.791^{+0.031}_{-0.044}$ & $ 0.264^{+0.067}_{-0.049}$\\
   					& 2PCFs (DIR-$w$CDM) & $0.752^{+0.042}_{-0.037}$ & $0.264^{+0.035}_{-0.054}$\\
					& 2PCFs ($\Lambda$CDM) & $0.761^{+0.027}_{-0.027}  $ & $0.272^{+0.031}_{-0.056}$\\
   					& 2PCFs (T18, $\Lambda$CDM) & $ 0.789^{+0.031}_{-0.019}$ & $0.248^{+0.065}_{-0.036}$\\
   					& 2PCFs (J20, $\Lambda$CDM) & $0.765^{+0.036}_{-0.031}$ & $0.252^{+0.041}_{-0.086}$\\
					& Peaks (cross-tomo, with IA) & $0.735^{+0.024}_{-0.032}$ & - \\
					& Peaks (cross-tomo, with baryons) & $0.750^{+0.026}_{-0.031}$ &- \\
					& Peaks (cross-tomo, with SLICS-HR) & $0.734^{+0.025}_{-0.032}  $ & - \\
					& Peaks (cross-tomo, no-boost) & $0.736^{+0.025}_{-0.032}$ & - \\
					& Peaks (cross-tomo) & $0.737^{+0.027}_{-0.031} $ & - \\
					& Joint (cross-tomo) & $0.743^{+0.024}_{-0.024} $ & - \\
					\hline
  \multirow{3}{*}{Mocks} 	& Peaks (cross-tomo, no syst) &  $0.787^{+0.024} _{-0.024}             $   &  $0.325^{+0.054}_{-0.067}$\\
    					& Peaks (cross-tomo) &$ 0.776^{+0.045}_{-0.045} $ &  $ 0.297^{+0.048}_{-0.066}$\\
    					& 2PCFs (FID) &$ 0.772^{+0.042}_{-0.042}$ & $0.314^{+0.049}_{-0.070}$ \\
  \hline
    \hline
    \end{tabular}
    \label{table:cosmo_pipeline_test}
\end{table}

We present in this section the results from our cosmological  inference analyses, beginning with the 2PCFs and the peak statistics pipelines. We next discuss our tests on the importance of various systematic effects, before introducing the results from our joint $[\xi_{\pm} ; N_{\rm peaks}]$ analysis.  All quoted parameters constraints correspond to the best-fit value $\pm$ the $1\sigma$ region of the marginalised posterior.

\subsection{2PCFs}
\label{subsec:results_2PCF}

\begin{figure}
\begin{center}
\includegraphics[width=3.3in]{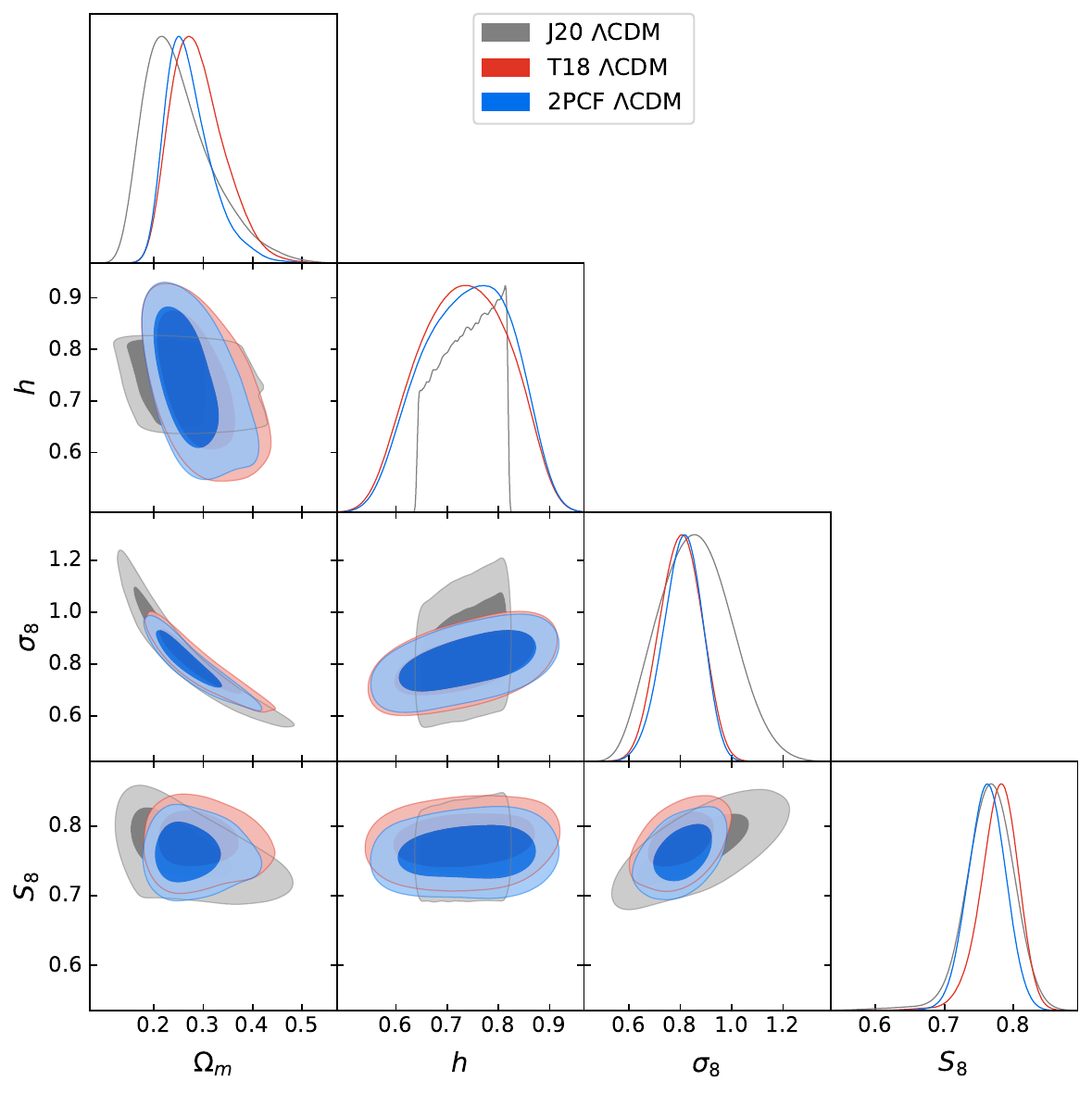}
\caption{Constraints on the $\Lambda$CDM cosmological parameters inferred from the 2PCFs, obtained from the DES-Y1 cosmology inference pipeline, our simulation-based covariance matrix and assuming the DIR $n(z)$ (blue). These results are compared to the $\Lambda$CDM constraints from T18 (in red) based on the same modelling and likelihood sampling strategy, and to those of J20 (grey),  which also use the DIR redshift distribution but adopt different modelling, prior ranges and likelihood sampling choices.}
\label{fig:cosmo_2pcf_LCDM}
\end{center}
\end{figure}

\begin{figure}
\begin{center}
\includegraphics[width=3.3in]{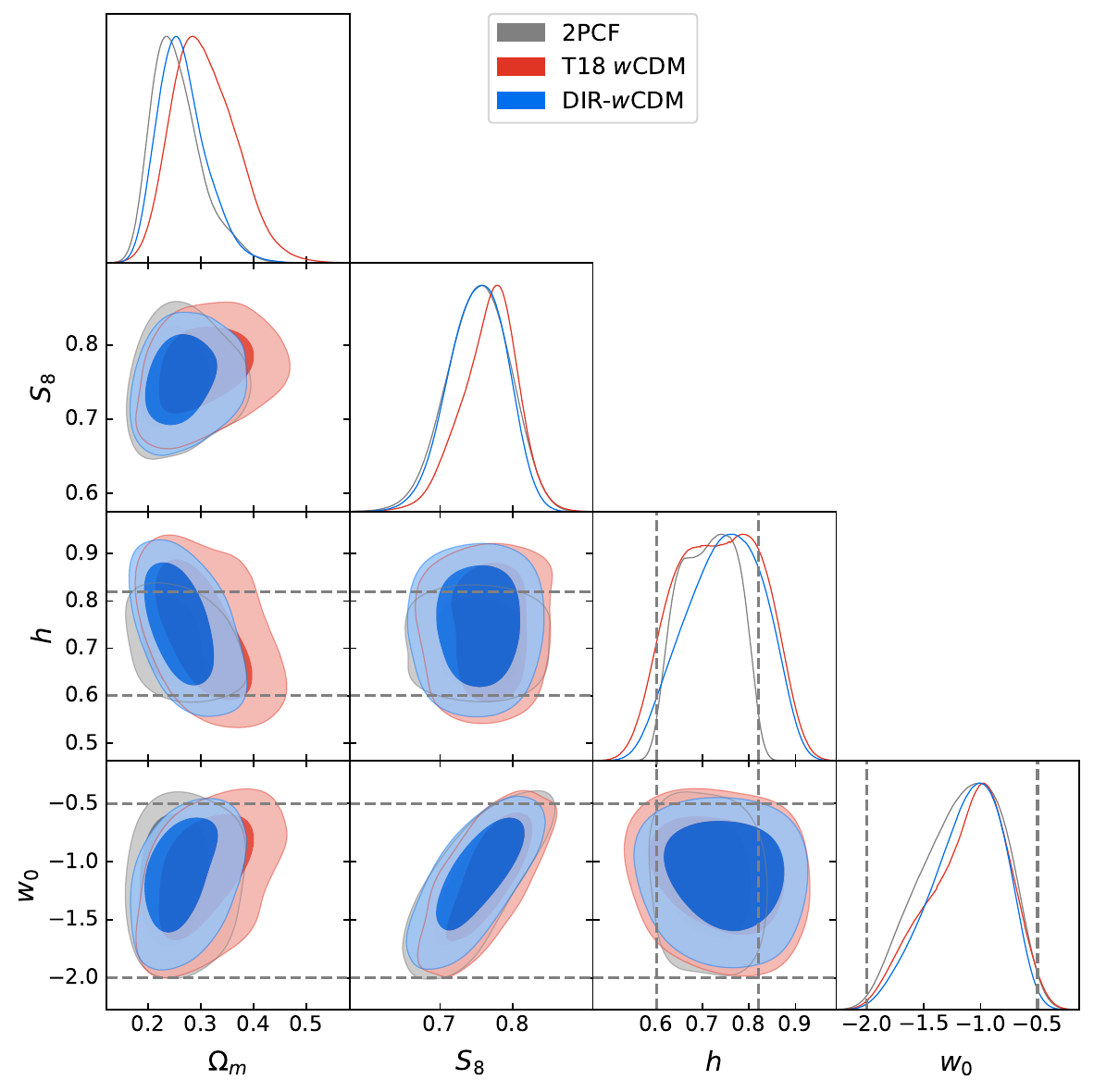}
\caption{Constraints on the $w$CDM cosmological parameters inferred from the 2PCFs with our fiducial pipeline (grey), from the T18 $w$CDM analysis (red) and from an intermediate pipeline, the DIR-$w$CDM, which uses the T18 parameter sampling and prior ranges  on our tiled data with our simulation-based covariance matrix, assuming the DES-DIR $n(z)$ (blue). Note that priors on $h$ and $w_0$ are significantly tighter in our fiducial pipeline, and that our fiducial 2PCFs analysis hits the priors (shown with the dashed lines) on these two parameters.}
\label{fig:cosmo_wCDM_2pcf}
\end{center}
\end{figure}

We first report in Fig. \ref{fig:cosmo_2pcf_LCDM} the constraint on $\Lambda$CDM parameters obtained from our 2PCFs measurement, over-plotted with  those from T18\footnote{In this comparison, we used the  values listed in their Table 3 using a fixed neutrino mass density, which better match our fiducial pipeline.}(red) and J20 (grey). Our constraints (in blue) assume the DES-DIR $n(z)$, it uses our simulation-based covariance matrix,   
we  marginalise over the 10 nuisance parameters listed in Table \ref{table:priors}, but the cosmological sampling follows that of T18. Another difference: as described in Sec. \ref{subsec:xipm}, our $\xi_{\pm}$ measurements are obtained from  the weighted mean $\xi_{\pm}$ obtained over the 19 tiles. As demonstrated in J20, the net effect of changing from the fiducial DES-Y1 $n(z)$ to the DIR $n(z)$ is to shift the amplitude of the modelled 2PCFs signal upwards, which translates into lower best-fit $S_8$ values. This can be seen by comparing the one-dimensional posteriors shown with the red and grey lines in the bottom right panel.  When analysed this way, we obtain:
 \begin{eqnarray}
S_8^{{\rm 2PCFs}, \Lambda} = 0.761^{+0.027}_{-0.027}. \nonumber
\end{eqnarray}
The priors and the parameter sampling in J20 are significantly different from T18 and are responsible for some of the differences between the blue and grey curves, notably the sharp edges in the $h$ posterior, and the more elongated contour in the $[\sigma_8 - \Omega_{\rm m}]$ plane. 
All parameter constraints are summarised in Table \ref{table:cosmo_pipeline_test}. Replacing the $\xi_{\pm}$ data extracted from the tiles with those measured on the full footprint results in a negligible change, with $S_8 = 0.762\pm 0.026$, thereby validating our mosaic methodology.
 Additionally, the resemblance between our confidence interval and those of T18 (the fractional errors on $S_8$ and $\Omega_{\rm m}$ agree to within 0.002 and 0.005, respectively) demonstrates the accuracy of our simulation-based covariance matrix. 
The constraints on the nuisance parameters are mostly prior-dominated. We provide a more complete comparison between T18, J20 and our 2PCFs $\Lambda$CDM analyses in Appendix \ref{subsec:T18_vs_2PCF_vs_J20}.

We next compare in Fig. \ref{fig:cosmo_wCDM_2pcf} the constraints on the four $w$CDM parameters inferred from our 2PCFs measurement (in grey),
 to the T18 $w$CDM results (in red). As explained previously, there are multiple difference between these two pipelines, which we can dissect here. We  show (in blue) an intermediate pipeline,  labelled  the DIR-$w$CDM, which uses the T18 parameter sampling and prior ranges, but assume the DIR $n(z)$, and uses our measurement on the tiled data and our simulation-based covariance matrix. Table \ref{table:pipelines} summarises the differences between these pipelines. By construction, differences between the blue and the grey contours are caused exclusively by the likelihood sampling strategy: the former uses the T18 priors and samples $A_{\rm s}$ over a flat prior, whereas the latter uses those listed in Table  \ref{table:priors} and samples $\sigma_8$.  In contrast, red and blue curves share the signal modelling as well as the parameter sampling, but differ in the $n(z)$  (which shifts down the best $S_8$ and $\Omega_{\rm m}$ values, clearly visible in Fig. \ref{fig:cosmo_wCDM_2pcf}), and in the covariance matrix (which weights slightly differently the various elements of the compressed statistics and therefore affects the size and shape of the contours).  
 
In our fiducial analysis the likelihood sampler hits the priors on $h$ and $w_0$, which are limited by the range of values probed by our {\it Cosmology training set}. We note, however, that this should have no impact on our analysis  due to the low sensitivity of lensing to these particular parameters, and that we marginalise over these anyway. Consequently,  we are unable to report constraints on $h$ and  $w_0$  in our 2PCFs pipeline with the current data\footnote{We expect this to change with the upcoming Stage-IV lensing surveys, as \citet{Martinet20} has shown  that peak statistics could provide a 6\% constraint on $\Omega_{\rm m}$ and a 13\% constraint on $w_0$ from 100 deg$^2$ of {\it Euclid}-like mocks built from the same SLICS and cosmo-SLICS suites. }. The best-fit parameters are reported in Table \ref{table:cosmo_pipeline_test}, notably:
\begin{eqnarray}
S_8^{{\rm 2PCFs}, w} = 0.753^{+0.043}_{-0.043}, \nonumber
\end{eqnarray}
which is consistent with, but has a larger uncertainty than   $S_8^{{\rm T18}, w} =  0.791^{+0.031}_{-0.044}$ reported in T18. The overall precision on the matter density is similar to that of the $\Lambda$CDM analysis (we measure $\Omega_{\rm m} = 0.254^{+0.033}_{-0.056}$), while the uncertainty on $S_8$ increases significantly, as expected from opening the parameter space.   When adopting the DIR-$w$CDM pipeline, we obtain:
\begin{eqnarray}
S_8^{{\rm DIR}- w{\rm CDM}} = 0.752^{+0.042}_{-0.037}, \nonumber
\end{eqnarray}
which best inferred value aligns with our 2PCFs analysis, with error bars slightly tighter.
The fact that the T18 constraints (4.7\% on $S_8$) are tighter than these (5.3\%), despite having larger uncertainty in their redshift distributions, indicates that the T18 priors and sampling strategy are informative about $S_8$ to some level, artificially decreasing the size of the reported error bars \citep[see the discussion on informative priors in][]{KiDS1000_Joachimi}.

\subsection{Peaks}
\label{subsec:results_GPR}

\begin{figure}
\begin{center}
\includegraphics[width=3.3in]{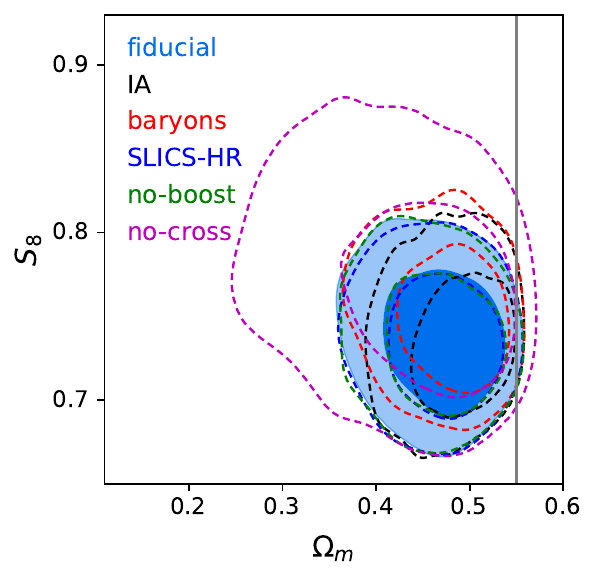}
\caption{Impact of the different correction factors on the constraint from the peaks statistics. In most cases the likelihood hits the upper prior edge on $\Omega_{\rm m}$ within $2\sigma$, as marked by  the vertical line, which prevents us from reporting constraints on that parameter.}
\label{fig:cosmo_biases}
\end{center}
\end{figure}

\begin{figure}
\begin{center}
\includegraphics[width=3.3in]{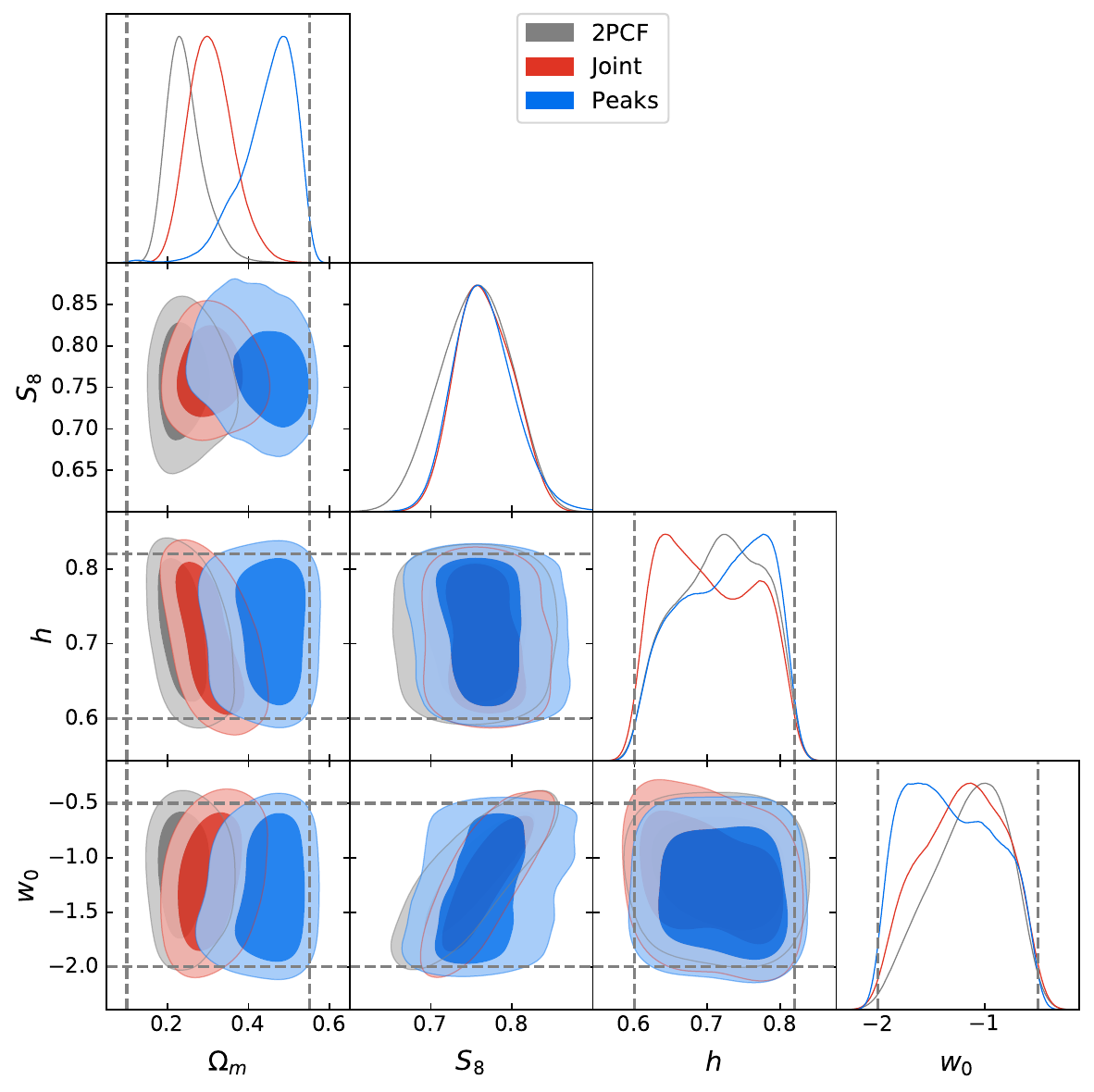}
\caption{Constraints on the four $w$CDM cosmological parameters obtained from 2PCFs (grey), peaks (blue) and from the joint analysis (red). The dashed lines indicate the prior ranges on $\Omega_{\rm m}$, $h$ and $w_0$, which are in most cases too narrow to provide meaningful constraints on these parameters (see main text for exceptions).}
\label{fig:cosmo_wCDM}
\end{center}
\end{figure}

We report in Fig. \ref{fig:cosmo_biases} the results of our peak count  analyses for models in which systematics are infused. All data presented from now on are obtained from the 12.5 arcmin filter; we investigated other choices of $\theta_{\rm ap}$ that yielded slightly less constraining results and hence we dropped them from the analysis. 

Before examining differences between the various cases, we first note that, for all of them, the inferred matter density values are unexpectedly high, with $\Omega_{\rm m}\gtrsim0.4$ at the $1\sigma$ level.  This is in  tension both  with the 2PCFs measurements and with external probes such as the {\it Planck} measurement of $\Omega_{\rm m}=0.3153 \pm 0.0073$ \citep[][see also a quantitative assessment of these tensions in Appendix \ref{subsec:tension}]{2018arXiv180706205P}.  In the case where the cross-redshift bins are excluded however, the tension is significantly reduced. Interestingly, this is also the only case where the $GI$ part of the intrinsic alignment signal is suppressed, which suggests that unmodelled IA systematics could be artificially pushing the likelihood analysis towards  high values of the matter density. We therefore adopt a conservative approach and remove the cross-redshift bins from our fiducial analysis. We still include them in one of our analysis variations, with the caveat that they might be contaminated by an unaccounted secondary signal. 

In order to provide the most accurate account of the other systematic effects, we include the cross-redshift bins in their measurements, shining light on their impact with the highest precision available. Starting with the IA, and  recalling that our alignment model applies to tomographic bins  1, 1$\cup$2 and 2 only with no free parameter, we include the systematic effect as an optional correction to the measured data vector (see  Sec. \ref{subsec:syst_IA} and Fig. \ref{fig:corr_factors}). The cosmology inference results, shown by the black dashed lines in Fig. \ref{fig:cosmo_biases},  indicate that  our low-redshift IA model has a relatively mild impact, affecting the best-fit $S_8$ value by less than $0.1\sigma$. Notably, we have:
\begin{eqnarray}
S_{8, \rm IA}^{\rm peaks} = 0.735^{+0.024}_{-0.032}, \nonumber
\end{eqnarray}
which is almost unchanged compared to the baseline analysis:
\begin{eqnarray}
S_{8, \rm cross-tomo}^{\rm peaks} = 0.737^{+0.027}_{-0.031}. \nonumber
\end{eqnarray}
We note that  the best-fit $\Omega_{\rm m}$ increases by $0.02$. This is consistent with the expectation that IA overall suppresses the lensing signal; therefore, given a  measurement (2PCFs or peaks), the inferred $S_8$ and/or $\Omega_{\rm m}$ values  increase with stronger IA model.

As mentioned earlier, the $GI$ term, that we are currently unable to model accurately, is most likely to impact the cross-tomographic bin measurements. The fact that the best-fit $S_8$ shifts up by as much as 0.037 (and $\Omega_{\rm m}$ shifts down by 0.02)  when removing the cross-tomographic bins suggests that unmodelled systematics affect these bins differently than the auto-tomographic bins. The $GI$ contribution does exactly this, and a similar $1\sigma$ effect is detected in T18 (see their Fig. 9). It is therefore plausible that unmodelled IA could be  leading to the observed preference for a high $\Omega_{\rm m}$ values when including the cross-tomographic bins, however we postpone a more robust analysis of IA to future work.

Adopting a similar strategy, we apply the baryon correction factor (see Sec. \ref{subsec:syst_baryons}, and  shown in  Fig. \ref{fig:corr_factors}), and re-run our cosmological inference pipeline. The impact is stronger than the IA, yielding:
\begin{eqnarray}
S_{8, \rm baryons}^{\rm peaks} =0.750^{+0.026}_{-0.031}, \nonumber
\end{eqnarray}
a $0.5\sigma$ shift in $S_8$ visualised by the red dashed lines in  Fig. \ref{fig:cosmo_biases}.
When correcting the signal for a possible bias due to limits in  mass resolution of the {\it Cosmology training sample}, the values of  $S_8$ is slightly reduced as expected, with:
\begin{eqnarray}
S_{8, \rm SLICS-HR}^{\rm peaks} =  0.734^{+0.025}_{-0.032}.
\end{eqnarray}
This demonstrates that our choice of $\mathcal{S}/\mathcal{N}$ bins is unaffected by  the known small-scales inaccuracies inherent to the $N$-body runs. 

Removing the boost factor correction can generally push the inferred cosmology to models with lower clustering, however  in our case,  the effect is marginal. We find 
\begin{eqnarray}
S_{8, \rm no-boost}^{\rm peaks} =  0.736^{+0.025}_{-0.032},
\end{eqnarray}
a negligible change with respect to the fiducial pipeline.

Overall, when including all redshift bins, we observe that the baryons and IA cause the largest shifts. Excluding the cross-redshift terms has the strongest impact, as it significantly offsets the best-fit value and degrades the constraining power. Nevertheless, the contours are fully consistent with the fiducial case, and this choice is necessary until an improved high-redshift IA model can be incorporated.

The fiducial constraints from the peak statistics are presented by the blue contours in Fig. \ref{fig:cosmo_wCDM} and compared to the 2PCFs (in grey, same as  in Fig. \ref{fig:cosmo_wCDM_2pcf}).
 The marginal constraints on $S_8$ are:
\begin{eqnarray}
S_{8, \rm fid}^{\rm peaks} = 0.780^{+0.019}_{-0.056} .\nonumber
\end{eqnarray}
These values are in excellent agreement with those reported by J20, T18 and with our 2PCFs method, with marginal constraints on $S_8$ being consistent to within $1\sigma$ in all cases. More importantly, the peak count analysis improves this fractional precision by 10\% compared to our fiducial 2PCFs pipeline, providing a 4.8\% measurement on $S_8$. 
Including the cross-tomographic bins results in $S_8^{\rm peaks} = 0.737^{+0.027}_{-0.031}$, which improves  the constraining power to a  $\sim$3.9\% measurement (see the pipeline `cross-tomo' in Table \ref{table:cosmo_pipeline_test}). Although these are excluded from the fiducial analysis due to unaccounted for systematics, this demonstrates that even with the current noise level and in presence of systematics, some 20\% additional information on $S_8$ can be recovered by including these bins compared to the fiducial case, potentially bringing  the gain to 30\% over the 2PCFs analysis. This aligns with the findings of \citet{Martinet20}, which observe a 50\% gain in a systematics-free Stage-IV weak lensing survey setup, using five tomographic bins and including cross-terms for all possible
combinations of slices (i.e. larger than pairs). We finally note that the posterior on  $\Omega_{\rm m}$ overlaps with the upper prior edge within $2\sigma$, while those on $h$ and $w_0$   are completely prior-dominated, hence we do not report constraints on these parameters.

\subsection{2PCFs+Peaks}
\label{sebsec:results_Joint}

The fact that both 2PCFs and peaks count methods individually prefer different values for the cosmological parameters is expected, given that they both probe slightly different physical scales, and react to the DES-Y1 shape noise in completely different ways. This is further supported by our measurements on the mean of the SLICS mocks presented in Appendix \ref{sec:syst_pipeline}, which correspond to noise-free data and show an excellent agreement in the best-fit values. We note that similar results have been observed in the literature already, notably when comparing the inferred cosmology from the 2PCFs and power spectra analyses of the HSC-Y1 data  \citep{HSCY1_2PCF,HSCY1_Cell} and of the  KiDS-1000 data \citep{KiDS1000_Asgari},  where it was shown that some fluctuations are expected and statistically consistent. We show with mock surveys in  Appendix \ref{subsec:tension} that the observed difference of $\Delta \Omega_{\rm m}=0.2$  is frequent, and that the tension between the two methods is low, enabling us to carry  out a joint 2PCFs+peaks likelihood analysis. For consistency, our fiducial case again excludes the cross-tomographic redshift bins in the peak count data, and results in a best-fit value of:
\begin{eqnarray}
S_{8}^{\rm joint} =  0.766^{+0.033}_{-0.038} ,
 \nonumber
\end{eqnarray}
with contours presented in red in  Fig. \ref{fig:cosmo_wCDM}. The best-fit value in this case is consistent with both the 2PCFs and the peak statistics, with a $\sim$20\% smaller fractional  uncertainty on $S_8$.  The joint analysis achieves 4.6\% precision, fast approaching  the 3.8\% precision achieved by the DES-Y1 $3\times$2-pts analysis \citep{DESY1_3x2}. We note that the same 20\% increase in precision is reported in M18 when comparing peaks and 2PCFs. However, their treatment of the systematics is simpler in comparison to this study, which likely affects their real precision gain.  The analysis variant in which the cross-tomographic terms are included yields a promising  3.2\% measurement, with:
\begin{eqnarray}
S_{8, \rm {cross-tomo}}^{\rm joint} = 0.743^{+0.024}_{-0.024} ,
 \nonumber
\end{eqnarray}
however a better modelling of systematics is required before we can exploit this configuration.

We repeated the full cosmology joint-probe inference with two other aperture filter sizes, and obtained similar  constraints, which leads us to the conclusion that given the current level of noise in the data, the choice of aperture filter size does not make a large difference, at least for the scales we tested. It is also clear that under these conditions a multi-scale analysis would bring no additional information. M18 found a mild improvement when combining the $12.5'$ and the $15.0'$ filters, but that was in the absence of tomography. In that case,  the noise levels are suppressed at the cost of losing the redshift information.

The dark energy equation of state can be constrained from the joint analysis, which has sufficient constraining power to not be prior-dominated\footnote{We adopt the criteria described in Appendix A of \citet{KiDS1000_Asgari}  to decide whether we can report a constrain: if the value of posterior at any edge of the prior is higher that 0.135 times its maximum value, the measurement is prior-dominated and the constraints should not be reported. Our measurement of  $w_0$ from the joint analysis marginally satisfies this criteria on the lower end only, with posterior values of 0.134 at the lower prior edge.}. We obtain:
\begin{eqnarray}
w_0^{\rm joint} > -1.55 
\nonumber
\end{eqnarray}
at $1\sigma$, which is consistent with $\Lambda$CDM cosmology but not yet competitive with T18, who find $w_0^{\rm T18} = -0.92^{+0.35}_{-0.40}$. As shown in \citet{Martinet20}, the cross-tomographic data points will greatly help with this measurement once properly calibrated.

The overall goodness-of-fit is evaluated with  the $p$-value, which can be interpreted as the probability that the model is not consistent with the data. This null hypothesis-rejection method depends notably on $\nu_{\rm eff}$, the effective number of degrees of freedom. This quantity is often estimated from the difference between the number of data points and the number of free parameters in the model, however this does not exactly hold in the more accurate context where these model parameters are not totally free but sampled from priors that often restrict the search, and are highly degenerate.  For example, the KiDS-1000 cosmic shear analysis presented in \citet{KiDS1000_Asgari} samples 12 parameters in their cosmological inference pipeline, but a careful study of the sampling reveals that the effective number of free parameter is closer to 4.5 \citep{KiDS1000_Joachimi}. 

In the current analysis,  the size of the data vector varies from 48  (peaks fiducial) to  305 (peaks + 2PCFs). The  $\chi^2$ for the three fiducial analysis pipelines, at their best-fit cosmologies, are respectively  $\chi^2_{\rm peaks} = 71$, $\chi^2_{\rm 2PCFs} =209$  and $\chi^2_{\rm joint} = 463$. If, following \citet{KiDS1000_Joachimi}, we also use 4.5 free parameters, this leads to  goodness-of-fit $p$-values of $0.005$, $0.978$ and $0.084$, respectively. The  $p$-values for the  peak count analysis is  low, indicating possible residual tensions that we could not completely identify nor resolve. We noted however from Fig. \ref{fig:N_peaks} that the data in first bin  has a  high level of scatter compared to the size of the error bars, which is partly responsible for dragging the $p$-value towards low values. When removing these points from the analysis, we obtain a $p$-value of $0.105$, which is higher than our goodness threshold of 0.05. We decided to keep the bin data in our fiducial analysis pipeline, even though it carries more noise than signal.

\section{Conclusions}
\label{sec:conclusion}

We analyse the DES-Y1 cosmic shear data with a cosmology inference pipeline exclusively calibrated on suites of $w$CDM weak lensing numerical $N$-body simulations. 
Our method is general and can be directly used with many non-Gaussian probes, however we opted for the  tomographic lensing peak count statistics. Our pipeline interfaces with the two-point functions via the public inference code {\sc cosmoSIS}, which allows us to conduct a joint $[\xi_{\pm};N_{\rm peaks}]$ analysis. We model the peak statistics signal with the cosmo-SLICS, the covariance with the SLICS and investigate the key cosmic shear systematics either by infusing them in our training set, or by extracting their impact from tailored external mock data. Notably, the impact of baryons is assessed with the {\it Magneticum} hydrodynamical simulations, the effect of finite particle mass is quantified from high-resolution light-cones, while the impact of intrinsic alignment  is investigated by infusing intrinsic shapes to mock galaxies following a physical model based on the shape of dark matter haloes. Source-lens clustering is also studied by comparing galaxy excess in peaks of different height, and found to have a negligible impact on our results given our peak selection criteria.

We validate our method on mock data vectors and against the DES-Y1 $\xi_\pm$ analyses of T18 and J20,  which we reproduce well given differences in our pipelines.
We identify a residual unknown systematic  in the cross-tomographic redshift bins of the peaks data, which appears to shift the inferred cosmological parameters towards high $\Omega_{\rm m}$ values. We identified the intrinsic alignment $GI$ term as one possible and likely cause of this effect, and in absence of an accurate IA modelling, we decided to remove these redshift bins from our fiducial analysis. 

We perform a joint  likelihood analysis and set constraints on   $S_8$, finding a $\sim$20\% gain in precision compared to the correlation function analysis. The joint analysis yields a 4.6\% measurement, with $S_{8}^{\rm joint} =0.766^{+0.033}_{-0.038}$, approaching the 3.8\% measurement reported by the DES-Y1 $3\times$2-pts analysis \citep{DESY1_3x2}, and closely approaching the 2.9\% $S_8$ measurement of \citet{KiDS1000_Asgari} with the fourth KiDS data release. We show that after an improved calibration of the cross-tomographic terms, the peak count method can achieve 3.9\% on $S_8$, and the joint analysis could reach  $3.2$\%, one of  the tightest measurements of this parameter so far.  One possible caveat is that we include no IA modelling in our cross-tomographic  analysis, and that its inclusion could possibly degrade our constraining power.

Our analysis pipeline builds from the infrastructure presented in M18, and is inspired by many aspects of the K16 data analysis. We have significantly upgraded the underlying simulation support, we include a better treatment of the photometric and shear calibration uncertainties, we improve the inference method with a full integration within a   likelihood sampler, and we demonstrate  the robustness of our results to uncertainty on baryonic feedback, to intrinsic alignments, to source-lens clustering  and  to limitations from the finite mass resolution of the $N$-body runs, which are lacking in previous analyses. Additionally, this paper is the first tomographic peak counts data analysis, and we confirm that including cross-redshift bins reinforces the constraints, once secondary signals  such as IA are properly calibrated.

For the peaks statistics to remain competitive with the 2PCFs, further development will be required in the modelling of baryon feedback. For example, a 20\% improvement on $S_8$ is obtained from the DES-Y1 3x2pt data when modelling and marginalising over the impact of baryons, as it enables to include additional small-scale  elements in the $\xi_{\pm}$ data vector \citep{2020arXiv200715026H}. A similar gain is observed in the cosmic shear-only DES-Y1 re-analysis by \citet{Asgari_DES_KiDS_cosebi}, where clean small scales are included via the COSEBIs estimators. Equivalent procedures with peaks statistics must be investigated, which would allow us to push back the   $\mathcal{S}/\mathcal{N}\le4$ limit. The `baryonification' method described in \citet{Baryonification2} is one possible avenue to achieve this, as well as the direct training on a variety of hydrodynamical light-cones, although the latter is a more expensive task and introduces several other uncertainties, e.g. how one changes the sub-grid physics for different cosmologies. The goal here would be to construct a response model in order to include the effect  as a nuisance parameter in a cosmology inference analysis.

Intrinsic alignment of galaxies is the second topic where further development is critical, and improving the IA modelling is mandatory for future analyses. The model we adopt here is too simple, and we demonstrate that IA likely impacts the inferred cosmology at the same level as the baryons and potentially up to $1\sigma$ level, making it one of the primary limiting factor in our analysis.
A better approach would involve a suite of dedicated  training samples where the modelling and the levels of IA can be adjusted, mimicking the varying $(A_{\rm IA},\eta)$ parameters that control the amplitude of the secondary signal  within  the NLA model. One could also relate the simulated intrinsic galaxy shapes with the halo-model based approach of \citet{Fortuna2020} directly from the HOD prescription. Even better, these approaches could be linked, allowing us to marginalise coherently over a unique set of common astrophysical parameters.

All of the improvements presented here further close the gap between analytical and simulation-based cosmological inference techniques. While the former is preferred in cosmic shear analyses performed by the large weak lensing collaborations, the latter is required by most measurement methods based on non-Gaussian statistics, for which a theoretical model of the signal and of the covariance matrix is generally not available.  Given that the performance of non-Gaussian estimators is now clearly shown  to significantly exceed that of two-point correlation functions, and that this trend will intensify  in future surveys \citep{MassiveNu1,Zuercher2020a, Martinet20}, we strongly advocate for a co-development of both approaches. 

Most of the work that has been put into the development of our method can now be directly exploited  with most of the alternative statistics mentioned in the introduction: one only needs to measure the desired statistic on our training sets ({\it Cosmology, Covariance} and {\it Systematics}), re-train the GPR emulator and adjust the linear models that describe the photometric redshifts and the shear calibration. The rest of the infrastructure  is already in place, and we intend to explore some of these alternative probes in the near future. For this reason we will make the mocks available upon request. We also intend to explore some extensions to the existing infrastructure, including for instance a new {\it Neutrino training set} derived from the {\it MassiveNuS} simulations \citep{MassiveNuS} and designed to measure the sum of the neutrino mass. Parallel pipelines tailored for the analysis of the KiDS-1000 and/or the HSC surveys could also be constructed directly, and we hope to see these novel methods take an important role in the upcoming Stage-IV lensing analyses.

\section*{Acknowledgements}

We would like to thank Marika Asgari for useful  discussions and comments,  Carlo Giocoli for his contribution to an earlier version of the {\it Magneticum} light-cone extractor,
as well as Shahab Joudaki making public the MCMC chains described in J20, which can be obtained from https://github.com/sjoudaki/kidsdes/. JHD acknowledges support from an STFC Ernest Rutherford Fellowship (project reference ST/S004858/1). NM acknowledges support from a CNES Fellowship, TC is supported by the INFN INDARK PD51 grant and by the PRIN-MIUR 2015 W7KAWC grant, and  KD by the Deutsche Forschungsgemeinschaft (DFG, German Research Foundation) under Germany's Excellence Strategy -- EXC-2094 -- 390783311. BG acknowledges the support of the Royal Society through an Enhancement Award (RGF/EA/181006). We also acknowledge support from the European Research Council under grant numbers 647112 (JHD, BG, CH \& QX) and 770935 (HH), as well as the Deutsche Forschungsgemeinschaft (HH, Heisenberg grant Hi 1495/5-1).  CH further acknowledges support from the Max Planck Society and the Alexander von Humboldt Foundation in the framework of the Max Planck-Humboldt Research Award endowed by the Federal Ministry of Education and Research. 
Computations for the $N$-body simulations were enabled by Compute Ontario (www.computeontario.ca), Westgrid (www.westgrid.ca) and Compute Canada (www.computecanada.ca).  
\\

This project used public archival data from the Dark Energy Survey (DES). Funding for the DES Projects has been provided by the U.S. Department of Energy, the U.S. National Science Foundation, the Ministry of Science and Education of Spain, the Science and Technology FacilitiesCouncil of the United Kingdom, the Higher Education Funding Council for England, the National Center for Supercomputing Applications at the University of Illinois at Urbana-Champaign, the Kavli Institute of Cosmological Physics at the University of Chicago, the Center for Cosmology and Astro-Particle Physics at the Ohio State University, the Mitchell Institute for Fundamental Physics and Astronomy at Texas A\&M University, Financiadora de Estudos e Projetos, Funda{\c c}{\~a}o Carlos Chagas Filho de Amparo {\`a} Pesquisa do Estado do Rio de Janeiro, Conselho Nacional de Desenvolvimento Cient{\'i}fico e Tecnol{\'o}gico and the Minist{\'e}rio da Ci{\^e}ncia, Tecnologia e Inova{\c c}{\~a}o, the Deutsche Forschungsgemeinschaft, and the Collaborating Institutions in the Dark Energy Survey.
The Collaborating Institutions are Argonne National Laboratory, the University of California at Santa Cruz, the University of Cambridge, Centro de Investigaciones Energ{\'e}ticas, Medioambientales y Tecnol{\'o}gicas-Madrid, the University of Chicago, University College London, the DES-Brazil Consortium, the University of Edinburgh, the Eidgen{\"o}ssische Technische Hochschule (ETH) Z{\"u}rich,  Fermi National Accelerator Laboratory, the University of Illinois at Urbana-Champaign, the Institut de Ci{\`e}ncies de l'Espai (IEEC/CSIC), the Institut de F{\'i}sica d'Altes Energies, Lawrence Berkeley National Laboratory, the Ludwig-Maximilians Universit{\"a}t M{\"u}nchen and the associated Excellence Cluster Universe, the University of Michigan, the National Optical Astronomy Observatory, the University of Nottingham, The Ohio State University, the OzDES Membership Consortium, the University of Pennsylvania, the University of Portsmouth, SLAC National Accelerator Laboratory, Stanford University, the University of Sussex, and Texas A\&M University.
Based in part on observations at Cerro Tololo Inter-American Observatory, National Optical Astronomy Observatory, which is operated by the Association of Universities for Research in Astronomy (AURA) under a cooperative agreement with the National Science Foundation.
\\
\\
{\footnotesize All authors contributed to the development of this paper. JHD (lead) conducted the analysis with significant scientific contribution from NM  (co-lead); the other authors are listed in alphabetical order. }

\section*{Data Availability}

The SLICS numerical simulations can be found at http://slics.roe.ac.uk/, while the SLICS-HR,  the cosmo-SLICS and the {\it Magneticum} can be made available upon request.
This work also uses public DES-Y1 data, which can be found at  https://des.ncsa.illinois.edu/releases/dr1.



\bibliographystyle{hapj}
\bibliography{peaks} 




\appendix

\section{Validation of the cosmology inference pipeline}
\label{sec:syst_pipeline}

In this section we present a series of validation tests we performed on our cosmology inference pipeline.

\subsection{Inference from 2PCFs} 
\label{subsec:2pcf_validation}

  \begin{figure}
\begin{center}
\includegraphics[width=3.0in]{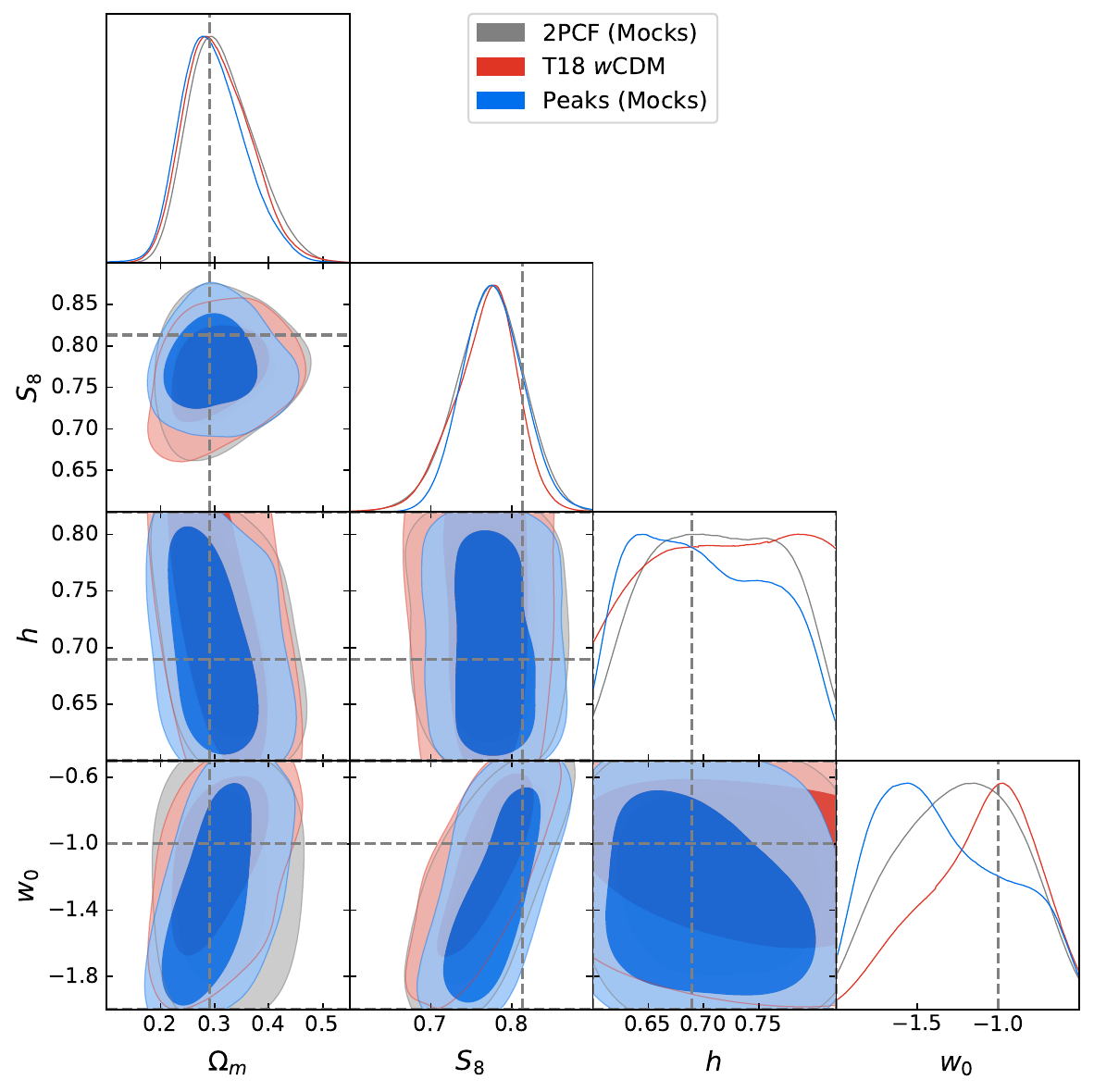}
\caption{Two-dimensional constraints on the $w$CDM cosmological parameters ($\Omega_{\rm m}, w_0, S_8, h$)  from 2PCFs (blue) and peak count analyses (grey) of the mock DES-Y1 data corresponding to the mean SLICS values. The red contours show the fiducial T18 results $w$CDM for reference. The dashed lines indicate the input SLICS cosmology. The inferred $S_8$ value appears lower than the input value, but this is a plotting artefact; maximum {\it a posterior} (MAP) value is $S_8^{\rm MAP}=0.793$ and $0.785$ for 2PCFs and peaks respectively, which are very close to the input of 0.813 (see main text for more details).}
\label{fig:pipeline_validation_wCDM_2pcf_GPR}
\end{center}
\end{figure}

\begin{figure*}
\begin{center}
\includegraphics[width=7.0in]{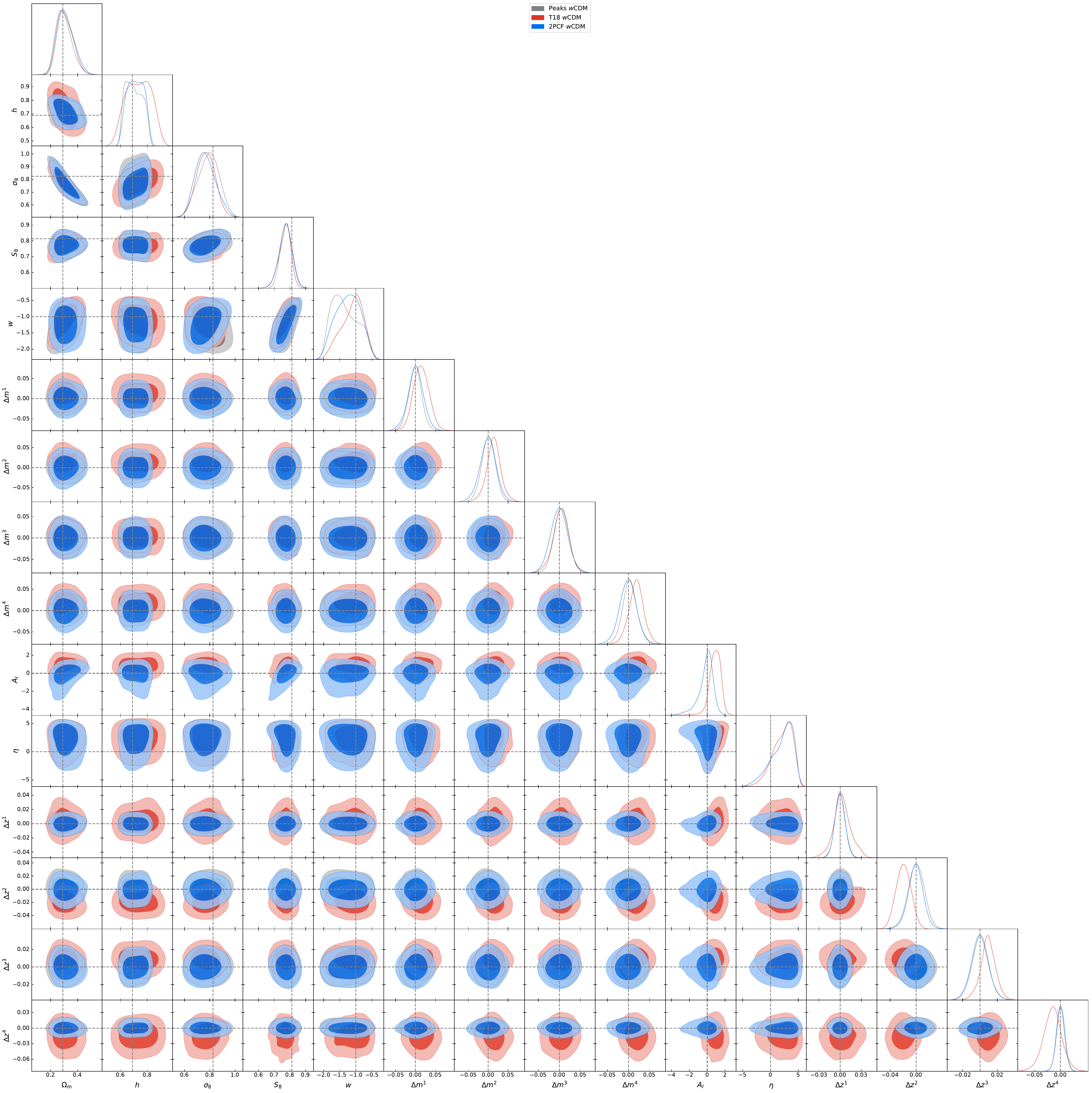}
\caption{Pipeline validation: Two-dimensional constraints on the $w$CDM cosmological parameters ($\Omega_{\rm m}, h, \sigma_8, S_8, w_0$) and on the  10 nuisance parameters ($4 \times \Delta z^i$, $4 \times \Delta m^i$, $A_{\rm IA}$ and $\eta$). Mean and covariance  are  obtained from the SLICS 2PCFs (blue) and peaks (grey), and compared to the T18 constraints  (red). The thin dashed lines indicate the input values in the simulations, which are well recovered by the blue and grey contours. Note that the T18 priors differ, most notably in $h$, $\Delta m^i$ and $\Delta z^i$.}
\label{fig:pipeline_validation_2pcf}
\end{center}
\end{figure*}

The $\xi_{\pm}$ measurements on the simulations, presented in Fig. \ref{fig:xi},  show an excellent agreement with the input cosmology, well within the statistical precision of the DES-Y1 data. There is a small loss of power in the simulations compared to the data that can be observed at large angular scales, a known finite-box effect that is caused by the absence of density fluctuation modes larger than the simulated volume. The $\xi_{\pm}$ statistics measured from the SLICS are  best modelled with a prediction that include a minimum $k$-mode of $k_{\rm min} = 2\pi/505 \,\,h{\rm Mpc}^{-1}$ in the matter power spectrum, which is readily implemented in {\sc cosmoSIS} with the {\tt kmin} option. A correction scheme for the sample variance also exists (HD15), however the contribution to the total error budget from these missing modes is negligible. We present in Fig. \ref{fig:pipeline_validation_wCDM_2pcf_GPR} the $w$CDM constraints from our 2PCFs  inference pipeline, when analysing the mean of the  {\it Covariance training set}. For the exercise presented in this section, the priors on nuisance parameters have all been centred on zero, since no systematic shifts are infused in the {\it Covariance training set}. 

Our inference method produces contours that are highly similar to those obtained by the $w$CDM analysis of T18 (blue vs. red), with a best fit value that is consistent with the input cosmology, albeit with a slightly lower $S_8$ value. This shift is caused by our parameter sampling, more precisely by the upper limit on $w_0$. A closer look  on the [$w_0 - S_8$] panel reveals that these two quantities are degenerate, and that the $w_0$ posterior is limited by the prior, especially on the upper bound. Had we access to higher values, the contours would extend further in the upper-right direction, bringing centroid to higher $S_8$ values. This is further supported  by the fact that  the maximum  {\it a posterior} (MAP) value is located at  $S_8^{\rm MAP}=0.793$, which is very close to the input of 0.813.

We next present in Fig. \ref{fig:pipeline_validation_2pcf} the two-dimensional marginal constraints  on the $w$CDM parameters ($\Omega_{\rm m}, h, \sigma_8, S_8, w_0$) and on the 10 nuisance parameters used in this analysis,  which are associated with photometric uncertainty ($4 \times \Delta z^i$), shear calibration ($4 \times \Delta m^i$) and intrinsic alignments of galaxies ($A_{\rm IA}$ and $\eta$).   The blue contours are obtained from our 2PCFs inference pipeline, executed on the mean 2PCFs  extracted from the SLICS and using the SLICS covariance matrix. 
We observe that the constraints on the four cosmological parameters are well centred on the input cosmology, and the sizes of the contours are fully consistent with those of T18.

We also note that our pipeline accurately recovers the input (null) amplitude of the intrinsic alignment in the SLICS, which also shows a degeneracy with $S_8$, whereas the data prefers values closer to $A_I=1$, as reported in T18 and seen in the red contours. The fact that the posterior of the $\eta$ parameter is almost identical in the data and in the IA-free simulation indicates that the evidence for a redshift evolution in the data is not strong. In fact, it instead suggests that the measurement of this parameter  is sensitive to something else, present in both simulations and data \citep[see the discussion on degeneracies between IA parameters and photometric redshift nuisance parameters in Appendix C of][]{KiDS1000_Heymans}.
All other nuisance parameters are prior dominated, and we observe a clear difference between our nuisance priors and those of T18, especially for the photometric bias; the DIR method has a smaller error, which reflects here in smaller contours for all $\Delta z^i$.

\subsection{Inference from peaks} 
\label{subsec:GPR_validation}

\begin{figure*}
\begin{center}
\includegraphics[width=5.0in]{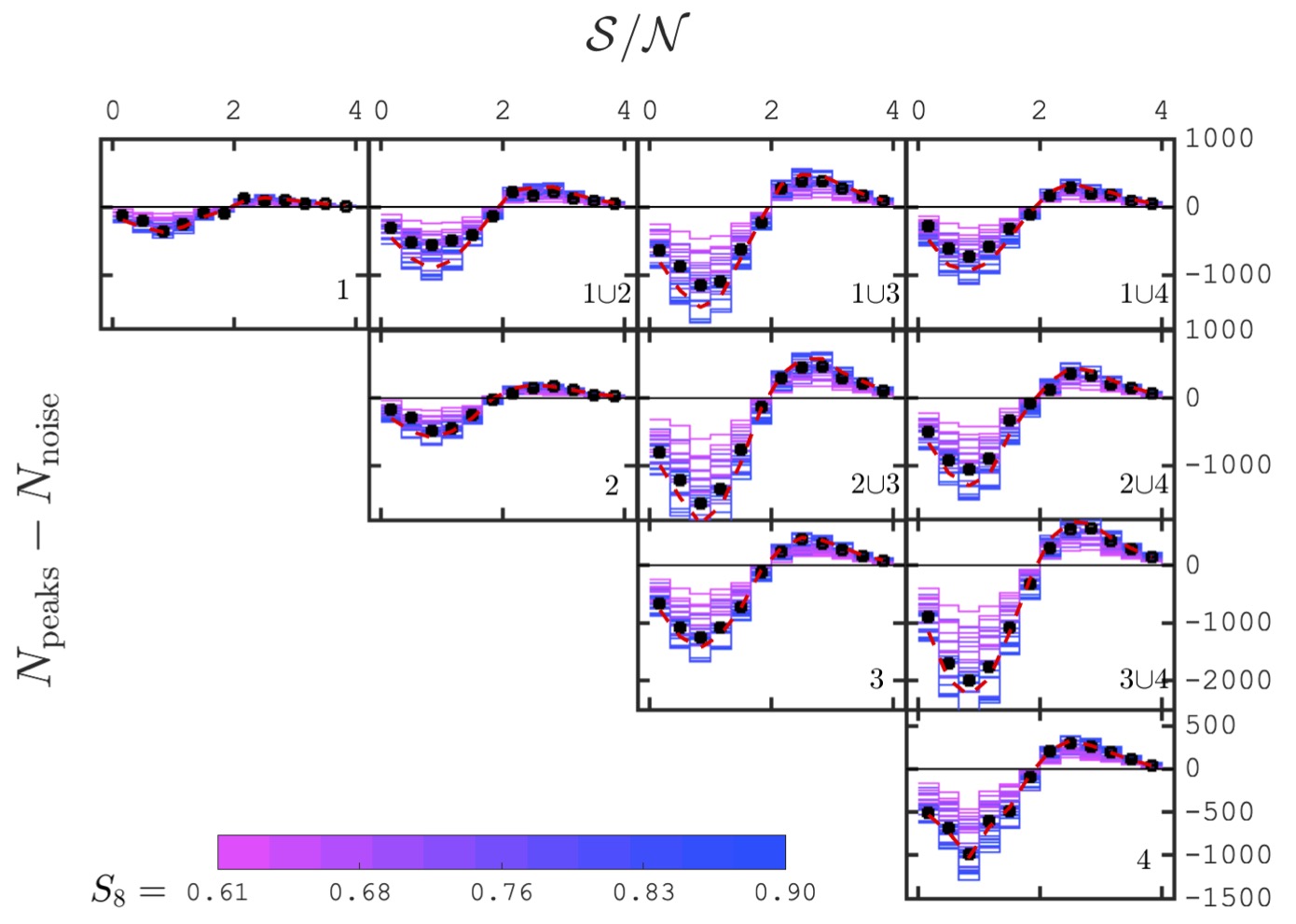}
\caption{Same as Fig. \ref{fig:N_peaks}, but the black points represent  the mean over the full SLICS sample instead of the data. The error bars are showing the error on the mean,  and the red dashed line indicates the best-fit cosmology. }
\label{fig:N_peaks_mocks}
\end{center}
\end{figure*}

We next run our peak statistics $w$CDM inference pipeline on the same data vector as in Appendix \ref{subsec:2pcf_validation}, e.g. mean of the {\it Covariance training set} measurements.
 The mean peak function is presented by the black symbols in Fig. \ref{fig:N_peaks_mocks}, showing the noise-free signal which lies well within the range covered by the {\it Cosmology training set}. 
Again, in this validation exercise, the priors on the nuisance parameters are centred on zero, and we marginalise over the 8 nuisance parameters ($4 \times \Delta z^i$, $4 \times \Delta m^i$, the two IA parameters are not included in the peaks analysis). We next feed this data vector into our {\sc cosmoSIS} pipeline,  and confirm in Fig. \ref{fig:pipeline_validation_wCDM_2pcf_GPR} that we recover the same cosmology as the 2PCFs analysis. Once again, the $S_8$ is slightly lower than expected, which can be explained by the asymmetric sampling of $w_0$ that causes a bias in the inferred contour plots. The maximum {\it a posterior} value is well  within $1\sigma$, with  $S_8^{\rm MAP}=0.785$. We recover similar contours on $\Omega_{\rm m}$ and $S_8$ from peaks and 2PCFs.  We finally repeated the exercise for the case where $\Delta z^i$ and $\Delta m^i$ are set to zero, and notice that the error on $S_8$ is reduced by almost a factor of two, which clearly indicates the importance of using accurate informative priors on these nuisance parameters. Results are summarised in  Table \ref{table:cosmo_pipeline_test}.

\subsection{Towards a joint analysis}
\label{subsec:tension}

\begin{figure}
\begin{center}
\includegraphics[width=3.0in]{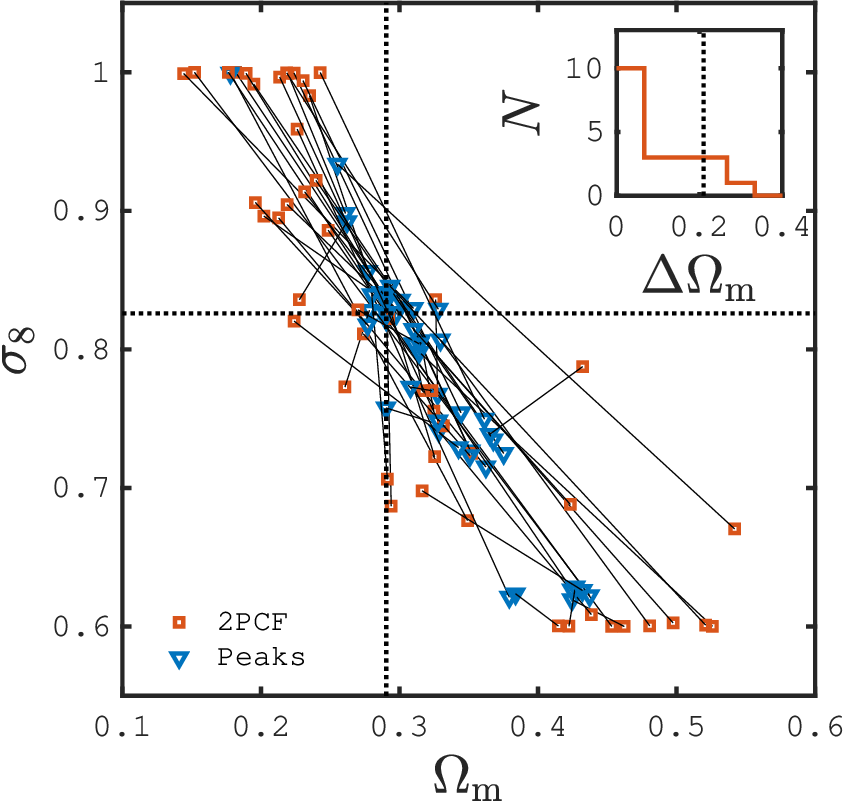}
\caption{Scatter in the best-fit parameters inferred from 50 simulations. The thin black lines link the solutions from peaks statistics (blue) and 2PCFs (orange) analyses associated with the same survey realisation. The distribution of the length of these lines is shown in the inset.}
\label{fig:tension}
\end{center}
\end{figure}

Peak statistics and 2PCFs are  affected differently by the noise in the data, and even though they converge in the noise-free limit (as shown in Fig. \ref{fig:pipeline_validation_wCDM_2pcf_GPR}), it is expected that a given noise realisation will scatter the best-fit cosmological parameters inferred by the two pipelines. We estimate the  level of scatter by comparing the best-fit parameters obtained at the maximum likelihood returned by 50 survey realisations taken from the {\it Covariance training set}. Following the recommendations of \citet{KiDS1000_Joachimi}, we improve the accuracy of the solution by repeating the process multiple times by varying the starting points in the $[\sigma_8 - \Omega_{\rm m}]$ plane and recording only the solution with the lowest $\chi^2$. In this exercise all nuisance parameters are held  to zero, and we include the cross-redshift bins in the peak count data to exacerbate the effect.

We report on Fig.  \ref{fig:tension} the best-fit parameters for the 50 peaks analyses (in blue, including the cross-tomographic terms) and 2PCFs analyses (in orange), and further link the pairs associated with the same simulations.
The inset shows the distribution of the difference in $\Omega_{\rm m}$, which extends beyond 0.3. The  value measured from the data is $\Delta \Omega_{\rm m}=0.21$, which is indicated by the vertical dotted line. In this metric, the difference in matter density inferred by both probes  is common and likely, given the noise levels present in  the DES-Y1 data. Repeating the same exercise for $S_8$ reveals a smaller scatter, with $\Delta S_8 \le 0.2$; the difference of 0.2 observed in the data is therefore a rare event. Excluding the cross-redshift bins significantly increases the width and the contours, which making the observed $\Delta S_8 = 0.3$ more likely.
 
\section{Infusion of Intrinsic Alignment}
\label{sec:syst_IA}

In this paper we estimate the impact of IA on peak count statistics by infusing an alignment between mock galaxies and properties of their host dark matter haloes, based on the method described in \citet{2006MNRAS.371..750H} and \citet{2013MNRAS.436..819J}. Since most of the simulated light-cones used in this paper do not use dark matter haloes to assign galaxy positions, we instead have to rely on a separate simulation suite. We use for this task the KiDS-HOD mock data described in \citet{SLICS},  in which dark matter haloes from the light-cones are populated with an HOD based on a conditional luminosity function, then sub-sampled to match the galaxy redshift distribution of the KiDS-450 data up to $z=1.5$. Since these mocks are slightly denser than the DES-Y1 data, we further downsample them such as to closely match the DES tomographic bins.  These DES-HOD mocks lack some of the very low redshift galaxies at $z<0.2$ and all galaxies with $z>1.5$, but these are all down-weighted by the lensing kernel, leaving the expected lensing signal almost unchanged. We match the mean redshift in each bins to within 0.04, and the galaxy density to within 0.07 gal/armin$^2$, which is sufficient for this exercise. 

A number of physical models are presented in  \citet{2013MNRAS.436..819J}, and in this first study we opted for the simplest possible case: we assume that all central galaxies are early-type red galaxies, and that all satellites are late-type blue galaxy. This is of course inaccurate, but provides a good staring point upon which we can build and improve the model in future work.

The ellipticity of the central galaxies is known to correlate with the shape of the host dark matter halos, in a complicated way that depends on galaxy type, redshift, and possibly merger history \citep[for a review see][]{2015SSRv..193...67K}. The galaxy catalogues that we construct from the DES-HOD mocks contain the inertia matrix of the host dark matter haloes, from which one can compute the eigenvalues and eigenvectors ($\omega_\mu, {\boldsymbol s}_\mu$). The projected ellipses reconstructed from these are described by the symmetric tensor ${\boldsymbol W}$ \citep{2013MNRAS.431..477J}:
\begin{eqnarray}
{\boldsymbol W}^{-1}=\sum_{\mu=1}^3 \frac{ {\boldsymbol s}_{\perp,\mu}\, {\boldsymbol s}_{\perp,\mu^T}}{\omega_{\mu^2}},
\end{eqnarray}
where ${\boldsymbol s}_{\perp,\mu}$ is the eigenvector projected along the line of sight,  and the semi-major axes are given by $\sqrt{1/\omega_\mu}$. Halo ellipticities ${\boldsymbol \epsilon}_h$ can be obtained from:
\begin{eqnarray}
     \epsilon_{h,1} = \frac{W_{1,1} - W_{2,2}}{ W_{1,1} + W_{2,2} +2\sqrt{{\rm det}(W)}}  \\
       \epsilon_{h,2} = \frac{2W_{1,2}}{W_{1,1} + W_{2,2} +2\sqrt{{\rm det}(W)}}.
\end{eqnarray}
Once these are determined, we opted for a  100\% alignment between the halo and the central galaxy ellipticities. This is likely to slightly over-estimate the effect, a deliberate choice that we make when  developing a relatively conservative approach. The absence of scatter between the two, combined with the approximation that all centrals are early-type galaxies both act as to maximise the IA signal in our model. We also find that this model does not work well at high redshift, since the haloes  are not fully relaxed, and their shapes are less well modelled.  We therefore concentrate on the lower redshift bins only.

\begin{figure}
\begin{center}
\includegraphics[width=3.3in]{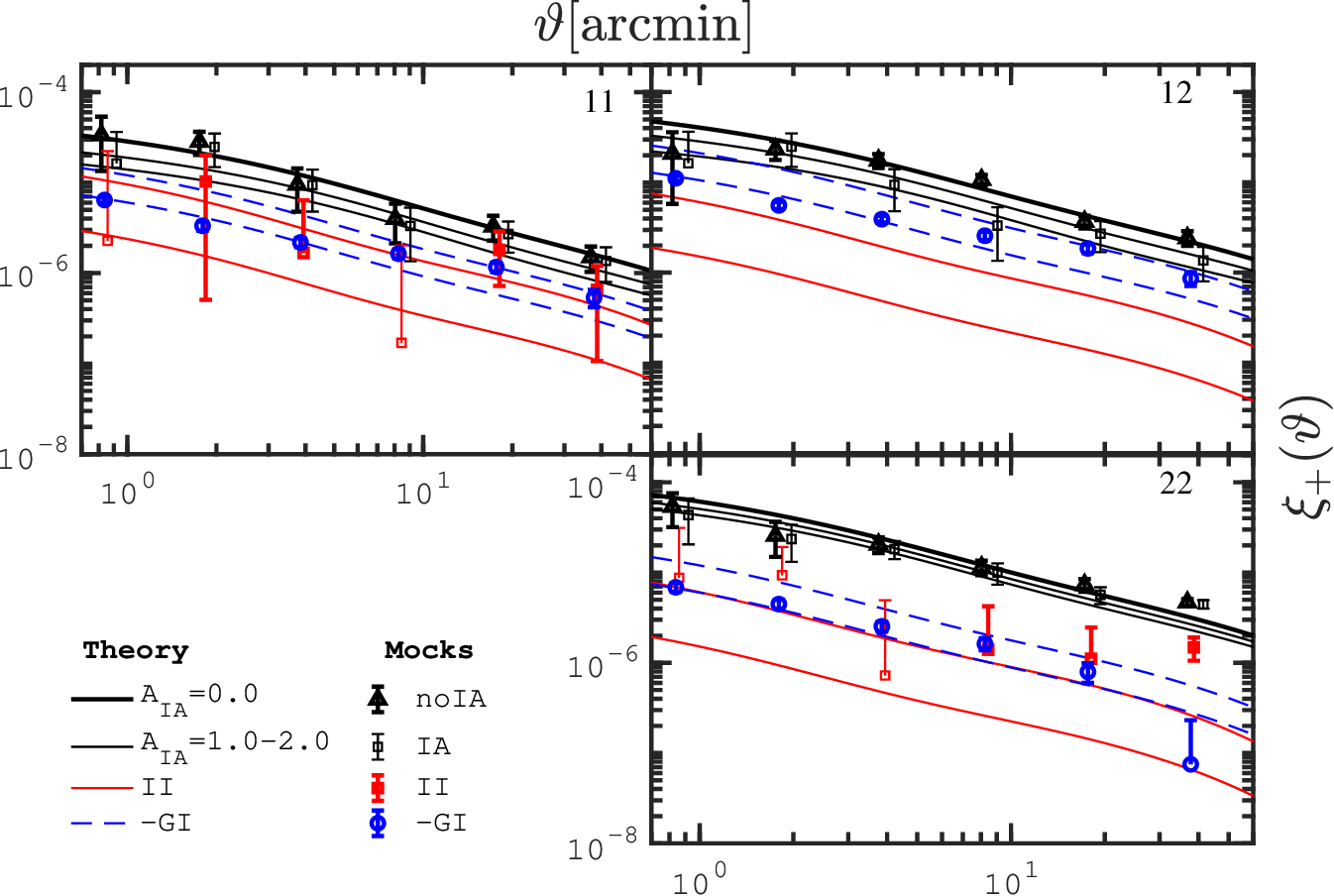}
\caption{Effect of the intrinsic alignment of galaxy on $\xi_{+}$. The upper and lower thin black lines show the  theoretical predictions based on the linear non-linear model of \citet{2007NJPh....9..444B} with $A_{\rm IA}=1.0$ and $2.0$, respectively, while the thick black line presents the predictions without IA. The solid red and the dashed blue lines show the `$II$' and `$-GI$' contributions.
The symbols represent the measurements on the dedicated {\rm IA} mocks. The thin red symbols present the absolute value of the `$II$' measurements, which becomes negative at small scales. Three panels show the combinations between the two lowest redshift bins, as indicated in the top right corner. Combinations involving higher redshifts show a similar agreement between mocks and models, albeit the effect is milder.}
\label{fig:xip_IA}
\end{center}
\end{figure}

We make a second important approximation by treating all satellite galaxies as blue, late-type, and assign them no intrinsic alignment, consistent with recent findings \citep{BlueIA, DESY1_IA_Samuroff, Johnston_IA}. We could assign the halo ellipticities to the centrals directly, but doing so strongly biases the ellipticity distribution to lower values compared with the data. Instead we keep the ellipticities drawn from the Gaussian distribution, and rotate them until they align with ${\boldsymbol \epsilon}_h$. Once this is done, we use Eq. (\ref{eq:eps_obs}) to shear these intrinsic ellipticities, and compute $\xi^{11}_+$, $\xi^{12}_+$ and $\xi^{22}_+$ with ${\boldsymbol \epsilon}$ given either from the IA model described above or from the no-IA case.  We additionally compute  the same 2PCFs statistics from the pure intrinsic shapes ${\boldsymbol \epsilon}_{\rm int}$, thereby estimating the $II$ term, as well as the combination $\langle{\boldsymbol \epsilon}_{\rm int}{\boldsymbol \gamma}\rangle$ to compute the $GI$ term. Results are presented in Fig. \ref{fig:xip_IA}, and compared with the theoretical predictions with $A_{\rm IA}=1.0$ and $2.0$. We see in all three tomographic bin combinations that the mocks reproduce reasonably well the IA, noIA and $GI$ models, however the $II$ terms remains very noisy. 

The IA model infused in these mocks is not accurate enough to be used for signal calibration, but is adequate for diagnostic tests such as those for which they are designed here. Future developments with more flexible options regarding early-types/late-types separation, inclusion of satellite alignment, and possibly different $N$-body runs, will be the subject of future work.

\section{Comparison between $\Lambda$CDM analyses}
\label{subsec:T18_vs_2PCF_vs_J20}

 Differences in signal modelling and in  the likelihood sampling strategy are responsible for the broader range of accepted $\Omega_{\rm m}$ and $\sigma_8$ values in J20, compared to T18. Notably, they replaced the {\sc halofit} predictions of the non-linear  matter spectrum by the halo-based model {\sc HMcode} \citep{MeadFit}. Furthermore, the sampling over the amplitude parameter $A_{\rm s}$ is replaced by a sampling over $\rm{ln}\left(10^{10} A_{\rm s}\right)$, the sum over the neutrino mass is fixed, and the ranges of priors are changed to that of \citet{KV450}. These differences are responsible for a loss of precision on the $S_8$ constraints: while T18 finds  $S_8^{{\rm T18, }\Lambda} = 0.792^{+0.032}_{-0.021}$, J20 reports $S_8^{{\rm J20}} = 0.763^{+0.037}_{-0.031}$, e.g. a $\sim 1\sigma$ shift compared to the original DES-Y1 results, and an error almost 30\% larger. 
 As explained in J20, the shift in $S_8$ is largely driven by differences in the $n(z)$ estimations.

\bsp	
\label{lastpage}
\end{document}